\documentclass[aps,prc,twocolumn,reprint,superscriptaddress]{revtex4}

\usepackage{graphicx} \usepackage{amsmath,amssymb,amsopn}
\usepackage{latexsym} \usepackage{bm} \usepackage{mathrsfs}
\usepackage{mathbbol} \usepackage[dvips]{color}
\usepackage{lmodern}
\usepackage{multirow}
\usepackage{float}

\def\eg{{\em e.g.}}
\def\ie{{\em i.e.}}

\newcommand{\beq}{\begin{equation}}
\newcommand{\eeq}{\end{equation}}
\newcommand{\bea}{\begin{eqnarray}}
\newcommand{\eea}{\end{eqnarray}}
\newcommand{\lsim}{\raisebox{-4pt}{$\,\stackrel{\textstyle <}{\sim}\,$}}
\newcommand{\gsim}{\raisebox{-4pt}{$\,\stackrel{\textstyle >}{\sim}\,$}}

\begin{document}
\title{Color screening and regeneration of bottomonia in high-energy heavy-ion collisions}

\author{X.~Du}
\affiliation{Cyclotron Institute and Department of
Physics and Astronomy, Texas A\&M University, College Station, TX 77843-3366, USA}
\author{M.~He}
\affiliation{Department of Applied Physics, Nanjing University of Science and Technology,
Nanjing 210094, China}
\author{R.~Rapp}
\affiliation{Cyclotron Institute and Department of
Physics and Astronomy, Texas A\&M University, College Station, TX 77843-3366, USA}

\date{\today}

\begin{abstract}
The production of ground-state and excited bottomonia in ultrarelativistic heavy-ion collisions is investigated within a kinetic-rate equation approach
including regeneration. We augment our previous calculations by an improved treatment
of medium effects, with temperature-dependent binding energies and pertinent reaction
rates, $B$-meson resonance states in the equilibrium limit near the hadronization
temperature, and a lattice-QCD based equation of state for the bulk medium.
In addition to the centrality dependence of the bottomonium yields
we compute their transverse-momentum ($p_T$) spectra and elliptic flow with
momentum-dependent reaction rates and a regeneration component based on
$b$-quark spectra from a nonperturbative transport model of heavy-quark
diffusion. The latter has noticeable consequences for the shape of the bottomonium
$p_T$ spectra. We quantify how uncertainties in the various modeling components
affect the predictions for observables. Based on this we argue that the $\Upsilon(1S)$
suppression is a promising observable for mapping out the in-medium properties
of the QCD force, while $\Upsilon(2S)$ production can help to quantify the
role of regeneration from partially thermalized $b$ quarks.
\end{abstract}

\maketitle
\section{Introduction}
\label{sec:intro}
Heavy quarkonia have long been recognized as a promising probe of the modifications
of the fundamental QCD force in hot and dense matter. In vacuum, the potential
between a heavy quark and antiquark is well established in terms of a
short-range Coulombic part and a long-range linear ``confining" part, which
allows for a robust phenomenology of the charmonium and bottomonium bound-state
spectra. This can serve as a controlled starting point for their in-medium
spectroscopy~\cite{Rapp:2008tf,BraunMunzinger:2009ih,Kluberg:2009wc,Mocsy:2013syh}.
Measurements of quarkonia in ultrarelativistic heavy-ion collisions (URHICs)
have much progressed over the last decade, mostly carried out via the dilepton
decay channel of the vector states $J/\psi$, $\psi'$ and the $\Upsilon(1S,2S,3S)$
family, cf.~Refs.~\cite{Rapp:2017chc,Scomparin:2017pno} for recent overviews.
While the observed invariant-mass spectrum does not include significant information
about their in-medium properties (since the vast majority of the dilepton decays
occurs long after the fireball has frozen out), systematic studies of quarkonium
production yields as a function of collision centrality, energy ($\sqrt{s}$), and
transverse momentum ($p_T$) have provided a rich source of information on how their
properties are affected in the presence of a hot QCD medium.

For charmonia, an interplay of thermal suppression and regeneration reactions
throughout the evolution of the fireball formed in nuclear collisions turned
out to provide a suitable framework to describe the observed production patterns
from CERN Super Proton Synchrotron (SPS) energy (0.017\,TeV)~\cite{Ramello:2003ig}, via BNL Relativistic Heavy Ion Collider (RHIC) energy (0.039-0.2\,TeV)~\cite{Adare:2006ns,Adamczyk:2016srz},
to CERN Large Hadron Collider (LHC) energy (2.76\,TeV)~\cite{Abelev:2013ila}, see also ~\cite{Grandchamp:2003uw,Andronic:2006ky,Liu:2009nb,Zhao:2010nk,Zhao:2011cv,Song:2011xi,Ferreiro:2012rq,Zhou:2014kka}.
In particular, the relative enhancement of $J/\psi$ yields when going from RHIC to
the LHC was anticipated as a consequence of regeneration processes which intensify
in the presence of larger charm-quark densities in the system. This interpretation
was corroborated by the measured $p_T$ spectra, which
confirmed the prediction that the regeneration yields are concentrated
at low $p_T\lsim m_\Psi$~\cite{Abelev:2013ila}.

For bottomonia, the role of regeneration processes is less obvious. In
Pb-Pb($\sqrt{s}$=2.76\,TeV) collisions at the LHC, the CMS Collaboration~\cite{cms:2012}
reported a ``sequential suppression" of bottomonia, characterized by an increasing
level of suppression for $\Upsilon(1S)$, $\Upsilon(2S)$ and $\Upsilon(3S)$ states,
following their ordering in vacuum binding energy. These data, along with the inclusive
$\Upsilon$ data from the STAR and PHENIX Collaborations in Au-Au($\sqrt{s}$=0.2\,TeV)
and U-U($\sqrt{s}$=0.193\,TeV) collisions at RHIC~\cite{Adamczyk:2013poh,Adare:2014hje,Adamczyk:2016dzv},
can indeed be reasonably well described by models which do not include regeneration
contributions~\cite{Krouppa:2015yoa,Krouppa:2016jcl,Hoelck:2016tqf}. This is more challenging for
recent ALICE data at forward rapidity, which exhibit stronger
suppression~\cite{Abelev:2014nua} than at mid-rapidity, even though a less dense
medium is expected to form at forward rapidity. Cold-nuclear-matter (CNM) effects may
play a role in this observation, as shadowing effects could be more pronounced at
forward rapidity. Since the typical ratio of $\Upsilon$ relative to total
$b\bar b$ production is only about 0.1\% in elementary $pp$ collisions (compared
to $\sim$1\% for charmonium to total $c\bar c$ production), even small
regeneration yields in URHICs may give a significant contribution to the observed
$\Upsilon$ production~\cite{Grandchamp:2005yw}.
In Ref.~\cite{Emerick:2011xu}, this was quantitatively investigated in a
kinetic-rate-equation framework. On the one hand, it was found that regeneration
contributions in 2.76\,TeV Pb-Pb collisions are moderate for the $\Upsilon(1S)$
state, at a $\sim$20\% level of the total yield in central Pb-Pb collisions
(including feeddown from higher states). On the other hand, with a
strong suppression of primordially produced $\Upsilon(2S)$ states [down to
$\lsim$5\% in central Pb-Pb(2.76\,TeV] collisions), the regeneration yield
emerged as the dominant source for semi-central and central collisions. The calculated
centrality dependence of the nuclear modification factors for both
$\Upsilon(1S)$ and $\Upsilon(2S)$
turned out to be in approximate agreement with the CMS data, provided a
so-called ``strong-binding scenario" (SBS) was employed, where the bottomonium
binding energies were assumed to be at their vacuum values. This was qualitatively
motivated by theoretical scenarios with a heavy-quark (HQ) potential taken as the internal
energy computed in lattice QCD (lQCD)~\cite{Riek:2010fk}. Similar findings where
also reported in other transport approaches~\cite{Krouppa:2015yoa,Zhou:2014hwa}.
The magnitude of the regeneration contribution for the $\Upsilon(1S)$ ,
however, does not suffice to account for the stronger suppression of the ALICE
data at forward rapidity, relative to mid-rapidity. Clearly, the decomposition into primordial and regenerated components
requires further studies. In the meantime, the CMS Collaboration has released
$p_T$ spectra for both $\Upsilon(1S)$ and $\Upsilon(2S)$~\cite{Khachatryan:2016xxp},
providing an excellent opportunity for additional tests and tuning of model
calculations~\cite{Krouppa:2015yoa}.

In the present work we extend our previous calculations of bottomonium kinetics
in the fireballs of URHICs in several respects. For a more realistic treatment of
the in-medium properties of bottomonia we implement in-medium binding energies as
extracted from microscopic $T$-matrix calculations~\cite{Riek:2010fk}. These
affect both the inelastic reaction rates and the equilibrium limit of bottomonium
abundances which figure in the regeneration reactions. The space-time evolution
of the fireball is updated by using a lQCD-based equation of state
(EoS)~\cite{He:2011zx}. We compute the production yields of $\Upsilon(1S,2S,3S)$
states as well as their $p_T$ spectra and the elliptic flow ($v_2$)
based on 3-momentum dependent dissociation rates and $b$-quark spectra for
regeneration processes which are taken from nonperturbative
transport simulations (which give a fair description of open-bottom observables
at the LHC~\cite{He:2014cla}). In contrast to $c$ quarks, $b$-quark spectra
are not expected to reach near thermalization at the RHIC and the LHC,
which has a significant impact on the $p_T$ dependence of bottomonium regeneration.
Since primordial $\Upsilon$ states are not expected to acquire a large $v_2$,
their measured total $v_2$ may provide a greater sensitivity to regeneration
processes than the inclusive yields or even $p_T$ spectra. We also calculate
bottomonium observables for 5.02\,TeV Pb-Pb collisions as recently measured at
the LHC.

Our paper is organized as follows.
In Sec.~\ref{sec:trans} we briefly recall the basic ingredients
of our kinetic-rate-equation approach, with emphasis on
its improvements over previous work~\cite{Emerick:2011xu}.
In particular, we scrutinize various mechanisms in the dissociation rates in
the presence of in-medium effects on the bottomonium binding
energies, improve the $\Upsilon$ equilibrium limits by
accounting for $B$-meson resonance states near $T_{\rm c}$
(Sec.~\ref{ssec:rate-eq}), and replace a massless-gas EoS in the
fireball evolution with a parametrization from lQCD
(Sec.~\ref{ssec:bulk}); we also discuss how we calculate $Y$
$p_T$ spectra and their elliptic flow (Sec.~\ref{ssec:ptspec}), and the
open-bottom and bottomonium input cross sections needed for phenomenology
(Sec.~\ref{ssec:xsec}). In Sec.~\ref{sec:rhic} we start the systematic
comparison of our updated results to available data with Au-Au and U-U
systems at the RHIC including both centrality (Sec.~\ref{ssec:rhic-centrality-tbs})
and $p_T$ dependencies (Sec.~\ref{ssec:rhic-pt-tbs}).
In Sec.~\ref{sec:lhc276} we turn to Pb-Pb(2.76\,TeV) collisions at the LHC,
studying centrality and rapidity dependencies for both the previously
employed SBS (Sec.~\ref{ssec:lhc276-centrality-sbs}) and our
updated approach (Sec.~\ref{ssec:lhc276-centrality-tbs}), conducting
a sensitivity study of model parameters (Sec.~\ref{ssec:lhc276-sensitivity-tbs}),
and then turning to $p_T$ spectra (Sec.~\ref{ssec:lhc276-pt-tbs})
and $v_2$ (Sec.~\ref{ssec:lhc276-v2-tbs}).
In Sec.~\ref{sec:502} we provide predictions for Pb-Pb(5.02\,TeV) collisions,
again contrasting the previous SBS (Sec.~\ref{ssec:lhc502-centrality-sbs})
with the updated approach (Sec.~\ref{ssec:lhc502-tbs}), including comparisons
to recently available data for $\Upsilon(1S)$, $\Upsilon(2S)$ and
$\Upsilon(3S)$ states. In Sec.~\ref{sec:excit} we summarize our
results in terms of an excitation function of the nuclear
modification factor for both $\Upsilon(1S)$ and $\Upsilon(2S)$
in comparison to data from the RHIC and the LHC.
In Sec.~\ref{sec:concl} we summarize and conclude.

\section{Bottomonium Transport in Medium}
\label{sec:trans}
In this work we utilize a kinetic-rate equation~\cite{Grandchamp:2003uw} as our
simulation tool for the time evolution of bottomonium abundances in
URHICs~\cite{Grandchamp:2005yw,Emerick:2011xu}.
We first introduce its basic framework and main transport parameters -- reaction
rate and equilibrium limit -- in Sec.~\ref{ssec:rate-eq}, review the bulk medium
evolution in Sec.~\ref{ssec:bulk}, describe the calculation of the $p_T$ spectra
and elliptic flow of bottomonia in Sec.~\ref{ssec:ptspec}, and summarize our
input cross sections to the rate equation for open bottom and bottomonia as
constrained by $pp$ data in Sec.~\ref{ssec:xsec}.

\subsection{Kinetic rate equation and transport coefficients}
\label{ssec:rate-eq}
The rate equation for a given $Y$ state is characterized by loss and gain terms as
\begin{equation}
\label{rate-eq}
\frac{\mathrm{d}N_{Y}(\tau)}{\mathrm{d}\tau}
=-\Gamma_{Y}(T(\tau))\left[ N_{Y}(\tau)-N_{Y}^{\rm eq}(T(\tau))\right] \ ,
\end{equation}
where the two transport coefficients are the inelastic reaction rate, $\Gamma_{Y}$,
and the equilibrium limit, $N_Y^{\rm eq}$. We include the bottomonium states
$Y=\Upsilon(1S), \Upsilon(2S), \Upsilon(3S), \chi_{b}(1P)$ and $\chi_{b}(2P)$,
where we combine the three states $\chi_{b0,1,2}$ into a single one,
as their vacuum mass splittings are within $\sim$60\,MeV. Since the vacuum binding
energies of most of these states, commonly defined as $E_B^Y = 2m_B-m_Y$, are
significantly larger than the pseudo-critical QCD transition temperature, $T_{\rm pc}$,
we neglect inelastic reactions in the hadronic phase (they may become important
for $E_B^Y\lsim T_{\rm pc}$, \ie, for the $\Upsilon(3S)$, and $\chi_{b}(2P)$,
similarly to the $\psi'$~\cite{Du:2015wha}) and focus on the kinetics in the quark-gluon
plasma (QGP) down to a (pseudo-) critical temperature of $T_{\rm pc}$=170\,MeV.

\subsubsection{In-medium binding energies and dissociation rates}
\label{sssec_rates}
\begin{figure}[!t]
        \centering
        \includegraphics[width=0.48\textwidth]{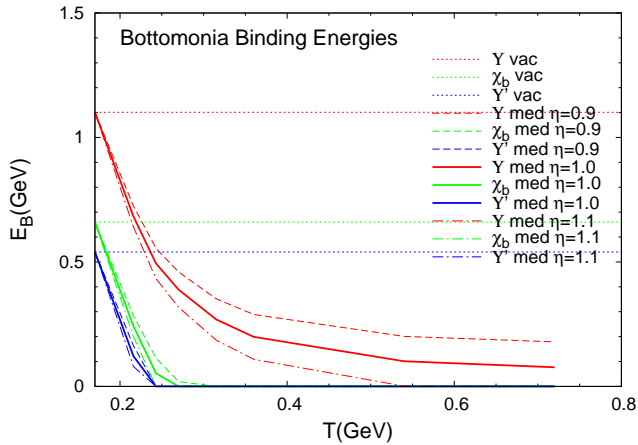}
\caption{Bottomonium binding energies for the SBS (vacuum $E_B$'s, dotted
lines)~\cite{Emerick:2011xu} and $T$-matrix binding scenarios (TBS) with
baseline value~\cite{Riek:2010fk} $\eta$=1.0 (solid lines), and a 10\%
smaller ($\eta$=0.9, dashed lines) or larger ($\eta$=1.1, dash-dotted lines)
reduction in $\Delta E_B(T)$; cf.~Eq.~(\ref{eta}). The red, green, and
blue lines are for $\Upsilon(1S)$, $\chi_b$, and $\Upsilon(2S)$ states,
respectively.}
        \label{fig_EB}
\end{figure}
The nature of the quarkonium dissociation rate in the QGP depends on the
interplay of bound-state scales (\eg, size and binding energy) and medium
scales (\eg, screening length (inverse Debye mass) and
temperature)~\cite{Grandchamp:2001pf,Grandchamp:2005yw,Brambilla:2008cx}.
Our starting point is different scenarios for the in-medium binding energies,
$E_B^Y(T)$, of the various bottomonium states. One may consider this as
fundamental information that one would like to extract from the experimental
data. This also includes the ``melting" temperatures at which the
states cease to exist, which generally do not coincide with a vanishing
binding energy due to finite dissociation widths. However, the latter already
affect the yields at temperatures (well) below the melting temperature, while
the binding energies affect the dissociation mechanisms.
In previous work~\cite{Grandchamp:2005yw,Emerick:2011xu}, the binding energies
were bracketed by a strong-binding scenario (SBS), where the vacuum
binding was simply assumed at all temperatures, and a weak-binding scenario (WBS),
which was based on a screened Cornell potential~\cite{Karsch:1987pv}
with a perturbative screening mass, $m_D\sim gT$ (see Fig.~3 in
Ref.~\cite{Grandchamp:2005yw} or Fig.~1 in Ref.~\cite{Emerick:2011xu}).
These scenarios were coupled with appropriate dissociation mechanisms, \ie,
gluo-dissociation ($g+Y\to b+\bar{b}$) for the SBS and quasifree dissociation
($p+Y\to b+\bar{b}+p$ with $p=q,\bar{q},g$) for the WBS.

In the present work we instead adopt in-medium binding energies predicted by
thermodynamic $T$-matrix calculations~\cite{Riek:2010fk} using internal-energy
potentials, $U_{\bar QQ}$, from lQCD. This choice for the underlying potential is
motivated by a better agreement with quarkonium correlators and charmonium
phenomenology~\cite{Zhao:2010nk} compared to more weakly coupled scenarios (such
as the free energy, $F_{\bar QQ}$), and also by yielding a much smaller (\ie,
more strongly coupled) heavy-quark diffusion coefficient which is preferred
by open heavy-flavor phenomenology~\cite{He:2014cla}. More rigorous determinations
of the in-medium potential are underway~\cite{Liu:2015ypa} and will be investigated
in future work.
We denote the $T$-matrix binding scenario by TBS, and replot
the temperature-dependent ground state binding energy by the red solid line
in Fig.~\ref{fig_EB}, as extracted from Fig.~27 left in Ref.~\cite{Riek:2010fk}.
We implement this together with the assumption of $Y$ bound-state masses fixed
at their vacuum values. This allows us to extract the in-medium $b$-quark mass
from the relation
\begin{equation}
m_{\Upsilon(1S)}=2m_b(T)- E_B^{\Upsilon(1S)}(T) \ ,
\label{m_ups}
\end{equation}
and subsequently use this expression to infer the binding energies,
$E_B^{Y}(T)$, of the excited states, which are also shown in Fig.~\ref{fig_EB}.
The use of internal energies from different lQCD computations induces uncertainties
of a few tens of percent in the $T$-matrix calculations of $E_B^{\Upsilon(1S)}(T)$.
To account for this, we will also allow for two scenarios where the in-medium reduction
of the $\Upsilon(1S)$ binding energy, $\Delta E_B(T) = E_B^{\rm vac} - E_B(T)$, is
decreased (increased) by 10\%, i.e.,
\begin{equation}
E_{B}^{\eta}(T)\equiv E_B^{\rm vac}-\eta \Delta E_B(T)
\label{eta}
\end{equation}
with $\eta$=0.9 ($\eta$=1.1).  This scenario is shown by the dashed (dash-dotted) lines in Fig.~\ref{fig_EB}.
In principle, one could consider $\eta$ as a parameter to be extracted from a best
fit to data. It turns out that the baseline TBS ($\eta$=1.0) transitions
from the SBS close to $T_{\rm pc}$ to the WBS at temperatures above $T\simeq350$\,MeV, where
the binding energies of the excited states have vanished and the ground-state binding
has dropped to about 200\,MeV.

\begin{figure}[!t]
        \centering
        \includegraphics[width=0.48\textwidth]{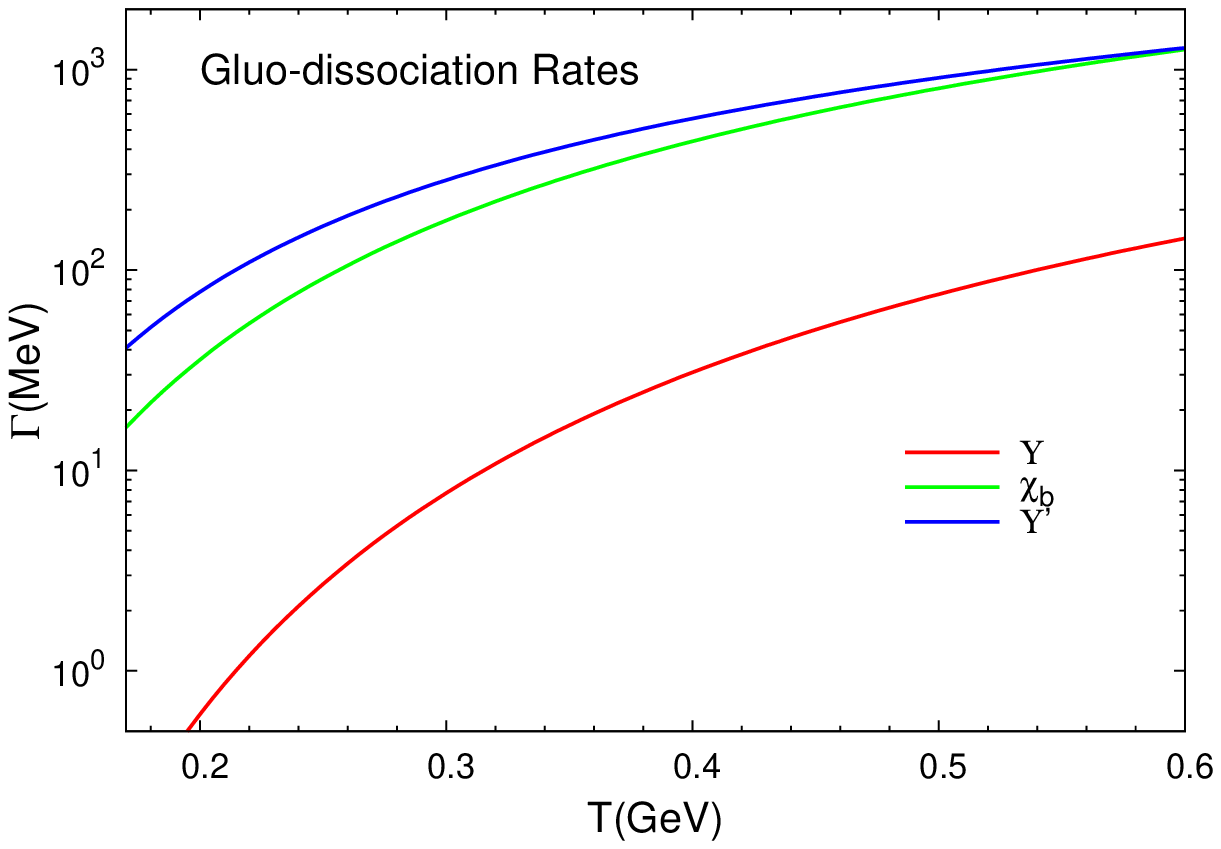}
        \includegraphics[width=0.48\textwidth]{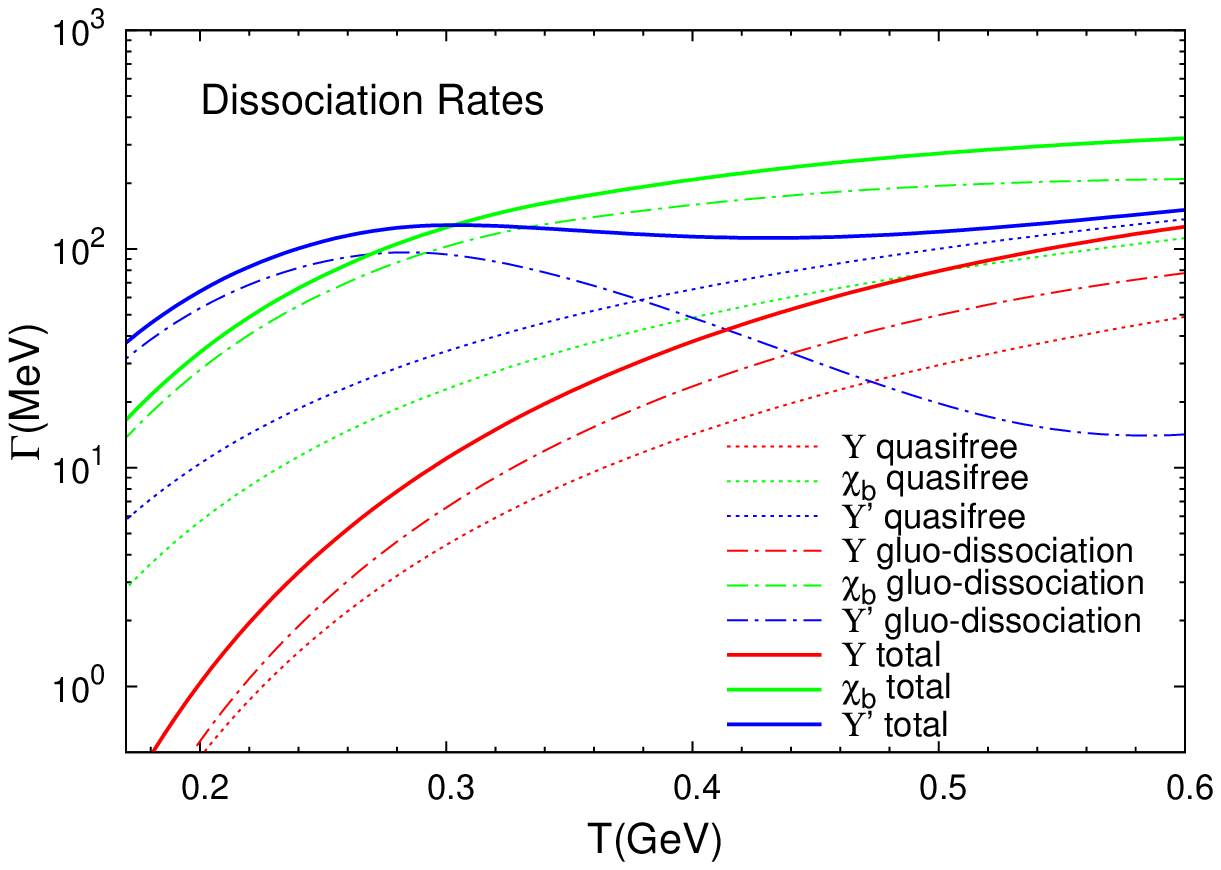}
        \includegraphics[width=0.48\textwidth]{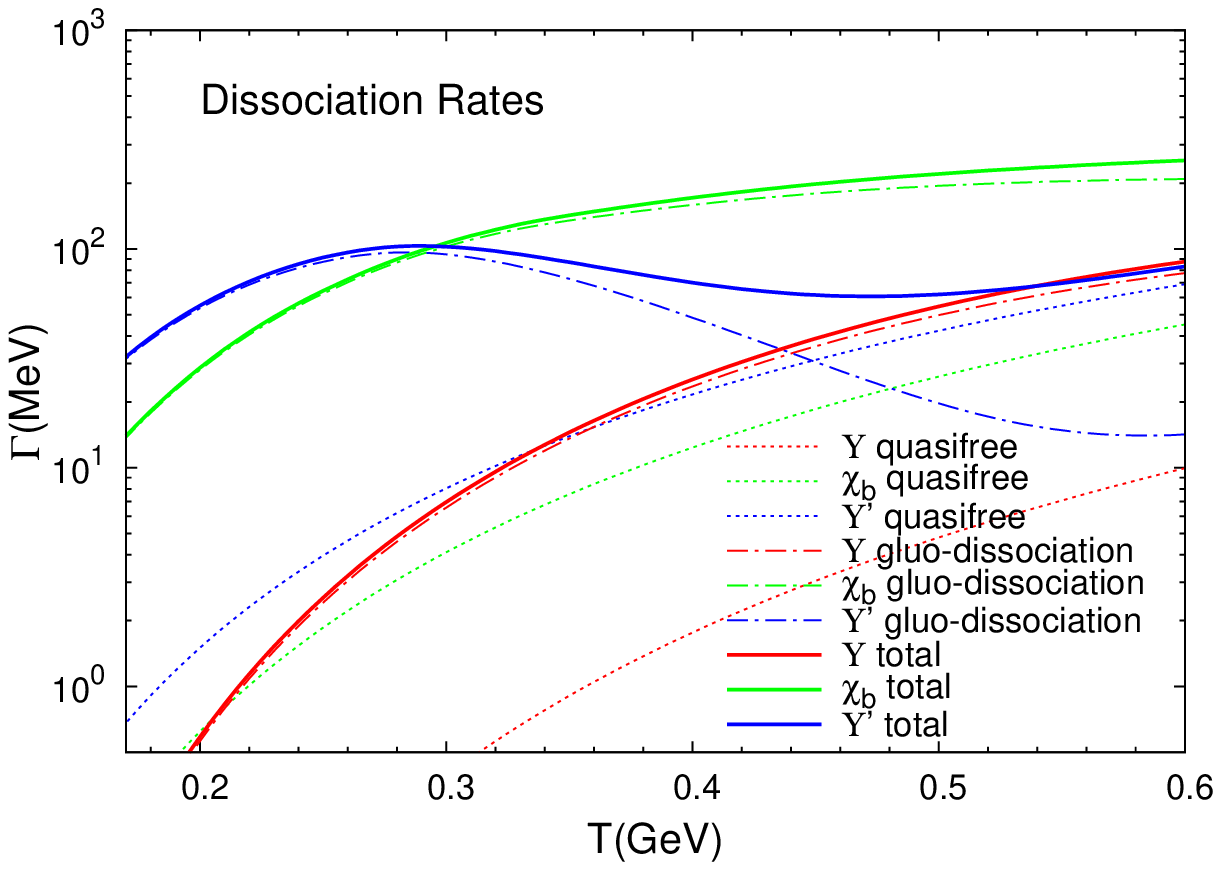}
        \caption{Bottomonium dissociation rates in QGP for the SBS using gluo-dissociation
with massless (upper panel) or massive gluons (dash-dotted lines in the middle and lower panel).
The middle and lower panels also show the rates from inelastic ``quasifree" scattering off
massive quarks and gluons (dotted lines) without (middle panel) and with (lower panel)
interference corrections, and their sum with massive gluo-dissociation rates (solid lines).
All rates are evaluated at zero $Y$ 3-momentum with a strong coupling constant of $g$=2.0.}
        \label{fig_gam-sbs}
\end{figure}

Next we turn to the bottomonium dissociation rates, starting with gluo-dissociation
for $Y+g\to b\bar b$ given by~\cite{Peskin:1979}
\begin{equation}
\Gamma_{Y}^{\rm gd}(p_{\Upsilon},T) =
\int\frac{\mathrm{d}^3p_{g}}{(2\pi)^3}d_{g}f_{g}(\omega_{g},T){v}_{\rm rel}
\sigma_{Yg\rightarrow b\bar{b}}(s) \ .
\label{eq-gluodissociation}
\end{equation}
Here, $f_g(\omega_{g},T)=(\mathrm{exp}\left(\frac{\omega_{g}}{T}\right)-1)^{-1}$ is
the Bose distribution of gluons (with degeneracy $d_g$=16), $s=(p^{(4)}+p_g^{(4)})^2$, and
\begin{equation}
v_{\rm rel}=\frac{\sqrt{(p_{1}^{(4)}\cdot p_{2}^{(4)})^2-m_{1}^2m_{2}^2}}{\omega_{1}\omega_{2}}
\end{equation}
is the relative velocity of incoming particles. The gluo-dissociation cross sections
for the different $Y$ states are detailed in Appendix~\ref{app1}. These rates have been utilized
within the SBS in a heat bath of massless partons in Ref.~\cite{Emerick:2011xu} and are
reproduced in the upper panel of Fig.~\ref{fig_gam-sbs}. They are quite large, especially
for the excited states, and were found to be compatible with the strong suppression of
the $\Upsilon(2S)$ observed at the LHC. However, massless partons overestimate the EoS
at given temperature, especially near $T_{\rm pc}$. Here we implement thermal gluon masses,
$m_{g}=\sqrt{1/2(1+N_f/6)}gT$, which suppress the rates not only for low temperatures
(where $m_{g} < E_B$), but even more so once the gluon mass becomes comparable to the
binding energy, For example, at $T$=300\,MeV, the rates
for the excited states are suppressed by around a factor of 2; see the dash-dotted lines
in the middle (or lower) panel of Fig.~\ref{fig_gam-sbs}.

In addition to gluo-dissociation, next-to-leading order inelastic parton scattering,
$p+Y\to b+\bar{b}+p$ with $p=q,\bar{q},g$, can suppress (or regenerate) $Y$ bound states.
The pertinent rate reads
\begin{equation}
\Gamma_{Y}^{\rm qf}(p,T)=\sum\limits_p\int\frac{\mathrm{d}^3p_{p}}{(2\pi)^3}d_{p}f_{p}(\omega_{p},T)
{v}_{\rm rel} \sigma_{Yp\rightarrow b\bar{b}p}(s) \ ,
\label{eq-quasifree}
\end{equation}
where $f_p$ is the Fermi or Bose distribution for $p$=$q$,$\bar q$ or $g$.
In previous work we have treated inelastic parton dissociation in ``quasifree" (qf)
approximation, applicable for weakly bound states, where the recoil of the spectator
heavy quark or antiquark is neglected while conserving 4-momentum~\cite{Grandchamp:2001pf}.
For binding energies comparable to, or larger than, the temperature sizable
corrections are expected due to interference effects between the parton
scattering off the heavy quark and antiquark~\cite{Laine:2006ns,Park:2007zza}.
In particular, in the limit of small bound-state size, $r\to0$, the width vanishes
since the colored medium parton does not resolve the color-neutral $Y$
configuration anymore.
These corrections amount to an interference factor $(1-{\rm e}^{i\vec{q}\cdot\vec{r}})$
in the expression for the width, where $\vec{q}$ is the 3-momentum of the exchanged
gluon. We implement the interference factor into the previously used quasifree
width expression with the identification $\vec{q}^{\,2}\simeq -t$.
The resulting $Y$ widths for inelastic scattering off massive partons without and with interference
correction are shown by the dotted lines in the middle and lower panels of Fig.~\ref{fig_gam-sbs},
respectively. As expected, for the SBS the interference effects give large corrections,
suppressing the rates by typically a factor of around 5 (more/less at low/high temperature).
The massive quasifree rates are generally well below the massive gluo-dissociation rates,
except for the $Y(2S)$ for $T\gsim450$\,MeV.

\begin{figure}[!t]
        \centering
        \includegraphics[width=0.48\textwidth]{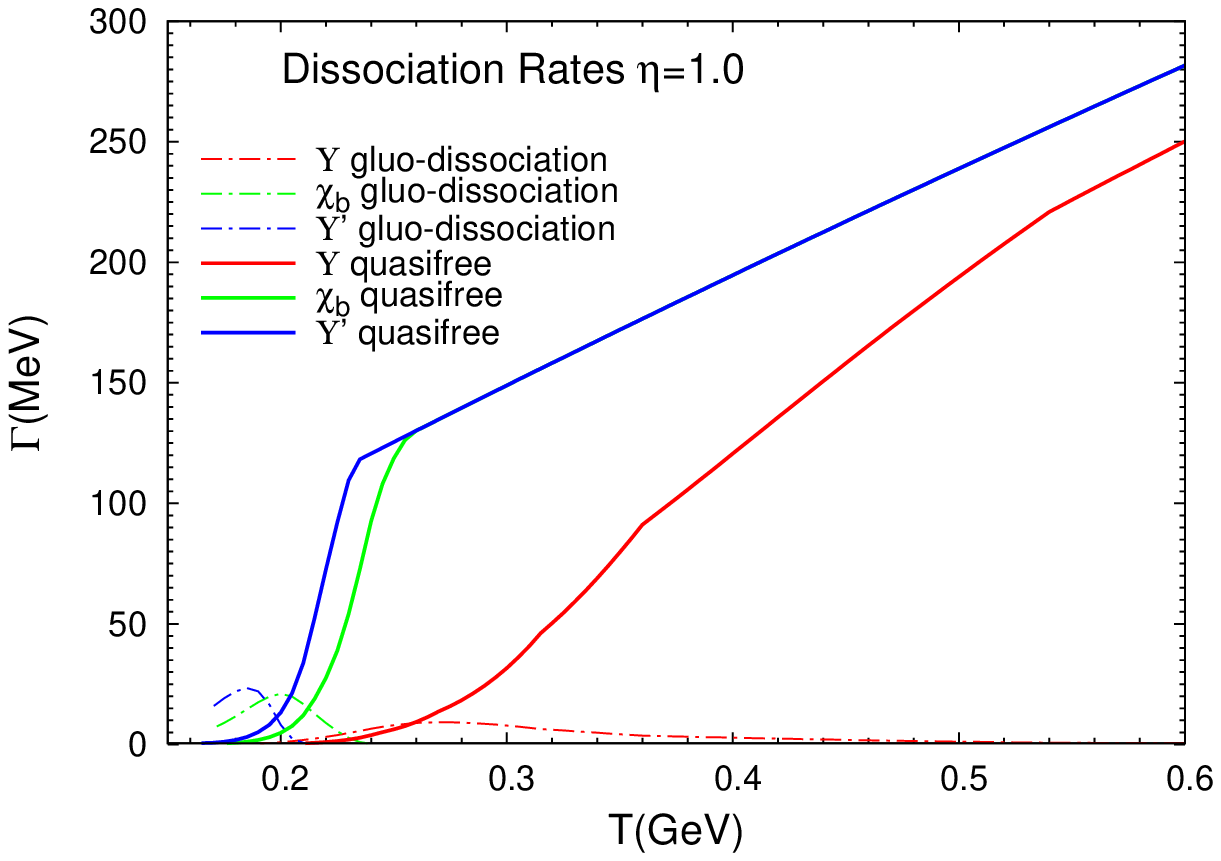}
        \includegraphics[width=0.48\textwidth]{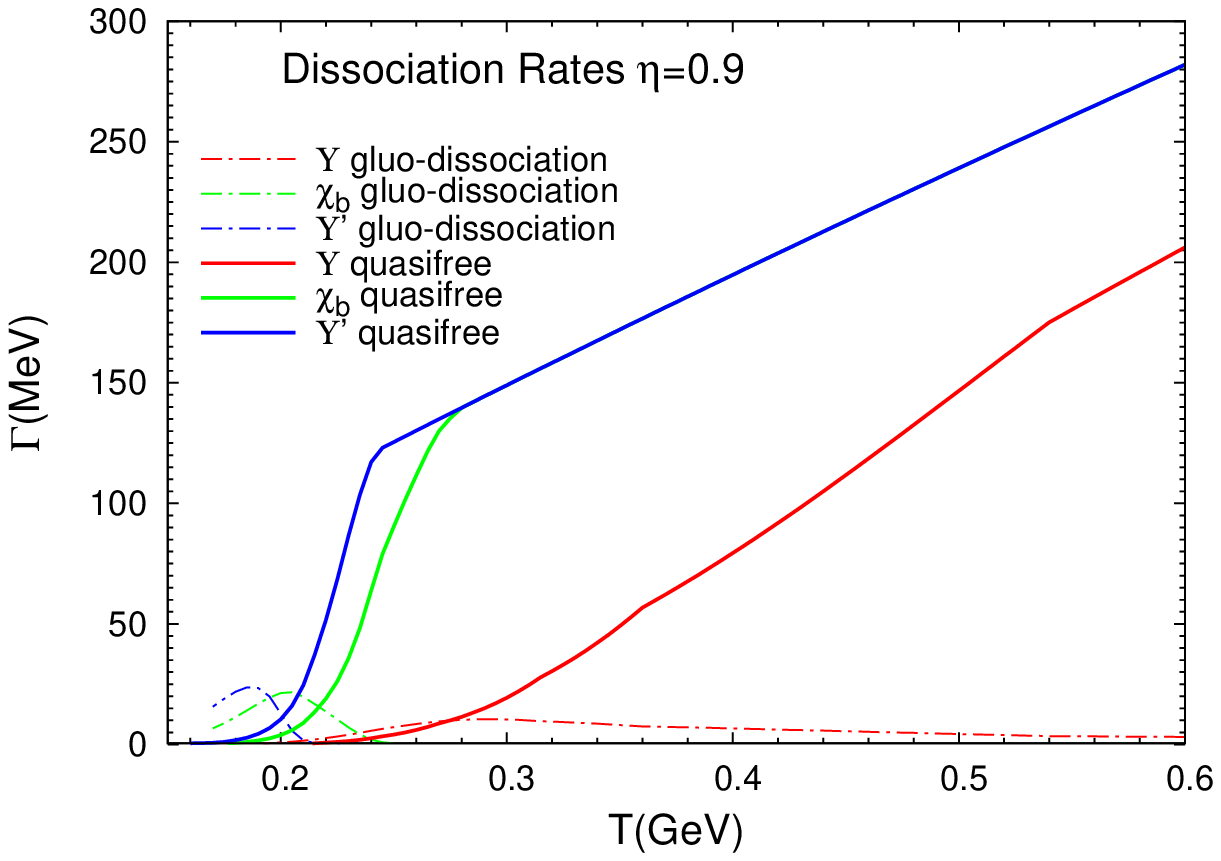}
        \includegraphics[width=0.48\textwidth]{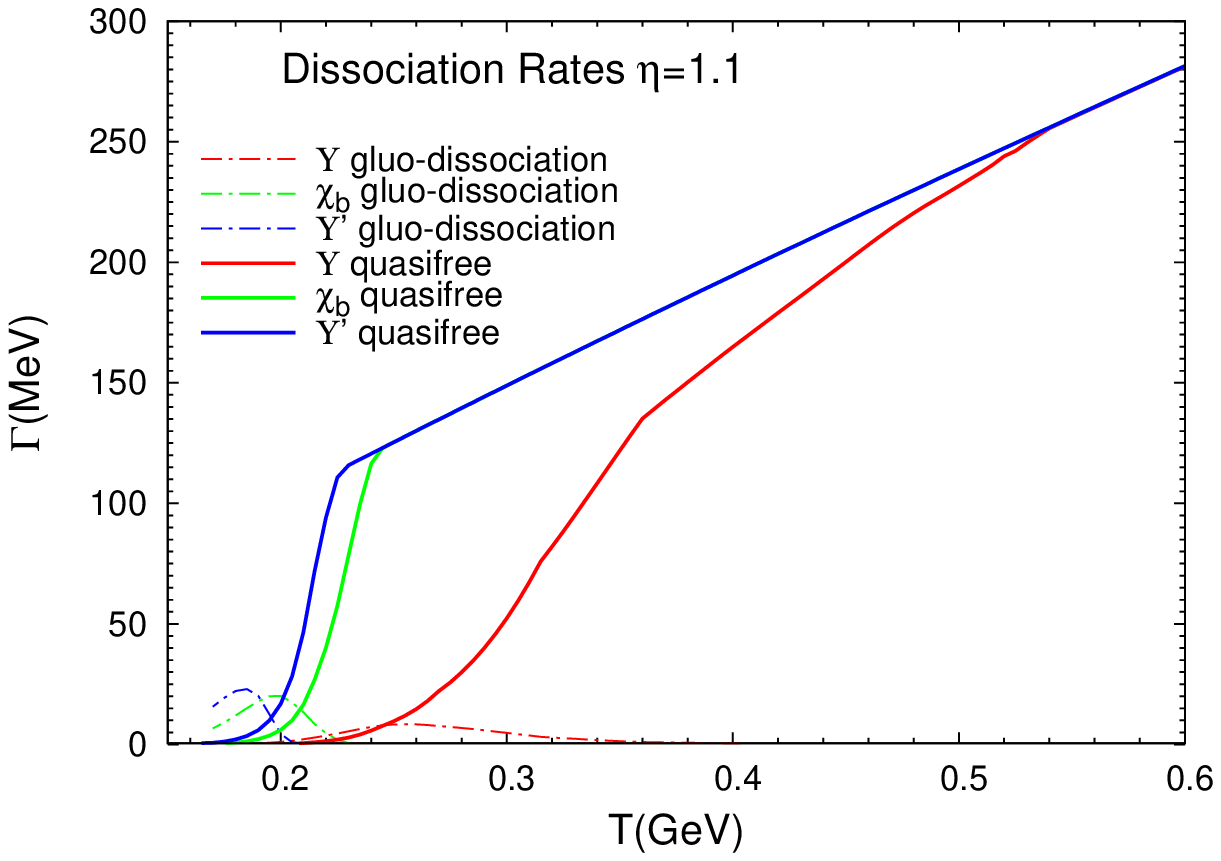}
\caption{Bottomonium dissociation rates for the in-medium $T$-matrix binding
scenario (TBS) in a massive thermal parton gas. Upper panel: baseline TBS (with
$\eta$=1.0 in Fig.~\ref{fig_EB}); middle (lower) panel: TBS with increased (decreased) binding energies
$\eta$=0.9 ($\eta$=1.1); note that $\eta$=0 recovers the SBS.
The dash-dotted and solid lines correspond to gluo-dissociation and inelastic
parton scattering, respectively, while red, green, and blue colors represent $\Upsilon(1S)$,
$\chi_b(1P)$ and $\Upsilon(2S)$ states, respectively. Dissociation rates are evaluated
at bottomonium 3-momentum $p$=0. Interference corrections are included in the
quasifree inelastic parton scattering.}
   \label{fig_TQFGD_TBS}
\end{figure}

\begin{figure}[!t]
        \centering
      \includegraphics[width=0.48\textwidth]{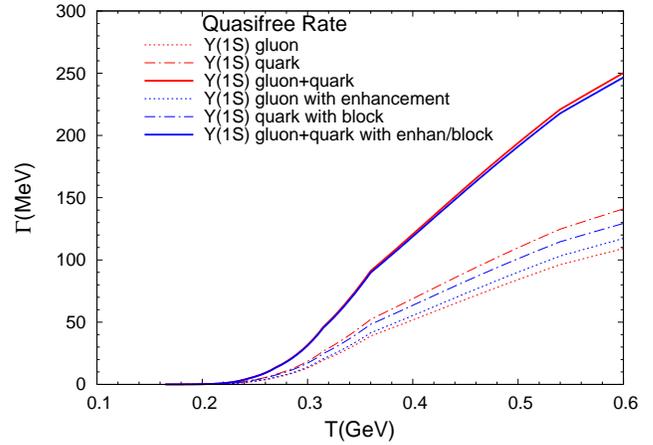}
\caption{Parton-induced quasifree dissociation rates (solid lines) for the baseline in-medium
$T$-matrix binding scenario (TBS with $\eta$=1.0) in a massive thermal parton gas with
(blue lines) or without (red lines) final-state Fermi blocking and Bose enhancement factors
for quarks (dash-double-dotted lines) and gluons (dotted lines), respectively. The rates are
evaluated at $p$=0 for $\Upsilon(1S)$. Interference corrections are included.}
   \label{fig_BLOCKENHANCE_TBS}
\end{figure}

Our final scenario implements in-medium binding energies based on $T$-matrix calculations
(TBS) of Ref.~\cite{Riek:2010fk}.
The reduced binding energies entail a substantial increase of the quasifree rates over the SBS,
especially for the $\Upsilon(1S)$. Within the TBS, the latter shows significant sensitivity to
the in-medium binding energy. For example, at $T$=350\,MeV, when going from the $\eta$=1.0 baseline
scenario to $\eta=0.9$ ($\eta=1.1$), where the binding energy varies from $\sim$200\,MeV in
the former to $\sim$300\,MeV ($\sim$100\,MeV) in the latter, the width decreases (increases)
by about 25\% (50\%), from 80 to 60\,MeV (120\,MeV), and
similarly at other temperatures, see Fig.~\ref{fig_TQFGD_TBS}. Variations in the already small
binding energy of the excited states have rather little impact on their rates. Furthermore,
the gluo-dissociation mechanism in the TBS is only relevant in a small temperature window
above $T_{\rm pc}$.

We note that the Bose-enhancement/Pauli-blocking factors, (1$\pm$$f_p$),  of the outgoing light partons
in the quasifree reaction rate, Eq.~(\ref{eq-quasifree}), have been neglected. Their effect
is an increase/decrease of the rate for outgoing gluons/quarks by less than 10\%, respectively,
which essentially cancel each other in the sum; see Fig.~\ref{fig_BLOCKENHANCE_TBS}.

\begin{figure}[!t]
\includegraphics[width=0.48\textwidth]{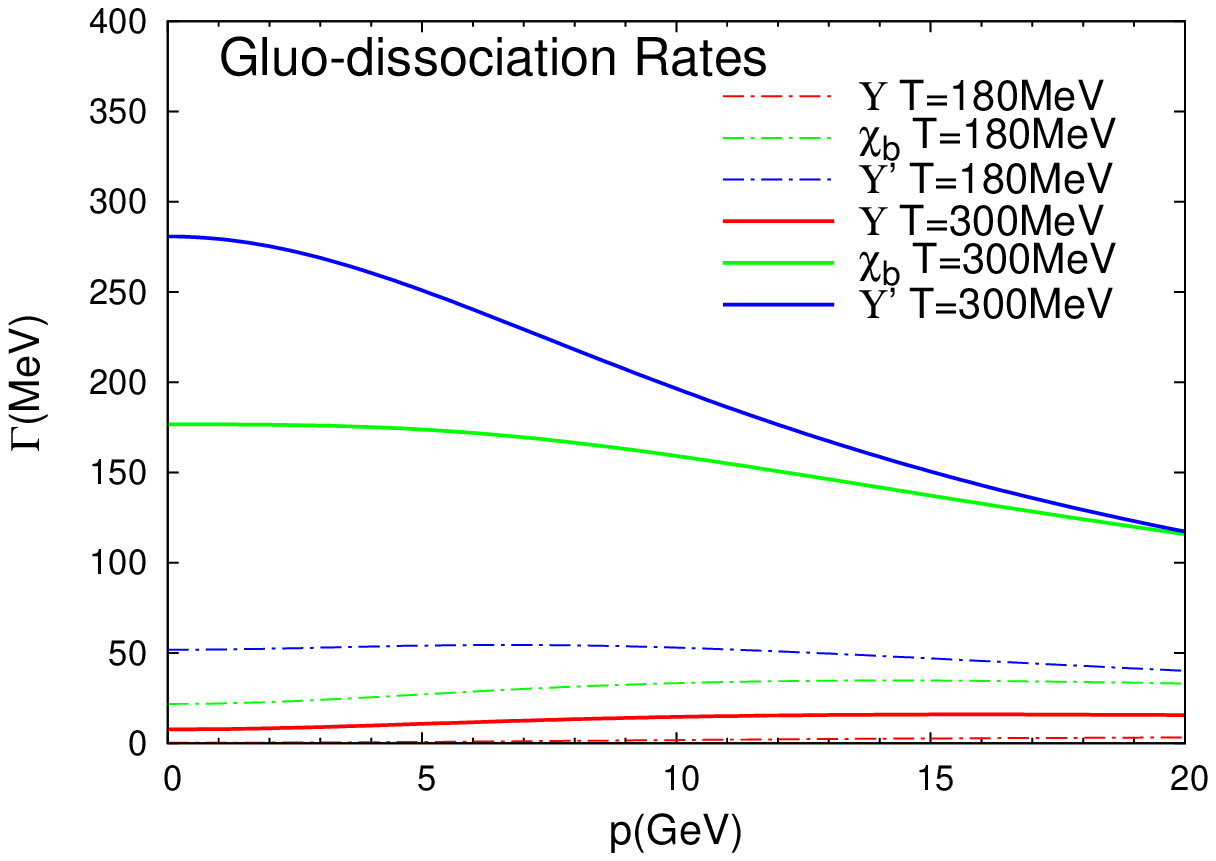}
\includegraphics[width=0.48\textwidth]{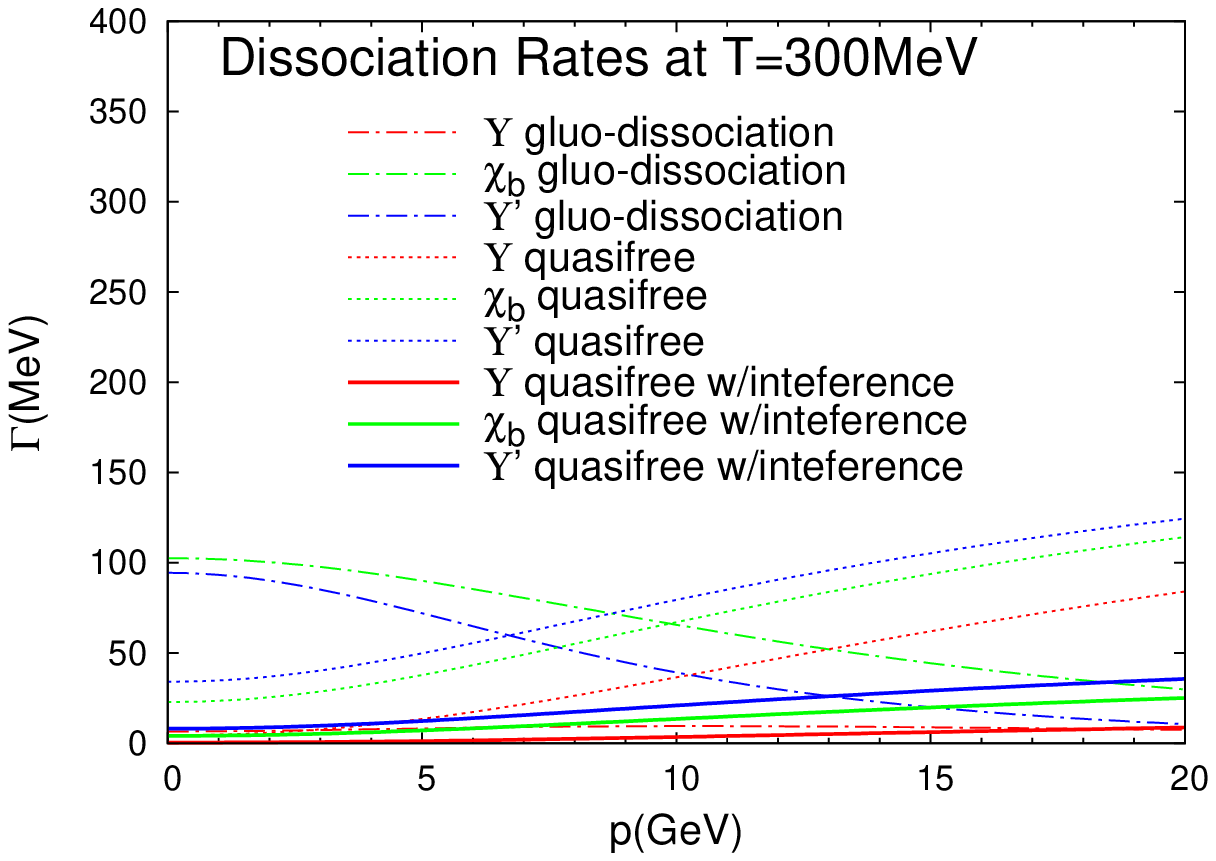}
\caption{Three-momentum dependence of bottomonium dissociation rates in the SBS (vacuum
binding). Upper panel: gluo-dissociation with massless partons at $T$=180\,MeV
(dash-dotted lines) and $T$=300\,MeV (solid lines).
Lower panel: gluo-dissociation (dash-dotted lines) and inelastic parton scattering
(solid lines) for massive partons, at a temperature of $T$=300\,MeV.
In both panels red, green and blue colors correspond to  $\Upsilon(1S)$, $\chi_b(1P)$, and
$\Upsilon(2S)$ states, respectively.}
\label{fig_ptQFGD_SBS}
\end{figure}
\begin{figure}[!t]
\includegraphics[width=0.48\textwidth]{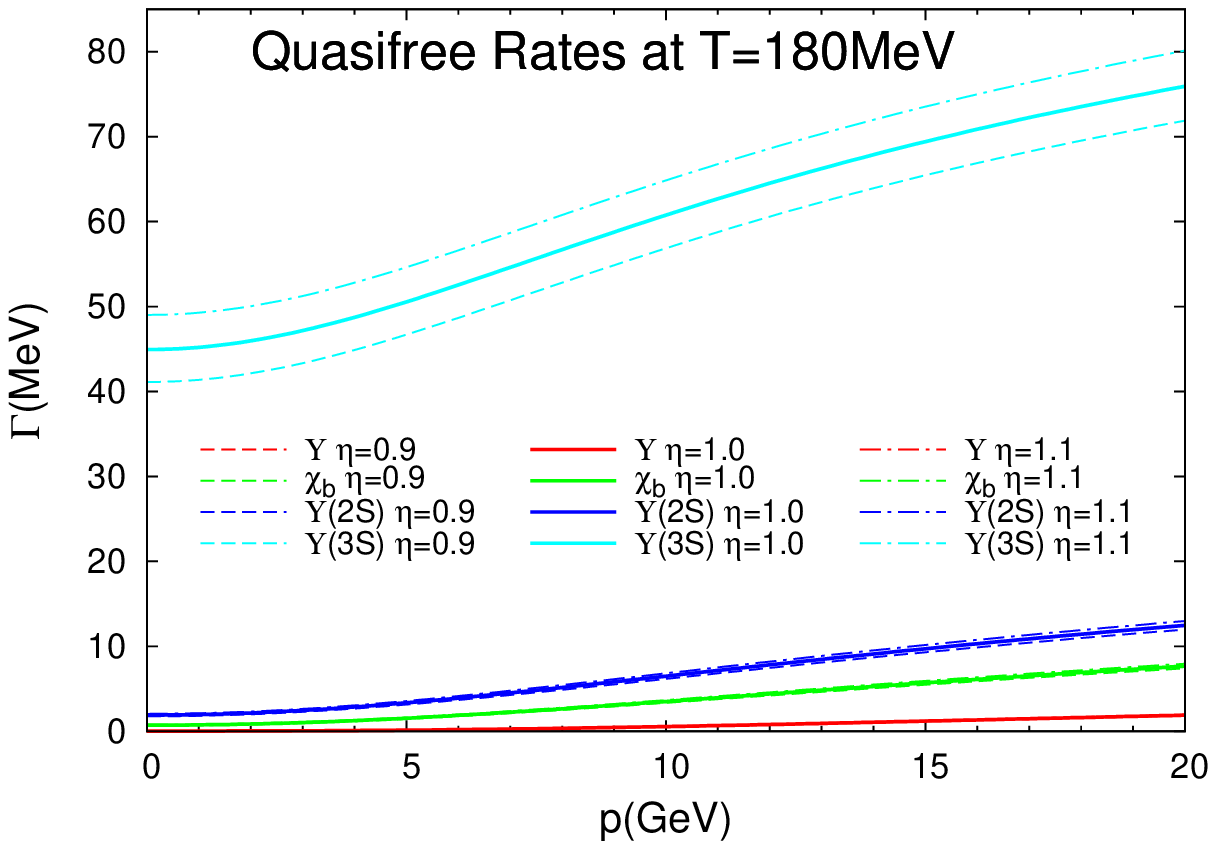}
\includegraphics[width=0.48\textwidth]{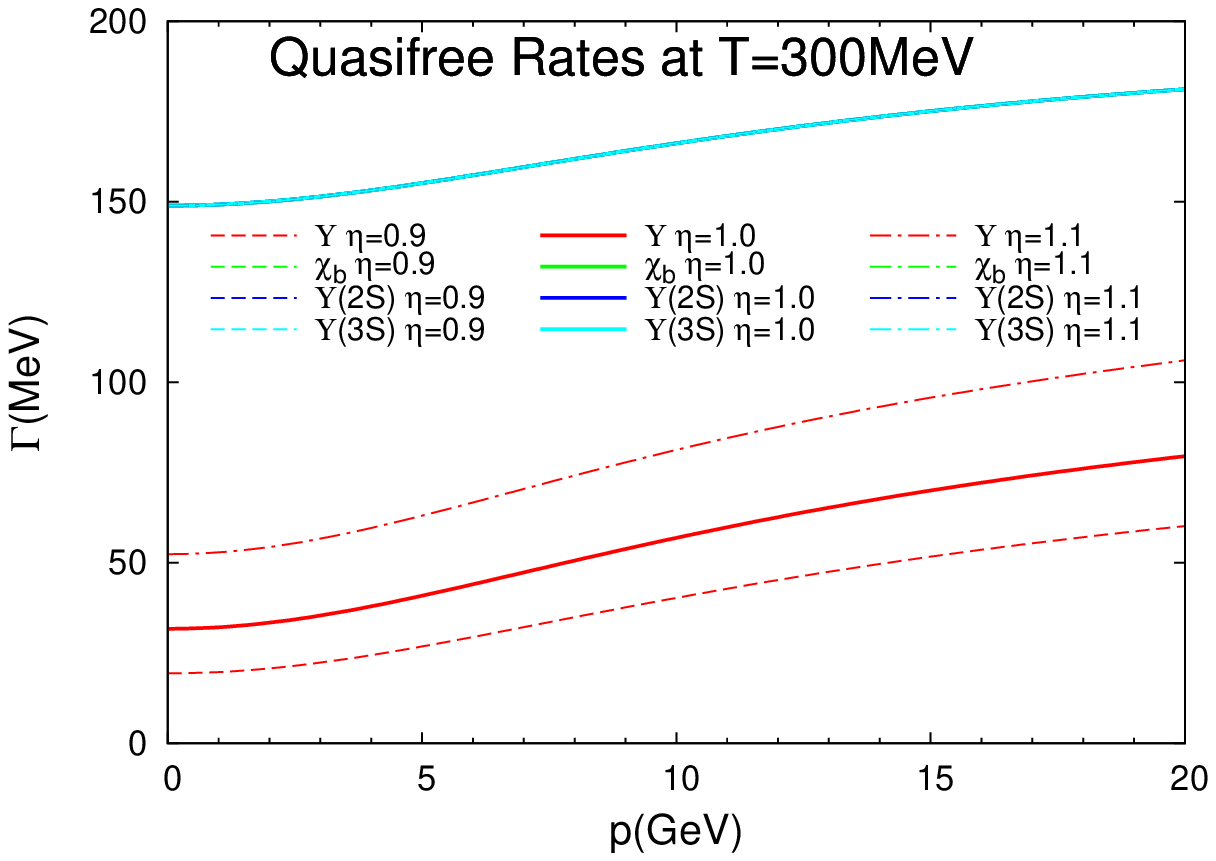}
\caption{Three-momentum dependence of bottomonium dissociation rates from inelastic
massive-parton scattering in the TBS for $T$=180\,MeV (upper panel) and $T$=300\,MeV
(lower panel). The solid and dash-dotted lines correspond to the baseline TBS ($\eta$=1.0)
and an increased (decreased) binding with $\eta$=0.9 ($\eta$=1.1), respectively. The red, green and blue lines
correspond to the $\Upsilon$, $\chi_b(1P)$, and $\Upsilon(2S)$ states, respectively.}
\label{fig_ptQF_TBS}
\end{figure}
\begin{figure}[!t]
\includegraphics[width=0.48\textwidth]{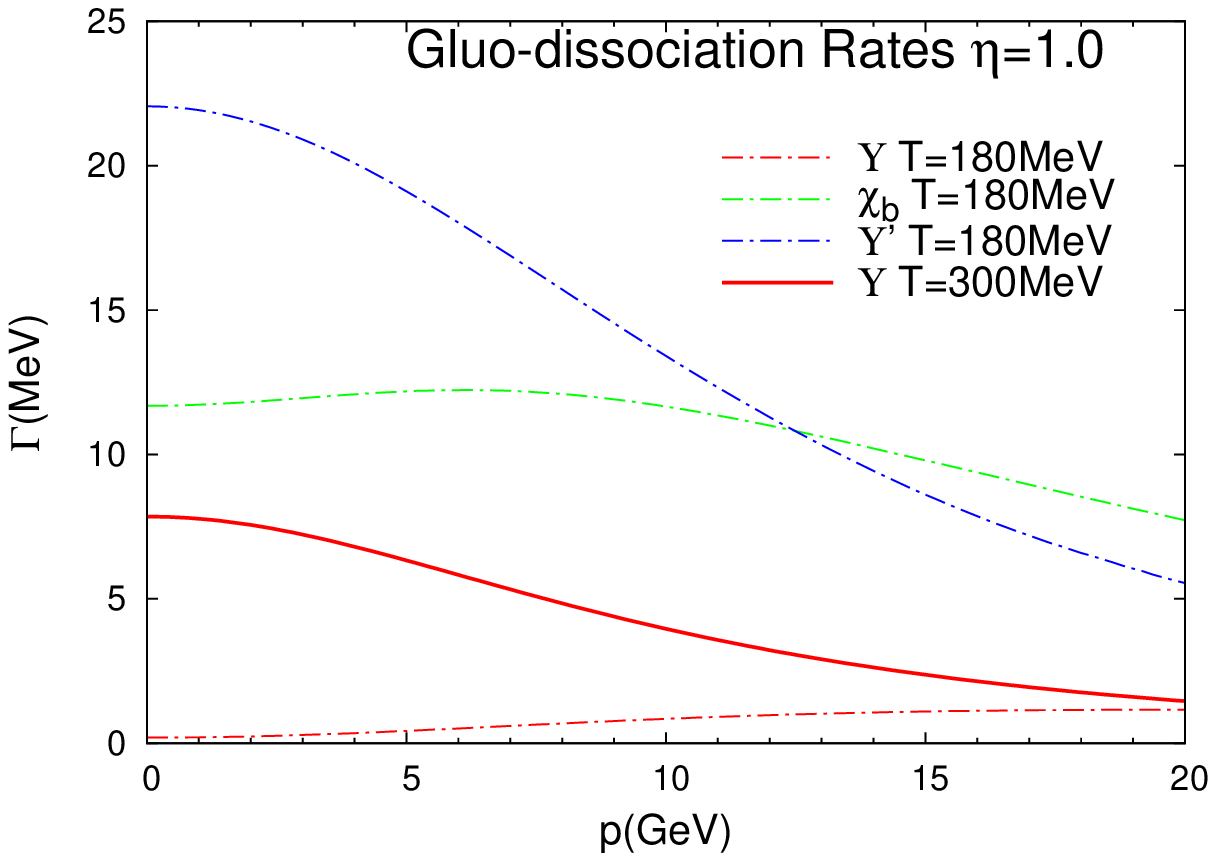}
\includegraphics[width=0.48\textwidth]{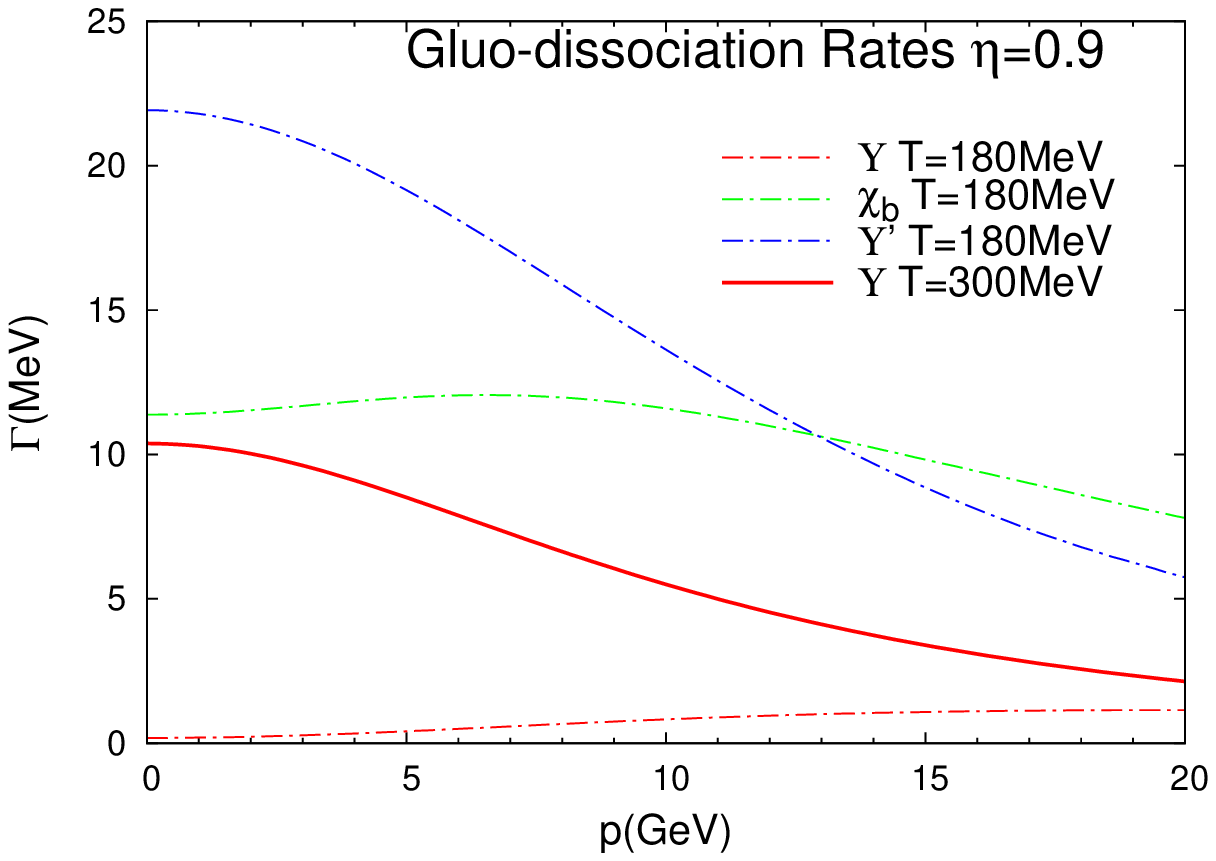}
\includegraphics[width=0.48\textwidth]{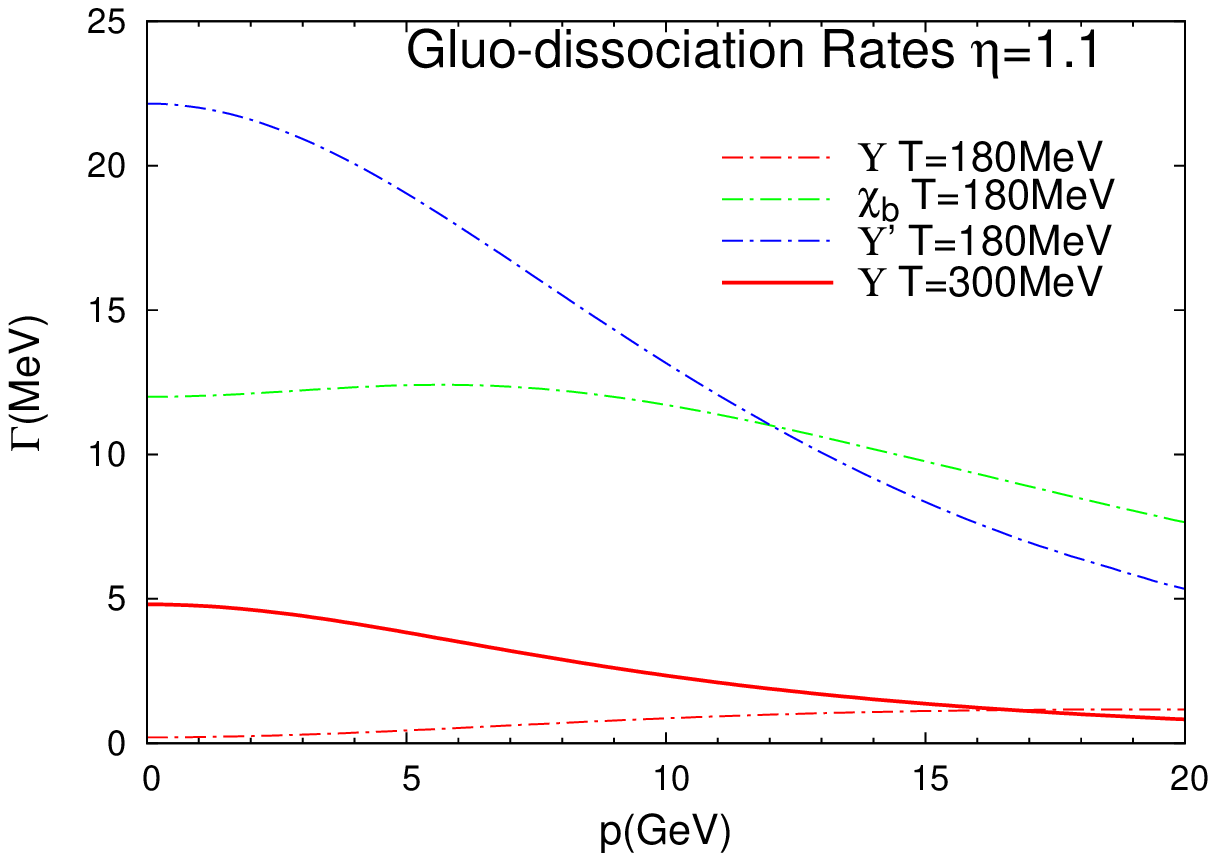}
\caption{Three-momentum dependence of bottomonium rates from gluo-dissociation in the
TBS for the baseline scenario ($\eta$=1.0, upper panel) and increased (decreased) binding $\eta$=0.9, middle panel ($\eta$=1.1,
lower panel) at $T$=180\,MeV (dash-dotted lines) and $T$=300\,MeV (solid lines).
The red, green and blue lines correspond to the $\Upsilon$, $\chi_b(1P)$, and $\Upsilon(2S)$ states, respectively.}
\label{fig_ptGD_TBS}
\end{figure}

In Figs.~\ref{fig_ptQFGD_SBS}, \ref{fig_ptQF_TBS}  and \ref{fig_ptGD_TBS} we display
the 3-momentum ($p$)
dependence of the rates for the binding energy scenarios discussed above. Generically,
gluo-dissociation differs from inelastic parton scattering in that the rate decreases
with 3-momentum while that of the inelastic parton scattering increases. This is a
direct consequence of the underlying matrix element (or cross section), which, as a
function of incoming parton energy, peaks slightly above the binding energy for gluo-dissociation
while it monotonically increases for inelastic parton scattering. The increase with $p$
of the latter is more pronounced for larger binding energies, primarily due to the opening
of phase space.
For the SBS with massless partons (upper panel of Fig.~\ref{fig_ptQFGD_SBS})
the $p$ dependence for gluo-dissociation rate is rather flat at low $T$ but starts to
develop a decreasing trend for the excited states with increasing $T$. For the massive
parton gas, this decreasing trend persists but is largely compensated once inelastic
parton scattering is included (lower panel of Fig.~\ref{fig_ptQFGD_SBS}).

For the TBS, the inelastic parton scattering at low $T$ results in a marked increase of
the rates with $p$ for all $Y$ states (upper panel of Fig.~\ref{fig_ptQF_TBS}), mostly due
to the phase space restrictions st low $p$ imposed by the still sizable binding energies.
At higher $T$, where the binding is much reduced, this trend weakens (lower panel of
Fig.~\ref{fig_ptQF_TBS}). Note that at $T$=300\,MeV, both $\Upsilon(2S)$ and $\chi_b(1P)$
have essentially become unbound so that the rate corresponds to twice the $b$-quark
scattering rate. At this temperature, the $\Upsilon(1S)$ still carries a significant
binding energy which induces a more pronounced $p$ dependence (as well as sensitivity
to the binding energy).
The gluo-dissociation rates in the TBS are shown in Fig.~\ref{fig_ptGD_TBS}. Except for
the $\Upsilon(1S)$ at low $T$ (where it is still strongly bound), they exhibit the usual
decreasing trend with $p$. They vanish for the excited states as soon as
they become unbound (for $T\lsim$300\,MeV), while they are quite significant close
to $T_{\rm pc}$ thus counter-balancing the increasing trend of the quasifree rate.
This also applies to the $\Upsilon(1S)$ as long as its binding
energy is larger than the temperature, \ie, for $T\lsim$300\,MeV.

\subsubsection{Equilibrium limit}
\label{sssec_equil}
Detailed balance between dissociation and formation reactions implies that the long-time
limit of the rate equation recovers the equilibrium abundances of quarkonia,
$N_Y^{\rm eq}$ in Eq.~(\ref{rate-eq}).
Assuming that the total number of $b\bar{b}$ pairs is conserved throughout the fireball
expansion, a pertinent conservation law is formulated,
\begin{equation}
\label{equil}
N_{{b\bar{b}}}=\frac{1}{2}\gamma_{b} n_{\rm{op}}V_{\rm{FB}}\frac{I_1(\gamma_{b}
n_{\rm{op}}V_{\rm{FB}})}{I_0(\gamma_{b} n_{\rm {op}}V_{\rm{FB}})}
+\gamma_{b}^2 n_{\rm{hid}} V_{\rm{FB}} \ ,
\end{equation}
where the sum of thermal densities of open ($n_{\rm{op}}$) and hidden ($n_{\rm{hid}}$)
bottom states in the system is matched to $N_{b\bar{b}}$ via
a temperature-dependent fugacity factor, $\gamma_b$, for each centrality of an AA
collision at given energy. The bottom densities are evaluated at each temperature
according to the phase of the fireball at volume $V_{\rm{FB}}$, \ie, with bottom
quarks in the QGP for $T>T_{\rm pc}$, with bottom hadrons in the hadronic phase for
$T<T_{\rm pc}$, and via a standard mixed-phase partitioning for $T=T_{\rm c}$
(if applicable, see Sec.~\ref{ssec:bulk} for a discussion on the mixed phase).
The number of $b\bar b$ pairs at given impact parameter is determined by the production
cross section $\sigma_{pp\rightarrow b\bar b}$, as
$N_{b\bar{b}}=(\sigma_{pp\rightarrow b\bar b}/\sigma_{pp}^{\rm inel}) N_{\rm coll}S_{\rm CNM}$,
where $N_{\rm coll}$ denotes the number of primordial $NN$ collisions upon first
impact of the incoming nuclei and $S_{\rm CNM}$ is a shadowing correction. The thermal
equilibrium value of a bottomonium state then follows as
\begin{equation}
N_Y^{\rm eq} (T) = V_{\rm FB} \gamma_b^2(T) n_Y(m_Y;T) \ .
\end{equation}

Three corrections to the equilibrium limit are in order for a more realistic implementation
in URHICs, two due to chemistry and one due to kinetics (sensitivity checks of the parameters
associated with these corrections will be elaborated in Sec.~\ref{ssec:lhc276-sensitivity-tbs}).

The first correction concerns a finite correlation volume, $V_{\rm corr}$, which accounts for the
finite distance by which a single $b\bar{b}$ can separate after essentially point-like
production~\cite{Hamieh:2000tk}. This limits the available phase space, which we model following
our previous treatment of charmonia~\cite{Grandchamp:2003uw}, by replacing the volume factor in
the canonical suppression factor represented by the modified Bessel functions in
Eq.~(\ref{equil}) by
\begin{equation}
V_{\rm corr}=\frac{4}{3}\pi\left(r_0+\langle v_b\rangle t\right)^3 \ ,
\label{Vcorr}
\end{equation}
that is,
\begin{equation}
\label{equil-corr}
N_{b\bar{b}}=\frac{1}{2}\gamma_{b} n_{\rm{op}}V_{\rm{FB}}\frac{I_1(\gamma_{b}
n_{\rm{op}}V_{\rm{corr}})}{I_0(\gamma_{b} n_{\rm {op}}V_{\rm{corr}})}
+\gamma_{b}^2 n_{\rm{hid}} V_{\rm{FB}} \ .
\end{equation}
The initial radius of the correlation volume, $r_0$$\simeq$0.8-1.2\,fm, characterizes
a typical strong interaction range, and the recoil velocity, $\langle v_b\rangle$=0.6-0.7,
is estimated from $B$-meson $p_T$ spectra (we use the central values unless otherwise noted).
For an increasing number of $b\bar{b}$ pairs, the individual correlation volumes may overlap,
eventually  merging into a single one to be used in the canonical suppression factor.

\begin{figure}[!t]
\includegraphics[width=0.48\textwidth]{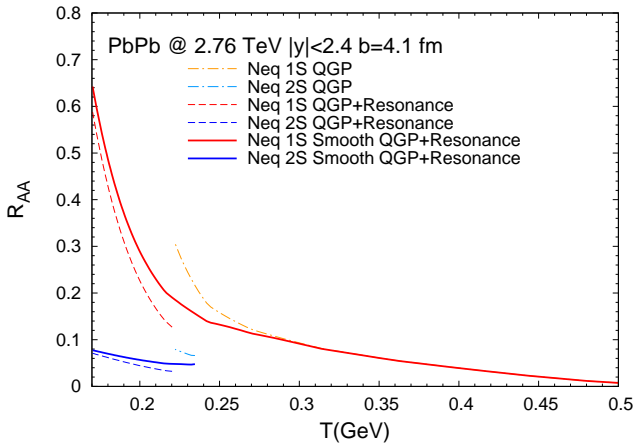}
\caption{Equilibrium limits of $Y$ states with (dashed lines) and without (dash-dotted lines) $B$-meson resonance
degrees of freedom,  and their smooth interpolation (solid lines) around a switching temperature of $T$=220\,MeV.
The red (blue) and yellow (light blue) curves are for $b$-quark only ($b$-quark plus resonance) degrees of
freedom for $\Upsilon(1S)$ and $\Upsilon(2S)$, respectively. The default TBS parameter $\eta$=1.0 is used.}
\label{fig_RVSTsmooth}
\end{figure}

The second correction, further following our previous treatment of charmonia~\cite{Zhao:2010nk},
concerns the emergence of open-bottom hadronic degrees of freedom as $T_{\rm pc}$ is approached
from above (this has recently been supported in an analysis of charm susceptibilities computed
in lQCD~\cite{Mukherjee:2015mxc}). Specifically, we allow for the existence of ground-state
($S$-wave) open-bottom mesons $B$, $B^*$, $B_s$, and $B_s^*$ with their respective spin-isospin
degeneracies. The presence of such states reduces the $b$-quark fugacity factor and thus the equilibrium
limit of the bottomonium states in the rate equation. Going up in temperature from $T_{\rm pc}$,
we continuously phase out the resonance states around a switching temperature of $T$=220\,MeV to
obtain a smooth connection to $b$-quark only degrees of freedom, cf.~Fig.~\ref{fig_RVSTsmooth}.
We will elaborate on the impact of this effect on the regeneration contribution to the $Y$
$R_{\rm AA}$'s in nuclear collisions in Sec.~\ref{ssec:lhc276-sensitivity-tbs}.

The third correction to the equilibrium limit arises from an incomplete kinetic equilibration
of $b$ quarks in URHICs, which affects the gain term in the rate equation (\ref{rate-eq}).
In particular, harder $b$-quark spectra than the thermalized limit
imply a reduced phase space overlap for bound-state formation. Following
Ref.~\cite{Grandchamp:2002wp}, we model this by implementing a thermal relaxation factor
into the $Y$ equilibrium limits,
\begin{equation}
\label{eq-rfactor}
R(\tau)=1-\mathrm{exp}\left( -\int_{\tau_0}^{\tau} \frac{\tau'}{\tau_{\rm b}}\mathrm{d}\tau' \right) \ ,
\end{equation}
with a $b$-quark relaxation time of $\tau_{\rm b}\simeq11$\,fm/$c$~\cite{Riek:2010fk}
at $\sim$$2\,T_{\rm c}$, slowly increasing with decreasing temperature. This approximation
has been supported by the studies in Ref.~\cite{Song:2012at}.

\subsection{Bulk medium evolution and solutions of the rate equation}
\label{ssec:bulk}
\begin{figure}[!t]
\includegraphics[width=0.48\textwidth]{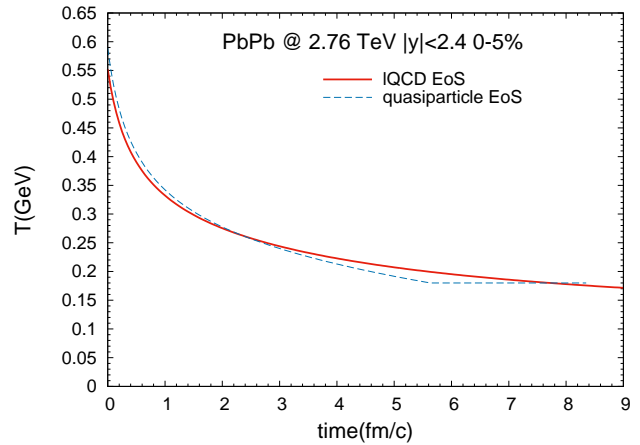}
\caption{Temperature evolution of the expanding firecylinder for central
Pb-Pb(2.76\,TeV) collisions using a lattice EoS with $T_{\rm pc}$=170\,MeV (red solid
line), compared to a massless quasiparticle EoS with mixed phase at $T_{\rm c}$=180\,MeV
(blue dashed line).}
\label{fig_eos}
\end{figure}

To solve the rate equation, the space-time evolution of the
medium is needed. We assume the conservation of total entropy in a cylindrical
isotropic fireball expansion of volume
\begin{equation}
\label{vfb}
V_{\rm{FB}}(\tau)=(z_0+v_z\tau)\pi \left(R_0+\frac{\sqrt{a_T^2\tau^2+1}-1}{a_T} \right)^2
\end{equation}
with a relativistic transverse acceleration and initial transverse radius $R_0$ estimated from
the Glauber model. The total entropy,
\begin{equation}
\label{entro-eq}
S_{\rm{total}}=s_{\rm{QGP}}(T)V_{\rm{FB}}(\tau) \ ,
\end{equation}
is determined from the final-state hadron multiplicities for a collision of given
energy and centrality [\eg, $S_{\rm tot}$=22000 for Pb-Pb(2.76TeV) covering
$\Delta y$=1.8 units in rapidity]. For the QGP entropy density, $s_{\rm{QGP}}(T)$,
we update our previous massless quasiparticle EoS with a fit to lQCD data~\cite{He:2011zx}
for the TBS calculation.
The initial longitudinal length in the Bjorken limit is the product of the rapidity
coverage of the fireball, $\Delta y$=1.8 and the QGP formation time, $\tau_0$ [for which
we use 0.2(0.6)\,fm at LHC (RHIC) energies], $z_0 = \Delta y \tau_0$. The relative
longitudinal velocity of the two fireball fronts for $\Delta y$=1.8 corresponds to $v_z$=1.4,
and the relativistic transverse acceleration is taken as $a_T$=0.1/fm.
For the case of the quasiparticle EoS, we define the QGP fraction in the mixed phase as
\begin{equation}
f_{\rm QGP}(\tau)=\frac{s(\tau)-s_{\rm HG}(T_c)}{s_{\rm QGP}(T_c)-s_{\rm HG}(T_c)} \
\end{equation}
where $s(\tau)=S_{\rm total}/V_{\rm FB}(\tau)$, $s_{\rm HG}(T_c)$ is the entropy density of
the hadron gas at $T_{\rm c}$=180\,MeV (as used previously), and $f_{\rm HG}(\tau)=1-f_{\rm QGP}(\tau)$.
The resulting time evolutions of temperature for central Pb-Pb(2.76\,TeV) collisions
for the massless quasiparticle EoS and the updated lQCD EoS are compared in Fig.~\ref{fig_eos}.
The nonperturbative effects lead to slightly higher (lower) temperatures in the
transition (high-temperature) region, as well as the absence of a mixed phase.
The lifetimes at the end of the QGP/mixed phase are within $\sim$10\%.

We now have all ingredients to solve the rate equation. For later purposes, we will
decompose it into two parts.  The suppression-only (or primordial) part is obtained from
\begin{equation}
\label{rate-prim}
\frac{\mathrm{d}N_{Y}^{\rm{prim}}(\tau)}{\mathrm{d}\tau}=-\Gamma_{Y}(\tau)N_{Y}^{\rm{prim}}(\tau)
 \ ,
\end{equation}
which has the solution
\begin{equation}
\label{sol-prim}
N_{Y}^{\rm{prim}}(\tau)=N_{Y}^{\rm{prim}}(\tau_0) \
\mathrm{exp}\left(-\int\limits_{\tau_0}^{\tau}\Gamma_{Y}(\tau')\mathrm{d}\tau'\right) \ ,
\end{equation}
characterizing the primordially produced bottomonia which survive the fireball evolution.
On the other hand, subtracting the rate equation of the primordial component from
the total one yields an equation for the regenerated component,
\begin{equation}
\label{rate-reg}
\frac{\mathrm{d}N_{Y}^{\rm{reg}}(\tau)}{\mathrm{d}\tau}=-\Gamma_{Y}(\tau)
\left[N_{Y}^{\rm reg}(\tau)-N_{Y}^{\rm eq}(\tau)\right],
\end{equation}
whose solution can also be written in a closed form as
\begin{eqnarray}
N_{Y}^{\rm{reg}}(\tau)&=&\frac{\int\limits_{\tau_{\rm diss}}^{\tau}\Gamma_{Y}(\tau')N_{Y}^{\rm eq}(\tau') \
\mathrm{exp}\left(\int\limits_{\tau_{\rm diss}}^{\tau'}\Gamma_{Y}(\tau'')\mathrm{d}\tau''\right)\mathrm{d}\tau'}
{\mathrm{exp}\left(\int\limits_{\tau_{\rm diss}}^{\tau}\Gamma_{Y}(\tau'')\mathrm{d}\tau''\right)}
\nonumber\\
&=&\int\limits_{\tau_{\rm diss}}^{\tau}\Gamma_{Y}(\tau')N_{Y}^{\rm eq}(\tau')
{\rm e}^{-\int\limits_{\tau'}^{\tau}\Gamma_{Y}(\tau'')\mathrm{d}\tau''}\mathrm{d}\tau'
\label{sol-reg}
\end{eqnarray}
where the total lifetime of the fireball, $\tau$=$\tau_{\rm f}$, is given by the end of the QGP/mixed
phase.
The exponential factor ${\rm e}^{-\int_{\tau'}^{\tau}\Gamma_{Y}(\tau'')\mathrm{d}\tau''}$
in the last line represents the in-medium suppression of the regenerated quarkonia. The lower
integration bound $\tau_{\rm diss}$ characterizes the time in the fireball evolution where
the temperature has dropped to the dissociation temperature of a given $Y$ state, below which
regeneration becomes operative. For the TBS, we have $T_{\rm diss}$$\simeq$260\,MeV, 240\,MeV and 190\,MeV
for $\chi_b(1P)$, $\Upsilon(2S)$ and $\Upsilon(3S)$, respectively.
The initial condition, $N_{Y}^{\rm{prim}}(\tau_0)=N_{\rm coll}\frac{\sigma_{pp\rightarrow Y}^{\rm tot}}
{\sigma_{pp}^{\rm inel}}S_{\rm CNM}^Y$ for the $Y$ numbers includes CNM
effects, in particular nuclear shadowing, calculated from the Glauber model (nuclear absorption
is included at the RHIC but neglected at the LHC due to the short nuclear passage time).

\subsection{Transverse-momentum spectra and elliptic flow}
\label{ssec:ptspec}
The rate equation approach above provides the 3-momentum inclusive yields of the
produced bottomonia. The explicit 3-momentum dependence of the yields can be recovered
in an approximate way by utilizing the decomposition into primordial and regenerated
components discussed above, following Ref.~\cite{Zhao:2007hh}. For the primordial
component, one straightforwardly solves the space-time dependent Boltzmann equation for
the bottomonium phase space distribution function while for the regeneration component a
coalescence model is employed. This is elaborated in more detail in the respective
Secs.~\ref{sssec_prim} and \ref{sssec_reg}, while Sec.~\ref{sssec_v2} discusses our
evaluation of the bottomonium elliptic flow.

\subsubsection{Transverse-momentum spectra of surviving primordial bottomonia}
\label{sssec_prim}
Without a gain term (and without a mean field), the Boltzmann transport equation
for the bottomonium phase space distribution, $f_Y$, reads
\begin{equation}
\begin{split}
\frac{\partial f_Y(\vec{x},\vec{p},\tau)}{\partial \tau}+
\vec{v}\cdot\frac{\partial f_Y(\vec{x},\vec{p},\tau)}{
\partial \vec{x}}=-\Gamma_{Y}(\vec{p},T(\tau))f_Y(\vec{x},\vec{p},\tau)
\end{split}
\end{equation}
where $\vec{v}=\vec{p}/E_p$ denotes the bottomonium velocity ($E_p^2=m_Y^2+p^2$)
and $\Gamma_{Y}(p,T)$ the 3-momentum dependent dissociation rate (as displayed
in Figs.~\ref{fig_ptQFGD_SBS}, \ref{fig_ptQF_TBS}, and \ref{fig_ptGD_TBS}).
Its solution can be cast in the form
\begin{equation}
\begin{split}
f_Y(\vec{x},\vec{p},\tau)=f_Y(\vec{x}-\vec{v}(\tau-\tau_0),\vec{p},\tau_0)
{\rm e}^{-\int\limits_{\tau_0}^{\tau}\Gamma_{Y}(\vec{p},T(\tau'))\mathrm{d}\tau'} \ ,
\end{split}
\end{equation}
from which $p_T$ spectra can be extracted assuming boost invariance as
\begin{equation}
\frac{\mathrm{d}^2N_Y(p_T,\phi)}{\mathrm{d}^2p_T}=\int f_Y(\vec{x}_T,\vec{p}_T,\tau)d^2x_T
 \ .
\end{equation}
The initial phase-space distribution,
$f_Y(\vec{x},\vec{p}_T,\tau_0)=f_Y^{\rm Glb}(\vec{x})f_Y^{\rm AA}(\vec{p}_T)$,
is factorized into $p_T$ spectra taken from experimental data in $pp$ collisions
and a Glauber model for the spatial distribution,
\begin{equation}
\begin{split}
f_Y^{\rm Glb}(\vec{x}_T)=\int \rho_A(\vec{x}_T+\vec{b}/2,z)\rho_A(\vec{x}_T-\vec{b}/2,z')\mathrm{d}z\mathrm{d}z'
\end{split}
\end{equation}
for an AA collision at impact parameter $b$.

We furthermore include formation time effects~\cite{Blaizot:1988ec,Karsch:1987zw,Gavin:1990gm}
to account for the finite time for the bound state to develop from the
primordially produced $b\bar b$ wave package. This evolution  tends to reduce the
suppression rate, intuitively associated with a geometric expansion of the wave
package from its near point-like production to the bound-state size~\cite{Gerland:2003wy}.
Accordingly, we assume the formation time $\tau_{\rm form}$ to depend on the vacuum binding energy,
and correct the dissociation rate for $\tau\leq\tau_{\rm form}\gamma$ as
\begin{equation}
\alpha_Y(\vec{p},T(\tau))\equiv\Gamma_Y(\vec{p},T(\tau))\frac{\tau}{\tau_{\rm form}}\frac{m_Y}{\sqrt{p^2+m_Y^2}} \ .
\end{equation}
in the (early) evolution of the primordial $p_T$ spectra. The explicit formation time
values for $\Upsilon(1S)$, $\Upsilon(2S)$ and $\Upsilon(3S)$ are chosen as 0.5, 1.0 and 1.5\,fm,
respectively. The latter two are close to typical values used for the $J/\psi$ and $\chi_c$,
as they have comparable binding energies [$E_B(J/\psi)$$\simeq$640\,MeV vs.~$E_B(\Upsilon(2S))$$\simeq$540\,MeV,
and $E_B(\chi_c)$$\simeq$230\,MeV vs.~$E_B(\Upsilon(3S))$$\simeq$200\,MeV].
The inverse Lorentz-$\gamma$ factor, $\gamma^{-1}=\frac{m_Y}{\sqrt{p^2+m_Y^2}}$, suppresses
the high-$p_T$ reaction rates especially for excited states which have larger formation times.
The reduced primordial suppression at high $p_T$ counterbalances the $p_T$ dependence in the
thermal dissociation rates.

\subsubsection{Transverse-momentum spectra from regeneration}
\label{sssec_reg}
The momentum spectra of regenerated quarkonia carry the imprint of the momentum
distributions of the recombining heavy quarks (or mesons). For charmonia, the
regeneration typically occurs several fm/$c$ into the evolution of the fireball, where
charm-quark spectra, with a thermal relaxation rate of a few fm/$c$, are probably
not far from their equilibrium distribution. Thus, the regenerated charmonia can be
rather well approximated by a blast-wave description close to $T_{\rm pc}$, which is
supported by the momentum spectra measured at the LHC~\cite{Abelev:2013ila}.
The situation changes for bottom(onium), primarily because the $\sim$3 times larger
$b$-quark mass, relative to $c$ quarks, implies a factor 3 longer thermal relaxation
times, and, to a lesser extent, because bottomonia are formed earlier in the fireball
evolution, due to their larger binding energies. Therefore, approximating regenerated
bottomonia with a thermal blast-wave expression cannot be expected to be accurate.
Instead, we here resort to an instantaneous coalescence model~\cite{Greco:2003mm},
which allows us to use more realistic nonequilibrium transverse-momentum spectra of
$b$ quarks as input. We take these spectra from relativistic Langevin simulations
of heavy quarks~\cite{He:2014cla} in a hydrodynamic background medium (akin to the
fireball evolution used for the rate equation) with nonperturbative heavy-quark
transport coefficients which are computed from the same underlying $T$-matrix
interactions~\cite{Riek:2010fk} as the bottomonium binding energies discussed in
Sec.~\ref{sssec_rates}.

The expression for the 2-differential $p_T$ spectra of an $Y$ meson formed through
instantaneous coalescence from bottom quark and antiquarks with $p_t$ distributions,
$\mathrm{d}^2N_{b,\bar b}/\mathrm{d}^2p_T$ (or $\mathrm{d}^2N_{Y}^{\rm coal}/\mathrm{d}^2p_T$),
is given by~\cite{Greco:2003mm}
\begin{equation}
\begin{split}
\frac{\mathrm{d}^2N_{Y}^{\rm coal}(p_T,\phi)}{\mathrm{d}^2 p_{T}} &= C_{\rm reg}
\int \mathrm{d}^2 p_{1t} \mathrm{d}^2 p_{2t} \frac{\mathrm{d}^2N_{b}}{\mathrm{d}^2 p_{1t}}
\frac{\mathrm{d}^2N_{\bar{b}}}{\mathrm{d}^2 p_{2t}}
\\
&\!\times\delta^{(2)}(\vec{p}_{T}-\vec{p}_{1t}-\vec{p}_{2t})
\\
&\!\times\Theta \left[\Delta^{2}_{p}-\frac{(\vec{p}_{1t}-\vec{p}_{2t})^2}{4}+
\frac{(m_{1t}-m_{2t})^2}{4}\right]
 .
\label{coal-1}
\end{split}
\end{equation}
Here, $C_{\rm reg}$ denotes a normalization constant which is matched to the
regeneration yield obtained from the rate equation, and $m_t=\sqrt{p_t^2+m_b^2}$
is the transverse mass of the $b$ quarks. The $\Theta$ function
characterizes the momentum space wave function of the formed $Y$,
suppressing high relative momenta of the coalescing $b$ quarks. The
covariant momentum space radius, $\Delta_p$, is inversely proportional
to the coordinate-space radius via the uncertainty relation,
$\Delta_p \Delta_x\simeq1$. We use
$\Delta_x \simeq r_{\Upsilon[\Upsilon(2S),\chi_b(1P)]} =0.2[0.5]$\,fm.
The $p_T$ spectra are obtained by integrating the 2-differential
spectrum in Eq.~(\ref{coal-1}) over the azimuthal angle,
\begin{equation}
\frac{\mathrm{d}N_{Y}^{\rm coal}(p_T)}{\mathrm{d}p_{T}}=
\int\limits_{0}^{2\pi}\frac{\mathrm{d}^2N_{Y}^{\rm coal}(p_T,\phi)}{\mathrm{d}^2 p_{T}}
p_{T}\mathrm{d}\phi \ .
\label{coal-2}
\end{equation}
As an estimate of the uncertainty in the regeneration time, we will adopt snapshots
of the evolving $b$-quark distributions from the Langevin simulations at different
local temperatures with pertinent flow velocities in the underlying hydro evolution of Ref.~\cite{He:2014cla}.

Finally, to account for the $p_T$ dependence of the formation rate, which is not captured
by the instantaneous coalescence approximation, we weight the coalescence spectrum,
Eq.~(\ref{coal-2}), by the $p_T$ dependence of the inelastic reaction rate,
\begin{equation}
\frac{dN_{Y}^{\rm reg}}{dp_T}= \hat{\Gamma}_Y(p_T,\bar{T}_{\rm reg})
\frac{\mathrm{d}N_{Y}^{\rm coal}}{\mathrm{d}p_T}
\end{equation}
where
$\hat{\Gamma}_Y(p_T,\bar{T}_{\rm reg})\equiv{\Gamma}_Y(p_T,\bar{T}_{\rm reg})/\bar{\Gamma}_Y(\bar{T}_{\rm reg})$
includes a normalization $\bar{\Gamma}_Y$ such that the norm of the regeneration
component as obtained from the rate equation is preserved.

\subsubsection{Elliptic flow}
\label{sssec_v2}
Another observable with a potential to disentangle primordially
produced and regenerated quarkonia is their elliptic flow. It is quantified by
the second coefficient, $v_2(p_T)$, in the the Fourier expansion of their
azimuthal-angle differential distribution,
\begin{equation}
\frac{\mathrm{d}^2N}{\mathrm{d}^2p_{T}}=
\frac{1}{2\pi}\frac{\mathrm{d}N(p_T)}{p_{T}\mathrm{d}p_{T}}[1+2v_{2}(p_T)\mathrm{cos}(2\phi)+\dots],
\end{equation}
where $\phi$ is defined relative to the $x$ axis, which lies in the
reaction plane aligned with the impact parameter. At mid-rapidity,
odd harmonics are suppressed, while higher even harmonics ($v_4, v_6, \dots$)
for bulk hadron production are typically much smaller than $v_2$.
From the above expansion one projects out the second coefficient via
\begin{equation}
v_2(p_T)=
\frac{\frac{1}{2\pi}\int\limits_0^{2\pi}
\frac{\mathrm{d}^2N(p_T,\phi)}{p_{T}\mathrm{d}p_{T}\mathrm{d}\phi}\mathrm{cos}(2\phi)\mathrm{d}\phi}
{\frac{1}{2\pi}\int\limits_0^{2\pi}\frac{\mathrm{d}^2N(p_T,\phi)}{p_{T}\mathrm{d}p_{T}\mathrm{d}\phi}\mathrm{d}\phi} \ .
\end{equation}
For the primordial component, we explicitly track the bottomonium paths through an elliptically
expanding fireball; the pertinent $v_2^{\rm prim}$ is generated entirely due to path length
differences and usually rather small in magnitude~\cite{Wang:2002ck} (contributions from elastic
scatterings are not accounted for; little is known about such processes).
For the regeneration component, the coalescence expression, Eq.~(\ref{coal-1}), incorporates
the $v_2$ information through the convolution of the underlying $b$- and $\bar b$-quark flows.
The total elliptic flow follows as the weighted sum of the two contributions,
\begin{equation}
v_2(p_T)=\frac{R_{AA}^{\rm prim}(p_T)v_2^{\rm prim}(p_T)+R_{AA}^{\rm coal}(p_T)v_2^{\rm coal}(p_T)}
{R_{AA}^{\rm prim}(p_T)+R_{AA}^{\rm coal}(p_T)} \ .
\end{equation}

\subsection{Open-bottom and bottomonium input cross sections}
\label{ssec:xsec}
The basic quantity to compute below is the nuclear modification factor, defined as the ratio of
yields in an AA collision at a given centrality divided by the $N_{\rm coll}$-scaled yield in $pp$,
\begin{equation}
R_{\rm AA}=\frac{N_Y^{\rm AA}} {N_{\rm coll}\frac{\sigma_{pp\rightarrow Y}}{\sigma_{pp}^{\rm inel}}}
\ .
\label{RAA}
\end{equation}
This has been measured as a function of several variables, \ie, nucleon participant number
($N_{\rm part}$) as a measure of centrality (which we estimate from the optical Glauber model),
transverse momentum ($p_T$), rapidity ($y$), and collision energy ($\sqrt{s}$).
The numerator in Eq.~(\ref{RAA}) contains the primordial component, which is also
proportional to the product of $N_{\rm coll} \sigma_{pp\rightarrow Y}$ (times a suppression
factor), and the regeneration component, which is largely controlled by the open-bottom
cross section, $\sigma_{pp\rightarrow b\bar{b}}$, independent of the denominator (although
in practice we will assume a proportionality between open- and hidden-bottom cross sections).

\begin{table}[!t]
\begin{tabular}{|c|c|c|c|c|}
\hline
$p_T$[GeV] & $\chi_b(1P)$[\%] & $\chi_b(2P)$[\%] & $\chi_b(3P)$[\%]
& total $\chi_b(nP)$[\%] \\
\hline
6$\sim$8 & 14.8 & 3.3 &  & 18.1  \\
\hline
8$\sim$10 & 17.2 & 5.2 &  & 22.4  \\
\hline
10$\sim$14 & 21.3 & 4.0 & 1.7 & 27.0  \\
\hline
14$\sim$18 & 24.4 & 5.2 & 1.8 & 31.4  \\
\hline
18$\sim$22 & 27.2 & 5.5 & 1.9 & 34.6  \\
\hline
22$\sim$40 & 29.2 & 6.0 & 2.9 & 38.1  \\
\hline
\end{tabular}
\caption{Feeddown fractions from $\chi_b(nP)$ states to $\Upsilon(1S)$ from LHCb~\cite{Aaij:2014caa}.}
\label{tab-fd}
\end{table}
Feeddowns from higher excited states contribute to the inclusive production of an observed meson.
A detailed summary of feeddown fractions can be found in Appendix~\ref{appf}.
For the $p_T$-dependent $R_{\rm AA}$'s at both RHIC and LHC energies, we include the $p_T$ dependence
of the feeddown in our calculations. By using harder $p_T$ spectra for excited states, but with
the same branching fraction as at low $p_T$, we automatically obtain a larger feeddown at high $p_T$
which is essentially consistent with LHCb Collaboration data~\cite{Aaij:2014caa}; cf.~Table~\ref{tab-fd}.

We slightly update several input cross sections for the TBS calculations relative to
Ref.~\cite{Emerick:2011xu}, as summarized in Table~\ref{tab_xsec}, but keep the ratio
$\frac{\sigma_{Y}}{\sigma_{b\bar b}}\simeq$\,0.176\,\% as in our previous work which is
within the uncertainty of measured values. This ratio only affects the regeneration component of $R_{\rm AA}$.
In general, the small $pp$ cross sections for $b\bar b$ pairs render their number less than 1 even in
AA collisions at the LHC. Therefore, the corresponding $Y$ equilibrium limits are in the canonical limit,
so that $R_{\rm AA}^{\rm reg}$ is essentially linear in $\frac{\sigma_{Y}}{\sigma_{b\bar b}}$, and
thus approximately constant for not too peripheral collisions (for the latter, the small QGP lifetime
implies that the relaxation time approximation for $b$-quark diffusion will lead to a noticeable suppression).
Overall, our results for the $R_{\rm AA}$'s will be influenced insignificantly by the update of the
input $pp$ cross sections.

For the bottomonium input cross sections at $\sqrt{s}$=200\,GeV we adopt the
STAR measurement~\cite{Ye:2017fwv} of $\Upsilon(1S+2S+3S)$,
$\frac{\mathrm{d}\sigma}{\mathrm{d}y}\cdot Br(\mu\bar{\mu})=81\pm5\pm8\,pb$. We reconstruct
$\sigma_{1S}^{tot}$ using the di-muon branching ratios $Br(1S\rightarrow\mu\bar{\mu})=2.48\,\%$,
$Br(2S\rightarrow\mu\bar{\mu})=1.93\,\%$, and $Br(3S\rightarrow\mu\bar{\mu})=2.18\,\%$ with cross
sections $\sigma_{2S}=0.33\sigma_{1S}^{tot}$ and
$\sigma_{3S}=0.15\sigma_{1S}^{tot}$ (see Appendix~\ref{appf}), so that
$\frac{\mathrm{d}\sigma_{1S}^{tot}}{\mathrm{d}y}=\frac{81\,pb}{1\cdot2.48\,\%+0.33\cdot1.93\,\%+0.15\cdot2.18\,\%}\simeq2.35\,nb$.
We use the same values for open-bottom cross section as in previous work~\cite{Emerick:2011xu}, \ie,
$\sigma_{pp\to b\bar{b}}$=3.2\,$\mu$b, with a factor of 0.52 to obtain
$\sigma_{pp\to b\bar{b}}$=1.67$\mu$b in one fireball ($\Delta y$=1.8), or
$\frac{\mathrm{d}\sigma_{pp\to b\bar{b}}}{\mathrm{d}y}$=0.92\,$\mu$b. This is consistent with
the most recent PHENIX results~\cite{Adare:2017caq}.
For simplicity, we use the same input values for uranium-uranium (U-U) collisions at 193\,GeV.

For $pp$ collisions at 2.76\,TeV, we use the the inclusive $\Upsilon(1S)$ cross section of
$\frac{\mathrm{d}\sigma_{pp\to \Upsilon(1S)}}{\mathrm{d}y}$=30.3\,nb for $|y|$$<$2.4 based on
CMS $pp$ data~\cite{Khachatryan:2016xxp}, which is $\sim$25\% smaller than in
Ref.~\cite{Emerick:2011xu}.
With $\frac{\sigma_{\Upsilon(1S)}}{\sigma_{b\bar b}}\simeq$\,0.176\,\% this gives
$\frac{\mathrm{d}\sigma_{pp\to b\bar{b}}}{\mathrm{d}y}$=17.2\,$\mu$b at 2.76\,TeV
for $|y|$$<$2.4 which is consistent
with the upper range of FONLL calculations~\cite{Cacciari:2012ny}, 15$\pm$6.2\,$\mu$b.
We estimate a 10\,\% reduction from $|y|$$<$0.9 to $|y|$$<$2.4.
From a comparison of 7\,TeV ALICE~\cite{Abelev:2014qha} and CMS~\cite{Chatrchyan:2013yna}
data, we estimate the forward-rapidity (2.5$<$$|y|$$<$4.0) cross section at about 45\,\% of the mid-rapidity
($|y|$$<$0.9) value, or 50\,\% of the ($|y|$$<$2.4) value.
The $\Upsilon(2S)$ cross section is about 33\,\% of inclusive $\Upsilon(1S)$ as
discussed in Appendix~\ref{appf}.

At 5.02\,TeV, we adopt for the inclusive $\Upsilon(1S)$ cross section the recent CMS $pp$
reference~\cite{Flores:2017qmcms,CMS:2017ucd},
$\frac{\mathrm{d}\sigma_{pp\to Y}}{\mathrm{d}y}$=64.0\,nb (57.6\,nb)
for $|y|<0.9$ ($|y|<2.4$), together with the $\Upsilon(1S)$ over open-bottom ratio of 0.176\,\%
and the same reduction of 55\,\% (50\,\%) from mid-rapidity $|y|<0.9$ ($|y|<2.4$) to forward
rapidity $2.5<y<4.0$.

The absolute input cross sections at different energies for different states for the TBS calculations are
summarized in Table.~\ref{tab_xsec}.

\begin{table}[!t]
\begin{tabular}{|l|c|c|c|}
\hline
Differential cross section $\frac{\mathrm{d}\sigma}{\mathrm{d}y}$
& 0.20\,TeV & 2.76\,TeV & 5.02\,TeV \\
\hline
pp$\rightarrow$ $\Upsilon(1S)$($|y|<0.5$)[nb] & 2.35 & - & - \\
\hline
pp$\rightarrow$ $\Upsilon(1S)$($|y|<2.4$)[nb] & - & 30.3 & 57.6 \\
\hline
pp$\rightarrow$ $\Upsilon(1S)$($2.5<y<4.0$)[nb] & - & 15.1 & 28.8 \\
\hline
pp$\rightarrow$ $\Upsilon(2S)$($|y|<0.5$)[nb] & 0.77 & - & - \\
\hline
pp$\rightarrow$ $\Upsilon(2S)$($|y|<2.4$)[nb] & - & 10.0 & 19.0 \\
\hline
pp$\rightarrow$ $\Upsilon(2S)$($2.5<y<4.0$)[nb] & - & 5.0 & 9.5 \\
\hline
pp$\rightarrow$ $b\bar{b}$($|y|<0.5$)[$\mu$b] & 0.92 & - & - \\
\hline
pp$\rightarrow$ $b\bar{b}$($|y|<2.4$)[$\mu$b] & - & 17.2 & 32.7 \\
\hline
pp$\rightarrow$ $b\bar{b}$($2.5<y<4.0$)[$\mu$b] & - & 8.6 & 16.4 \\
\hline
\end{tabular}
\caption{Summary of input cross sections extracted from $pp$ collisions used in our calculations.
The $\frac{\mathrm{d}\sigma_{pp\to Y}}{\mathrm{d}y}$ values at 200\,GeV are based on STAR
data~\cite{Ye:2017fwv}. The value for $\frac{\mathrm{d}\sigma_{pp\to b\bar{b}}}{\mathrm{d}y}$ is
adopted from previous work~\cite{Emerick:2011xu} which is consistent with PHENIX results~\cite{Adare:2017caq}.
The $\frac{\mathrm{d}\sigma_{pp\to Y}}{\mathrm{d}y}$ values at 2.76 and 5.02\,TeV
for $|y|<2.4$ are based on CMS data~\cite{Khachatryan:2016xxp,Flores:2017qmcms,CMS:2017ucd}, with a
fixed $\frac{\sigma_{pp\to Y}}{\sigma_{pp\to b\bar{b}}}$ ratio of 0.176\,\%, as in previous
work~\cite{Emerick:2011xu}. A 50\,\% reduction in the cross sections is assumed when going from
mid-rapidity ($|y|<2.4$) to forward rapidity ($2.5<y<4.0$)~\cite{Abelev:2014qha,Chatrchyan:2013yna}.}
\label{tab_xsec}
\end{table}

\section{Bottomonium Production at RHIC}
\label{sec:rhic}
We are now in position to present our numerical results for bottomonium observables
in comparison to experimental data, starting with RHIC energies.
Before presenting and discussing the results for the centrality and $p_T$ dependencies in
Secs.~\ref{ssec:rhic-centrality-tbs} and \ref{ssec:rhic-pt-tbs}, let us briefly outline
our implementation of CNM effects, which we estimate from d-Au collisions. We assume no
shadowing on open-bottom and bottomonium production and interpret the STAR measurement of
$R_{\rm dAu}(1S)=0.83 \pm 0.15 ({\rm dAu}_{\rm{stat}}) \pm 0.1 (pp_{\rm{stat}}) \pm$~0.03~(sys)
\cite{Adamczyk:2013poh} as being due to nuclear absorption with a $YN$ absorption cross section
in a range of $\sigma_Y^{\rm abs}$=0-3\,mb (identical for all bottomonia).

\subsection{Centrality Dependence at RHIC}
\label{ssec:rhic-centrality-tbs}
\begin{figure}[!t]
\includegraphics[width=0.48\textwidth]{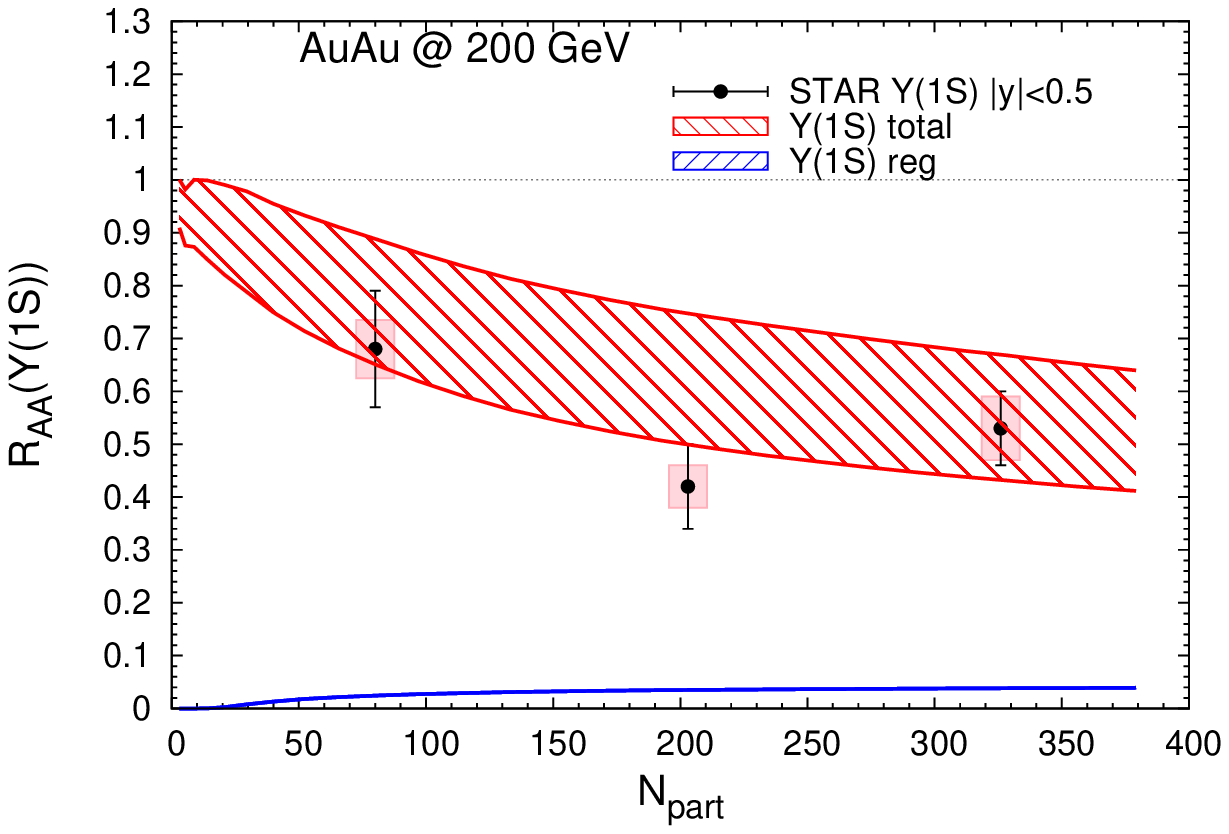}
\includegraphics[width=0.48\textwidth]{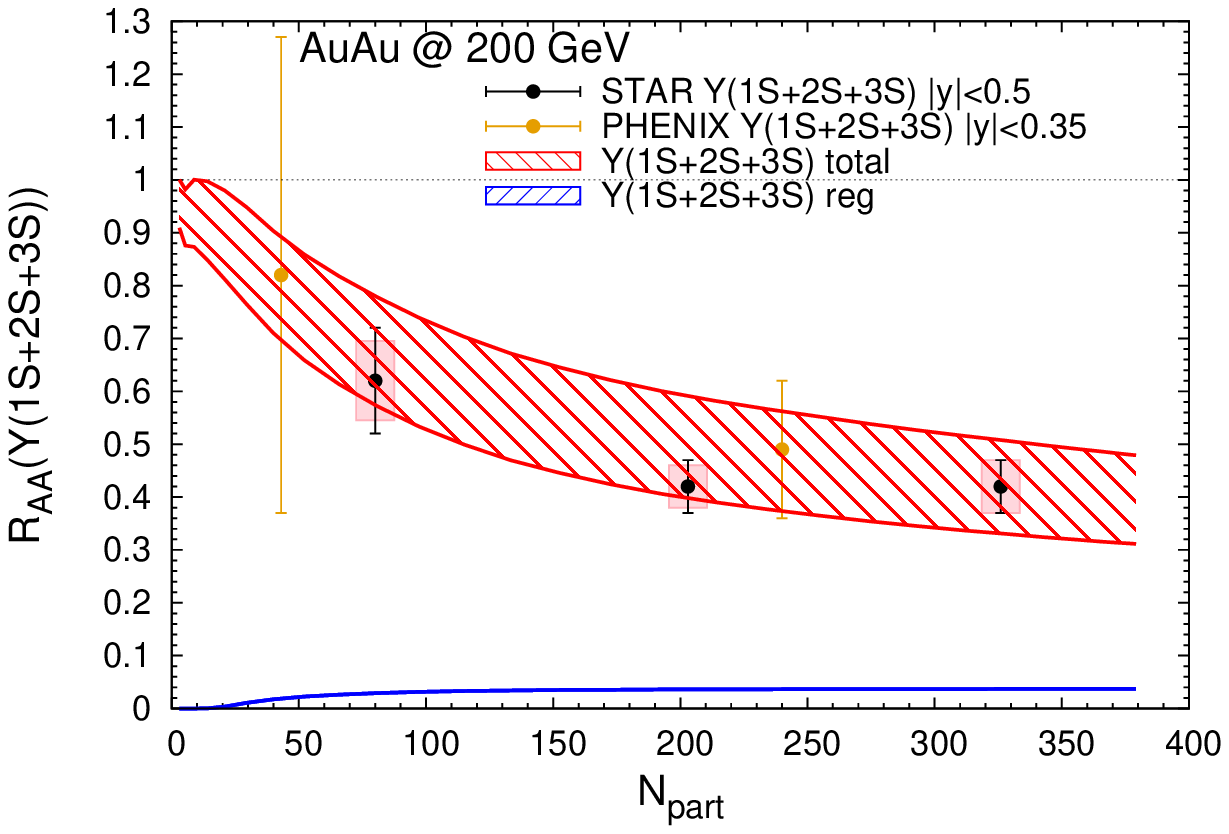}
\includegraphics[width=0.48\textwidth]{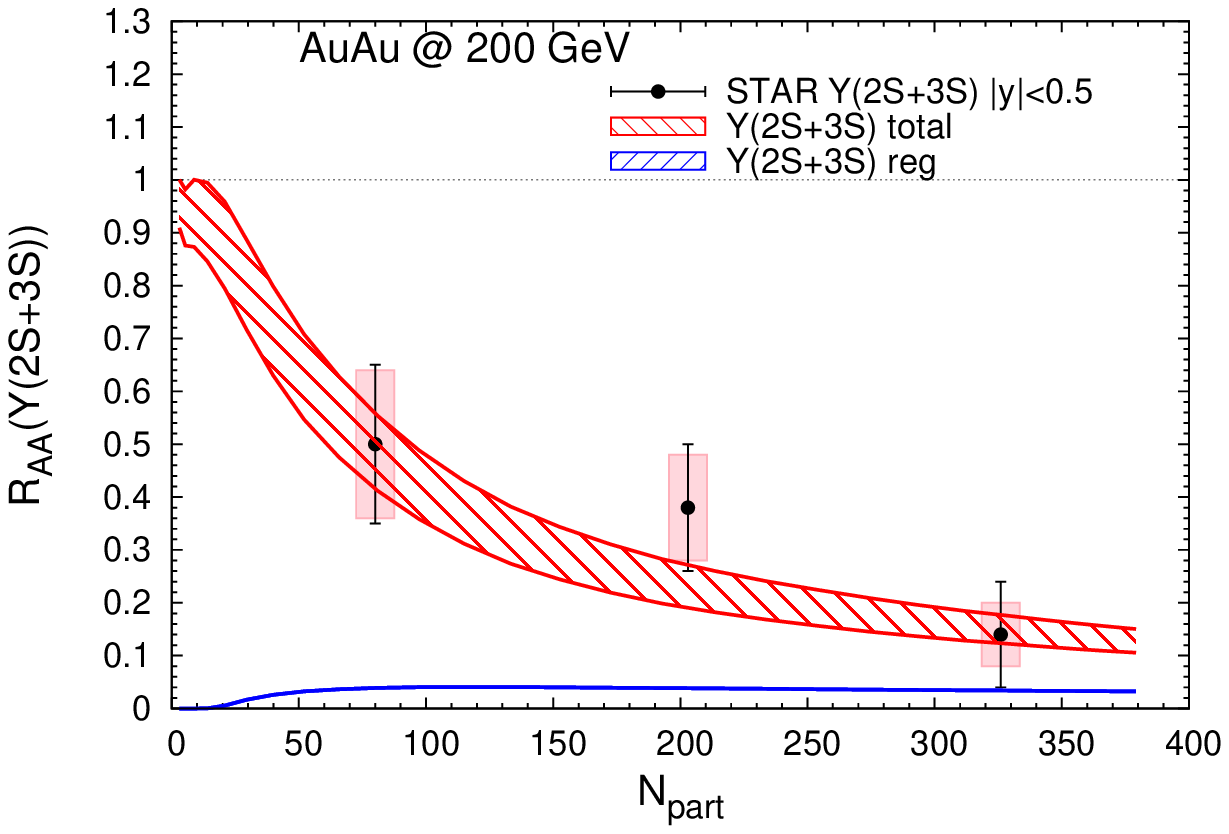}
\caption{Centrality dependence of bottomonium yields in Au-Au(200\,GeV) collisions using the baseline TBS
($\eta=1.0$) with updated feeddowns. The total (red band) and regenerated (blue lines) contributions are
shown for inclusive $\Upsilon(1S)$ (upper panel), $\Upsilon(1S+2S+3S)$ (middle panel), and $\Upsilon(2S+3S)$
(lower panel) production at mid-rapidity ($|y|$$<$0.5) and compared to STAR~\cite{Ye:2017fwv} and
PHENIX~\cite{Adare:2014hje} data. The band width of the total yields is due to CNM effects
with $\sigma_Y^{\rm abs}$=0-3\,mb~\cite{Adamczyk:2013poh}.}
\label{fig_TBS-200AuAu}
\end{figure}
\begin{figure}[!t]
\includegraphics[width=0.48\textwidth]{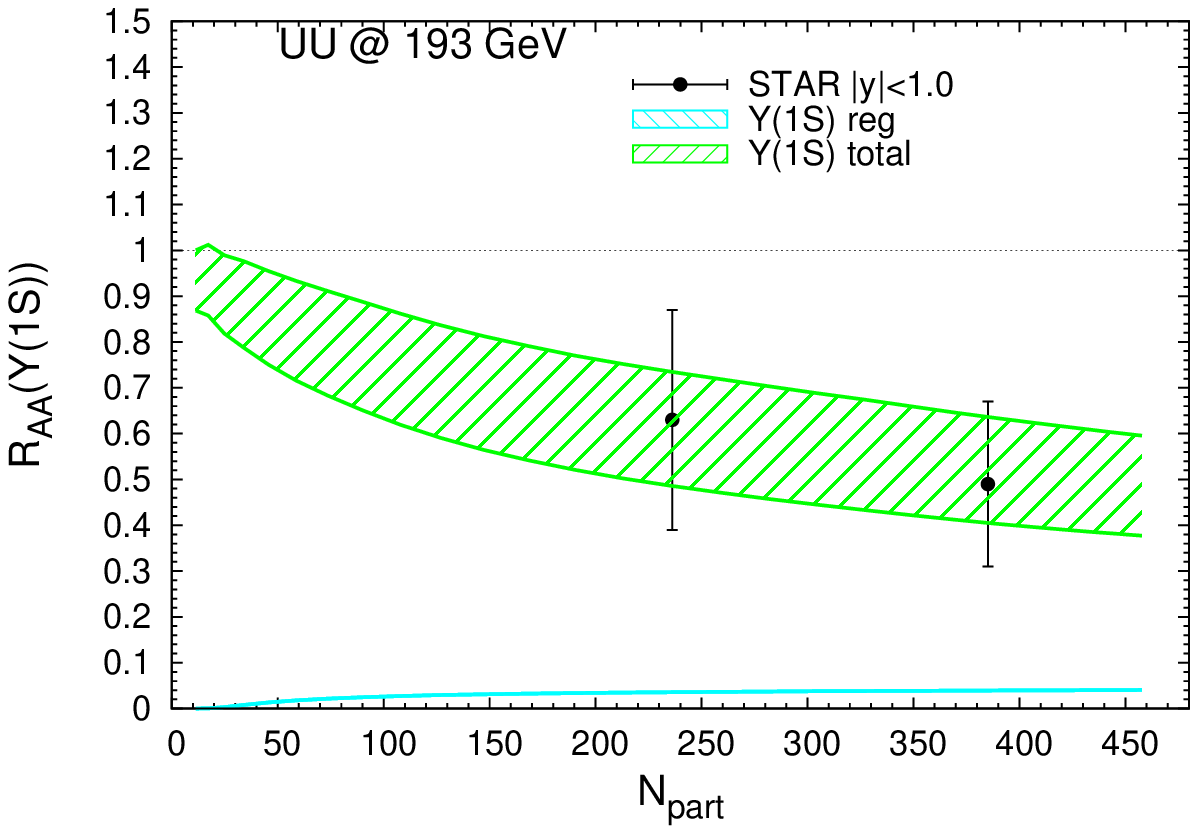}
\includegraphics[width=0.48\textwidth]{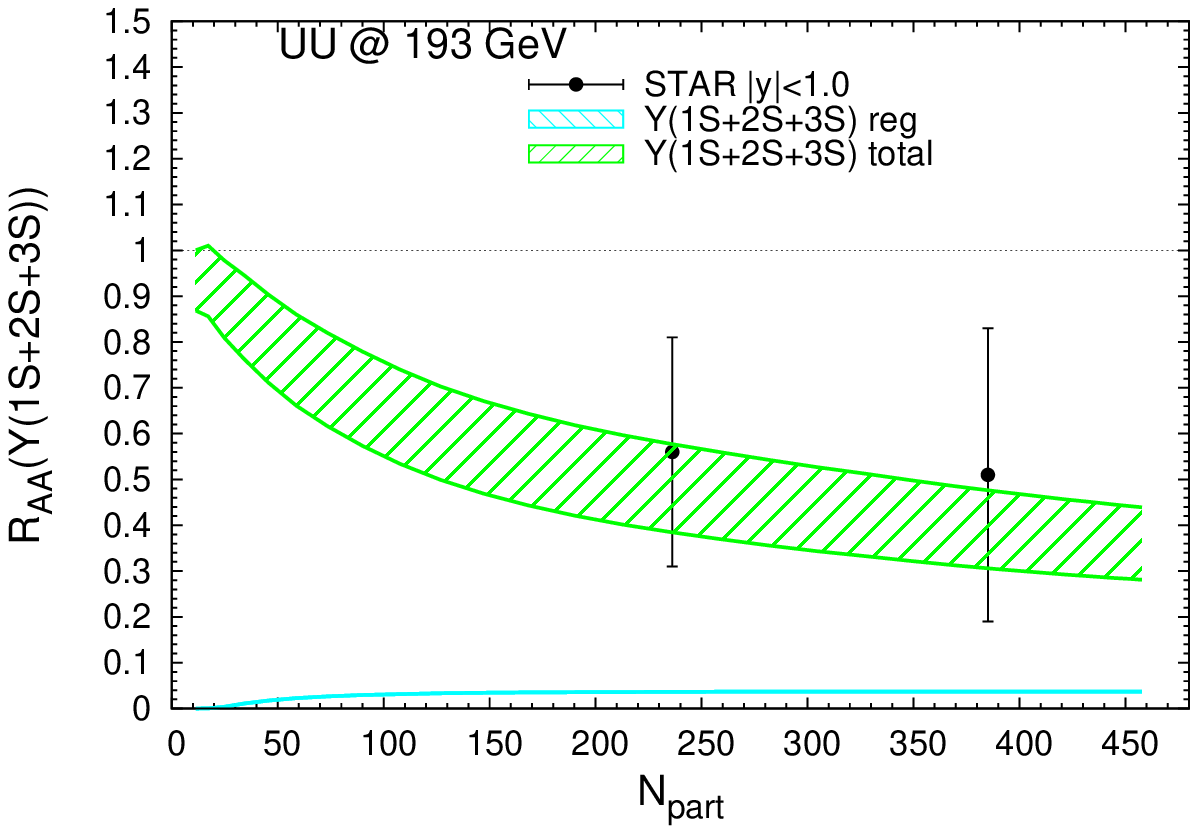}
\caption{Centrality dependence of bottomonium yields in U-U(193\,GeV) collisions using the
baseline TBS with updated feeddowns.
The total (green band) and  regenerated (light-blue line) contributions are shown for inclusive
$\Upsilon(1S)$ (upper panel) and $\Upsilon(1S+2S+3S)$ (lower panel) at mid-rapidity ($|y|$$<$1.0)
and compared to STAR data~\cite{Adamczyk:2016dzv}.
The band width of the total yields is due to CNM effects with
$\sigma_Y^{\rm abs}$=0-3\,mb~\cite{Adamczyk:2013poh}.}
\label{fig_TBS-200UU}
\end{figure}

Our results for the $R_{\rm AA}(N_{\rm part})$ for inclusive $\Upsilon(1S)$ and $\Upsilon(1S+2S+3S)$
states in Au-Au and U-U collisions are shown in Figs.~\ref{fig_TBS-200AuAu} and \ref{fig_TBS-200UU},
respectively, in comparison to RHIC data.
We focus on the $T$-matrix binding scenario (TBS) with baseline binding strength, $\eta$=1.0.
The suppression of the inclusive $\Upsilon(1S)$ yield (upper panels) is mostly due to the excited states
(as well as nuclear absorption), which manifests itself as a stronger suppression in the combined $R_{\rm AA}$
of $\Upsilon(1S+2S+3S)$ [middle (lower) panel in Fig.~\ref{fig_TBS-200AuAu} (\ref{fig_TBS-200UU})].
Primordial production dominates but regeneration, in the canonical limit with $N_{b\bar{b}}<1$ even in
central collisions, is non-zero. Our results for U-U show slightly more suppression than for Au-Au,
and both are generally consistent with the data.

\subsection{Transverse-momentum dependence at RHIC}
\label{ssec:rhic-pt-tbs}
\begin{figure}[!t]
\includegraphics[width=0.48\textwidth]{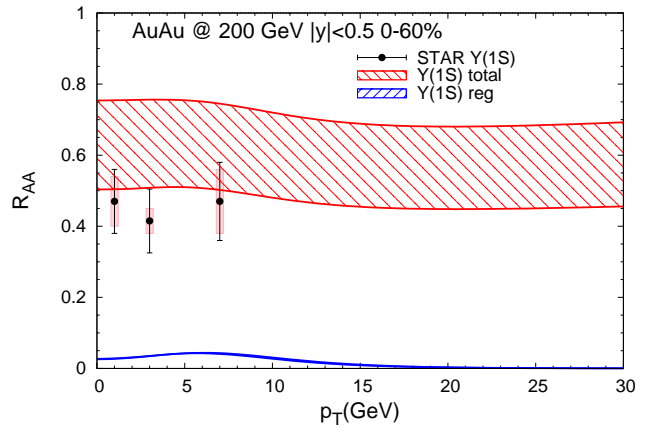}
\caption{The $p_T$-dependent $R_{\rm AA}$ for inclusive $\Upsilon(1S)$ in 0-60\,\% Au-Au(200\,GeV)
collisions within the baseline TBS,
compared to STAR data~\cite{Ye:2017fwv}. The red (blue) band is for the total (regeneration) yield, where the
band width of the former is due to CNM effects with $\sigma_Y^{\rm abs}$=0-3\,mb~\cite{Adamczyk:2013poh}.}
\label{fig_TBS-200AuAupt}
\end{figure}

We proceed to compute $Y$ $p_T$ spectra by utilizing their $R_{\rm AA}$'s for primordial and
regenerated components as obtained in the previous section to form the weighted sum
\begin{eqnarray}
R_{\rm AA}(p_T)&=& \frac{ \frac{dN_Y^{\rm prim}}{p_T\mathrm{d}p_T} + \frac{dN_Y^{\rm reg}}{p_T\mathrm{d}p_T} }
{ N_{\rm coll} \frac{dN_Y^{pp}}{p_T\mathrm{d}p_T} }
\nonumber\\
&=& R_{\rm AA}^{\rm prim} \frac{\frac{d\hat{N}_{Y}^{\rm prim}}{p_T\mathrm{d}p_T}} {\frac{d\hat{N}_Y^{pp}}{p_T\mathrm{d}p_T}}
+ R_{\rm AA}^{\rm reg} \frac {\frac{d\hat{N}_{Y}^{\rm reg}}{p_T\mathrm{d}p_T}} {\frac{d\hat{N}_Y^{pp}}{p_T\mathrm{d}p_T}}
\end{eqnarray}
where the ``hat" indicates a normalized distribution,
\begin{equation}
\begin{split}
\int\limits_0^\infty p_T\mathrm{d}p_T \frac{\mathrm{d}\hat{N}(p_T)}{p_T\mathrm{d}p_T} = 1  \ ,
\end{split}
\end{equation}
and the $R_{\rm AA}$ coefficients represent a given centrality class, \eg, 0-60\,\%.
For the normalized $pp$ spectra, we employ an empirical parametrization,
\begin{equation}
\frac{\mathrm{d}^2\hat{N}_{pp}(p_T)}{\mathrm{d}^2p_{T}}=\frac{N}{\left(1+\left(\frac{p_{T}}{D}\right)^2\right)^A} \
\label{pp-pTspec}
\end{equation}
with fit parameters $A$=3.0 and $D$=5.3\,GeV estimated from $m_T$ scaling from charmonium
$p_T$ spectra~\cite{Zhao:2010nk} as baseline. The $p_T$-dependent $R_{\rm AA}$ for 0-60\% Au-Au(200\,GeV)
is shown in Fig.~\ref{fig_TBS-200AuAupt}. It tends to slightly overestimate the STAR data, although the
lower end of the band (with maximal nuclear absorption) is close to the data, a trend which is also
reflected in the centrality-dependent $R_{\rm AA}$ (recall the upper panel of Fig.~\ref{fig_TBS-200AuAu})

\section{Bottomonium Production in 2.76\,TeV Pb-Pb Collisions}
\label{sec:lhc276}
Turning to Pb-Pb collisions at the LHC, we first focus on $\sqrt{s}$=2.76\,TeV. To make contact with the earlier
employed SBS~\cite{Emerick:2011xu}, we start by revisiting the inclusive $R_{\rm AA}$'s for $\Upsilon(1S)$
and $\Upsilon(2S)$ within the SBS approach (with vacuum binding energies, massless gluo-dissociation rates
and a quasiparticle EoS, and input cross sections as used in Ref.~\cite{Emerick:2011xu} with an up to 25\%
shadowing in central collisions for both open bottom and bottomonia), but with updated feeddown fractions
[albeit neglecting explicit feeddown from $\Upsilon(3S)$ and $\chi_b(2P)$ states].
We then turn to the TBS calculation with all updates included, also treating $\Upsilon(3S)$ and $\chi_b(2P)$
and their feeddown contributions explicitly.

\subsection{Centrality dependence for SBS}
\label{ssec:lhc276-centrality-sbs}
\begin{figure}[!t]
\includegraphics[width=0.48\textwidth]{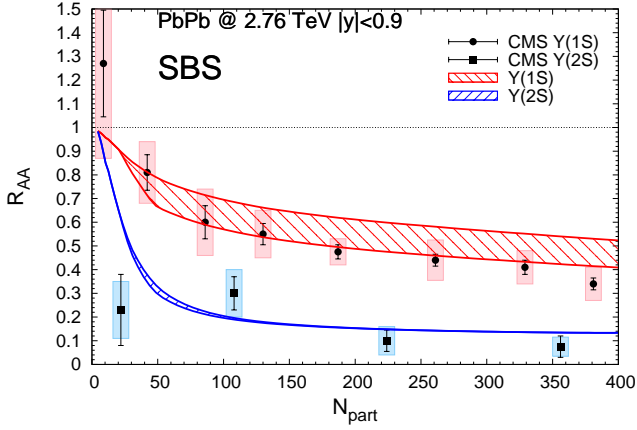}
\includegraphics[width=0.48\textwidth]{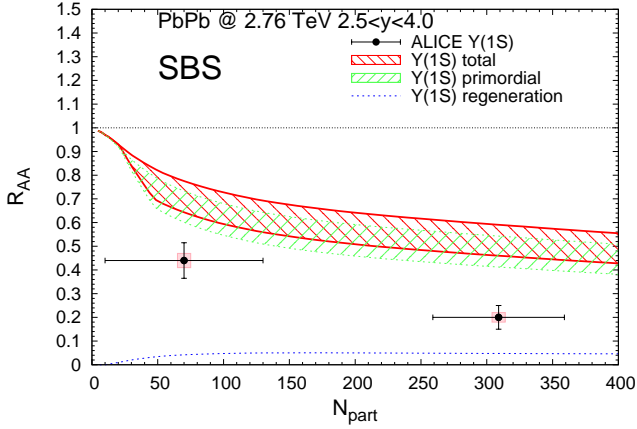}
\caption{Centrality dependence of bottomonium production in Pb-Pb(2.76\,TeV) collisions within
the SBS~\cite{Emerick:2011xu} with updated feeddowns.
Upper panel: $\Upsilon(1S)$ (red band) and $\Upsilon(2S)$ (blue band)
at mid-rapidity compared to CMS data~\cite{Khachatryan:2016xxp}.
Lower panel: $\Upsilon(1S)$ at forward rapidity compared to ALICE data~\cite{Abelev:2014nua},
where we also show the regeneration contribution (dotted line). The band widths of the totals in both
panels are due to a 0-25\% variation in the shadowing suppression of the initial bottomonium yields.}
\label{fig_sbs-276}
\end{figure}

We compare the feeddown-updated SBS to the recent CMS data~\cite{Khachatryan:2016xxp}
in Fig.~\ref{fig_sbs-276}; we find fair agreement with the strong $\Upsilon(2S)$ suppression
while the $\Upsilon(1S)$ yields tend to be somewhat overestimated, essentially due to the
now smaller feeddown. The stronger suppression exhibited by the forward-rapidity
ALICE data~\cite{Abelev:2014nua} cannot be reproduced, as before. The $b\bar b$ production cross
sections do not vary strongly enough with rapidity to generate the extra suppression. In particular,
the regeneration contribution at this energy does not provide a quadratic dependence on the
open-bottom cross section since the bottom yields are in the canonical limit,
\ie, with no more than one $b\bar b$ pair in the fireball.

\subsection{Centrality Dependence for TBS}
\label{ssec:lhc276-centrality-tbs}
Next we turn to the results of our updated approach based on the TBS.
Aside from the cross section inputs, CNM effects are implemented via a shadowing
suppression of both open bottom and bottomonia of up to 15\,\% at mid-rapidity, estimated from EPS09
NLO calculations~\cite{Eskola:2009uj} and ATLAS p-Pb data for $\Upsilon(1S)$~\cite{Atlas:2015conf},
and up to 30\,\% at forward/backward rapidity from p-Pb data from LHCb~\cite{Aaij:2014mza}
and ALICE~\cite{Abelev:2014oea}.
\begin{figure}[!t]
\includegraphics[width=0.48\textwidth]{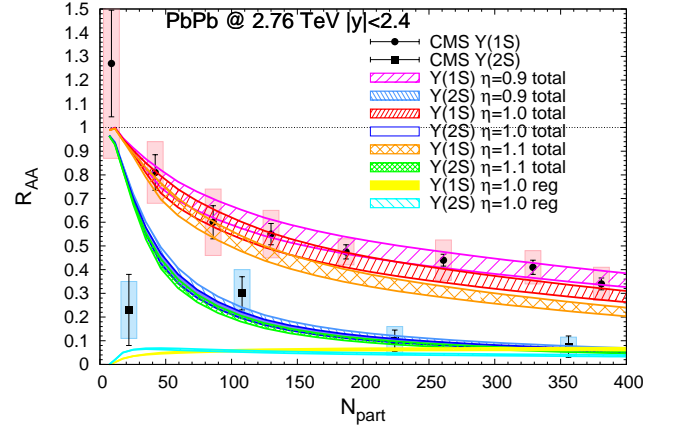}
\includegraphics[width=0.48\textwidth]{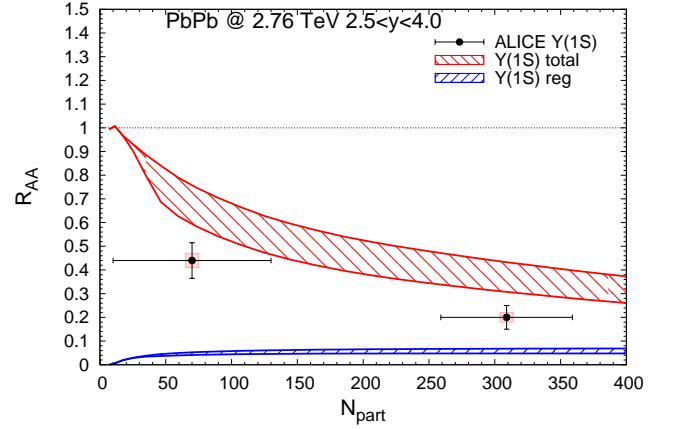}
\caption{Centrality dependence of $\Upsilon(1S)$ and $\Upsilon(2S)$ production in
Pb-Pb(2.76\,TeV) collisions within the TBS.
Upper panel: inclusive $\Upsilon(1S)$ and $\Upsilon(2S)$ results at mid-rapidity for
$\eta$=0.9, $\eta$=1.0 and $\eta$=1.1 scenarios compared to CMS data~\cite{Khachatryan:2016xxp}.
The red (blue) band is the total $\Upsilon(1S)$ ($\Upsilon(2S)$) $R_{\rm AA}$ for baseline $\eta=$1.0,
the pink (light blue) band is the total $\Upsilon(1S)$ ($\Upsilon(2S)$) $R_{\rm AA}$ for $\eta$=0.9,
the orange (green) band is the total $\Upsilon(1S)$ ($\Upsilon(2S)$) $R_{\rm AA}$ for $\eta$=1.1,
and the yellow (cyan) band is the $\Upsilon(1S)$ ($\Upsilon(2S)$) regeneration contribution with $\eta$=1.0.
The regeneration components for different scenarios have tiny differences.
The bands reflect the uncertainty due to shadowing between 0-15\%.
Lower panel: inclusive $\Upsilon(1S)$ (red band) and regenerated component (blue band) at forward
rapidity for $\eta$=1.0, compared to ALICE data~\cite{Abelev:2014nua};
the bands reflect the uncertainty due to a shadowing reduction between 0-30\%.}
\label{fig_tbs-276}
\end{figure}

The baseline TBS (with $\eta$=1.0) provides a fair description of the recent CMS
data~\cite{Khachatryan:2016xxp} for both $\Upsilon(1S)$ and $\Upsilon(2S)$ data; cf.~upper panel of
Fig.~\ref{fig_tbs-276}.
Compared to the (feeddown-updated) SBS shown in the previous figure, the
additional $\Upsilon(1S)$ suppression appears to be less than one might have expected given the much
reduced binding energies. The main reason for this is the now massive thermal quasiparticles
in the dissociation rates (as dictated by a more realistic EoS), which render gluo-dissociation
ineffectively. The addition of the quasifree rates within the TBS leads to an overall increase
of the rates compared to the SBS, but not by much. The inclusion of correlation volume effects leads
to an increase of the regeneration component, while the $B$-meson resonance states close to $T_{\rm pc}$
reduce it. Regeneration is relatively small for the ground state, but amounts to about $\sim$50\% of
the $\Upsilon(2S)$ yield in central collisions. This is somewhat
smaller than in the SBS where it is the dominant contribution, which improves the description of the
semi-central and central CMS data [we will elaborate on the quantitative role of the $B$-meson resonance states
in the $\Upsilon(2S)$ regeneration contribution in the next section].
The enhanced suppression of the $\Upsilon(1S)$, relative to the SBS, is welcome in comparison to the
forward-rapidity ALICE data (lower panel of Fig.~\ref{fig_tbs-276}), although the latter are still
significantly overpredicted.

To test the sensitivity of our results to a key in-medium property of the bottomonia, \ie, their
temperature-dependent binding energy as a measure of color screening, we additionally display in
the upper panel of Fig.~\ref{fig_tbs-276} the results of calculations where the baseline TBS
binding energies are less (further) reduced, by decreasing (increasing) the in-medium reduction
of $E_B$ relative to the vacuum by 10~\%; recall Eq.~(\ref{eta}). This is implemented by changing
the parameter $\eta$=1.0 to $\eta$=0.9 ($\eta$=1.1), displayed by the solid vs. dashed (dash-dotted)
lines in Fig.~\ref{fig_EB} (recall that $\eta$=0 recovers the vacuum binding SBS). One finds a
significant increase (decrease) of the inclusive $\Upsilon(1S)$ $R_{\rm AA}$, while the
$\Upsilon(2S)$ $R_{\rm AA}$ is little affected [since $E_B(T)$ is already small].
Thus the inclusive $\Upsilon(1S)$ can in principle serve as a measure of color screening, provided
other modeling uncertainties can be sufficiently controlled, as originally envisaged in
Ref.~\cite{Grandchamp:2005yw}. In the following section, we will therefore scrutinize several of
these uncertainties quantitatively.
Since $\eta$=1.0 provides a compromise between the CMS and ALICE data, we adopt this
value from hereon as our default (unless otherwise noted).

\subsection{Sensitivity to model parameters for TBS}
\label{ssec:lhc276-sensitivity-tbs}
This section is dedicated to quantify model dependencies unrelated to the in-medium binding energies of
the bottomonium states. Specifically, we will quantify uncertainties in the implementation of the following
components: (i) $B$-meson resonance formation, (ii) correlation volume, (iii) bottomonium formation time,
(iv) QGP formation time, (v) fireball expansion, and (vi) $b$-quark relaxation time.
We will discuss all these effects relative to our baseline TBS results (without shadowing), mostly
focusing on (but not limited to) the centrality dependence of the $\Upsilon(1S)$ $R_{\rm AA}$ at mid-rapidity in Pb-Pb (2.76\,TeV)
collisions, with selected results also for the $\Upsilon(2S)$ and $\Upsilon(3S)$.

\begin{figure}[!t]
\includegraphics[width=0.48\textwidth]{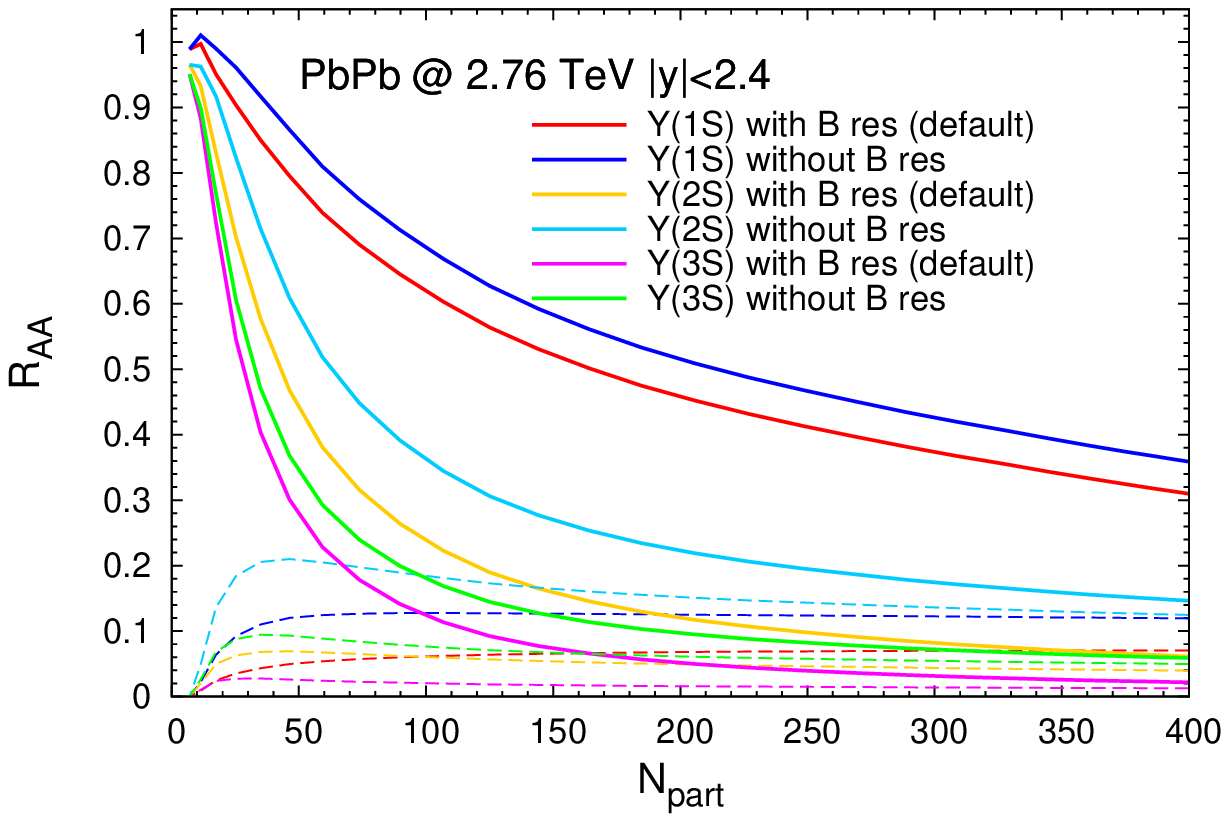}
\includegraphics[width=0.48\textwidth]{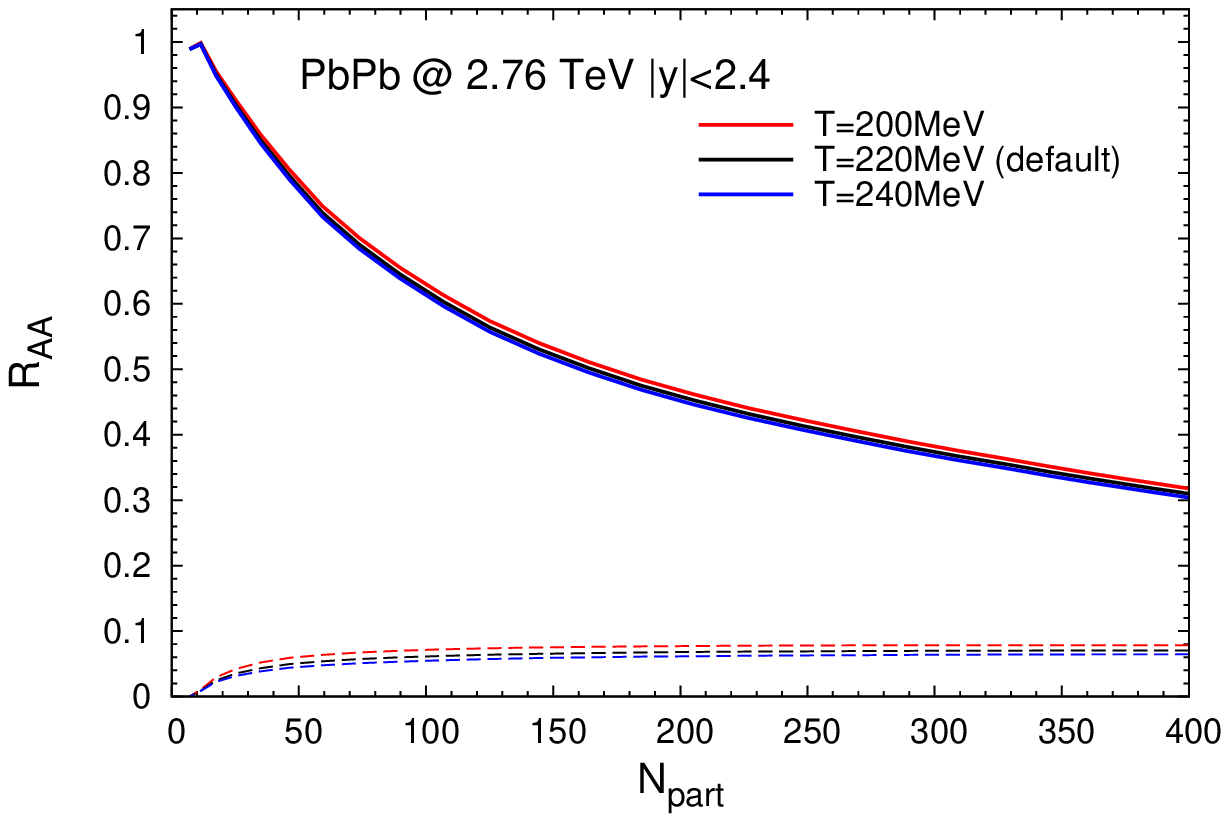}
\caption{Upper panel: comparison of $\Upsilon(1S)$, $\Upsilon(2S)$, and $\Upsilon(3S)$ $R_{\rm AA}$'s
(solid lines: total; dashed lines: regeneration contribution) for the TBS with and without
$B$-meson resonance states near $T_{\rm pc}$, assuming the default switching temperature of
$T$=220\,MeV. The red/blue curves are with/out resonances for $\Upsilon(1S)$, the
cyan/yellow curve are with/out resonances for $\Upsilon(2S)$, and
the magenta/green curves are with/out resonances for $\Upsilon(3S)$.
Lower panel: sensitivity of the default scenario to the switching temperature.}
\label{fig_tbs-model-res}
\end{figure}
In the upper panel of Fig.~\ref{fig_tbs-model-res} we display the comparison of total and regenerated
contributions when switching off the presence of the $B$-meson resonance states in the calculation of
the $Y$ equilibrium limits near $T_{\rm pc}$ (recall Fig.~\ref{fig_RVSTsmooth}). Without the resonance
states, the $b$-quark fugacity factor is significantly larger in this temperature range, leading to an
increase in the regeneration (while the primordial contribution is unaffected).
The impact is most significant for the $\Upsilon(2S)$,
where the regeneration contribution increases by close to a factor of 3 in central collisions.
The effect is much less for the $\Upsilon(1S)$ (at $\sim$50\%, translating into less than 20\% for
the total), since in the temperature
range where the enhancement of the fugacity factor is active, the inelastic reaction rate of the $\Upsilon(1S)$
is already rather small, \eg, $\Gamma_{\Upsilon(1S)}\lsim10$\,MeV at $T$=220\,MeV.
For the $\Upsilon(3S)$ the increase of the regeneration component is also close to a factor of 3, but the absolute
value of the regeneration contribution in the $R_{\rm AA}$ is smaller than for the $\Upsilon(2S)$ due to its larger mass (\ie, smaller equilibrium limit).
The relative enhancement of the regeneration components when neglecting $B$-meson resonance states is comparable at RHIC
energy (not shown here), but overall less significant due to the
generally larger primordial components compared to 2.76\,TeV.

We note that the calculations published in our recent papers~\cite{Du:2017hss,Rapp:2017chc} did not yet include the
$B$-meson resonance effects, which indeed led to problems with overestimating the $\Upsilon(2S)$
yields measured by CMS in semi-central and central Pb-Pb collisions at both 2.76 and 5.02\,TeV. This problem
is now largely resolved upon inclusion of this effect, which, as we mentioned above, is consistent
with recent analysis of lQCD results for $c$-quark susceptibilities~\cite{Mukherjee:2015mxc}, and was
predicted by $T$-matrix calculations with the $U$ potential in Refs.~\cite{vanHees:2007me,Riek:2010fk}.
Thus, the qualitative feature of heavy-light resonances above $T_{\rm pc}$ is by now well established,
but one still needs to further check its implementation. Toward this end we show in the lower panel of
Fig.~\ref{fig_tbs-model-res} the sensitivity of the
$\Upsilon(1S)$  $R_{\rm AA}$ to the onset temperature assumed for the $B$-meson formation; it turns out to be small.

\begin{figure}[!t]
\includegraphics[width=0.48\textwidth]{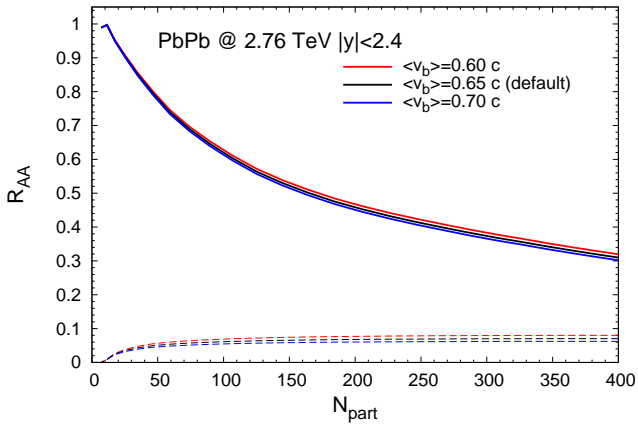}
\caption{Sensitivity of $\Upsilon(1S)$ production to the mean $b$-quark speed, $\langle v_b\rangle$, by which
the correlation volume
expands (red, black and blue lines are for $\langle v_b\rangle$=0.6$c$, 0.65$c$, and 0.7$c$, respectively).
Solid (dashed) lines are for the total (regeneration component of the) $\Upsilon(1S)$ $R_{\rm AA}$
regeneration component}
\label{fig_tbs-model-chem}
\end{figure}

Second, we test the sensitivity to the modeling of the $b$-quark correlation volume, Eq.~(\ref{Vcorr}),
by varying the mean speed, $\langle v_b\rangle$, with which the $b$ and $\bar b$ quark expand the
radius of the volume within which canonical (or ``diagonal") regeneration can occur. A larger speed
leads to a larger correlation volume which increases the available phase space for $b$ quarks and thus
decreases the $b$-quark fugacity, $\gamma_b$, and the pertinent
regeneration yield. This model component creates a small
uncertainty in the $\Upsilon(1S)$ $R_{\rm AA}$;
cf.~Fig.~\ref{fig_tbs-model-chem}.
\begin{figure}[!t]
\includegraphics[width=0.48\textwidth]{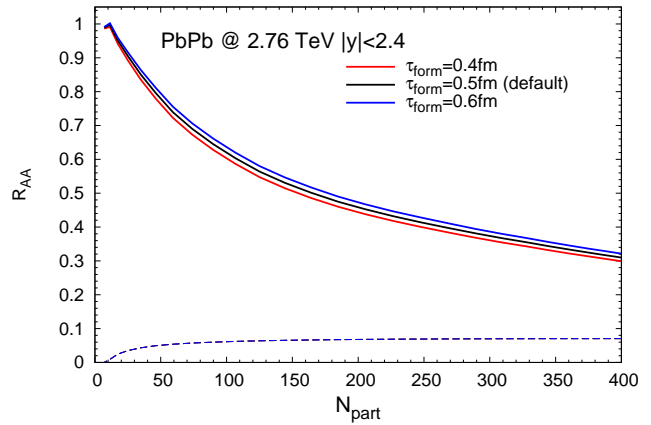}
\caption{Sensitivity of $\Upsilon(1S)$ production to a variation of the $Y$ formation times by $\pm$20\,\%.}
\label{fig_tbs-model-tauform}
\end{figure}

Third, we test the sensitivity to the formation times,
$\tau_{\rm form}$, of the $Y$ states, by varying the default
values of 0.5, 1.0 and 1.5\,fm for $\Upsilon(1S)$,
$\Upsilon(2S)$, and $\Upsilon(3S)$, respectively, by
$\pm$20\,\%.
Larger formation times reduce the dissociation rates in the
early stages thus resulting in less suppression of the primordial component. This is mostly relevant
for the $\Upsilon(1S)$ whose suppression has the largest sensitivity to the earliest phases.  However,
the pertinent variation of its total $R_{\rm AA}$ is below 5\%; cf.~Fig.~\ref{fig_tbs-model-tauform}.

\begin{figure}[!t]
\includegraphics[width=0.48\textwidth]{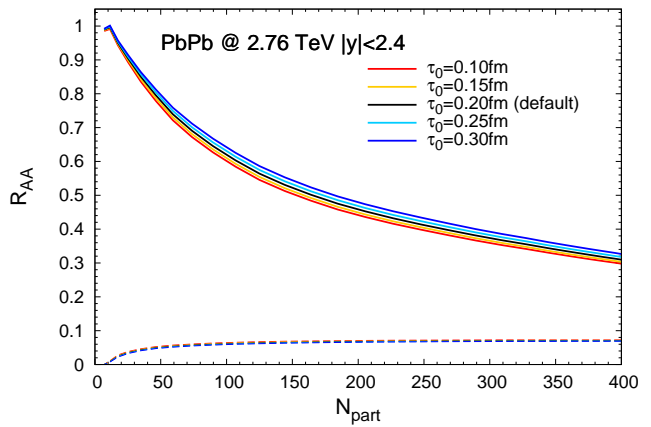}
\caption{Sensitivity of $\Upsilon(1S)$ production to a variation of the QGP formation time
over the range $\tau_0$=0.1-0.3\,fm/$c$.
The regeneration component (dashed lines) is virtually unaffected.}
\label{fig_tbs-model-tau0}
\end{figure}
Fourth, we test the sensitivity to the initial QGP formation time, $\tau_0$, which controls the initial
temperature, $T_0$. Varying $\tau_0$ by $\pm$0.1\,fm around the default value of 0.2\,fm, which implies
a formidable range of initial temperatures of $T_0$$\simeq$520-750\,MeV, produces relatively small
modifications in the $\Upsilon(1S)$ $R_{\rm AA}$; cf.~Fig.~\ref{fig_tbs-model-tau0}.
One of the reasons for this is that the $\Upsilon(1S)$ formation times ``protect" it from large dissociation
rates in the earliest phase of the medium evolution. Another reason is that, despite the large range in
temperature, the variation in the absolute time duration is actually rather small (since the default value
is already quite small), so that even rather large widths do not have a strong impact. This further implies
that pre-equilibrium evolution also has a small effect on the $Y$ production yields.

\begin{figure}[!t]
\includegraphics[width=0.48\textwidth]{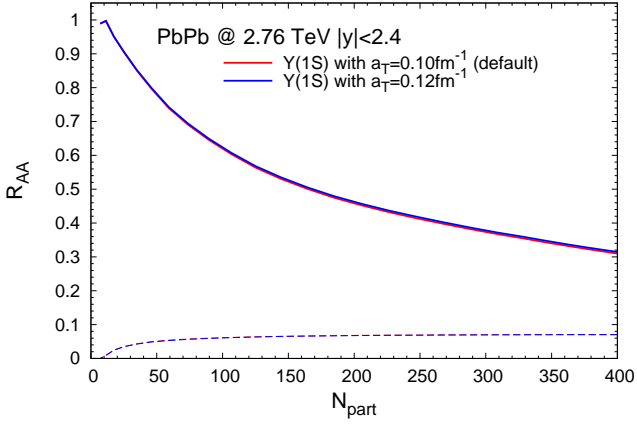}
\caption{Sensitivity of $\Upsilon(1S)$ production to a variation of the fireball expansion
acceleration over the range $a_{\rm T}$=0.10-0.12\,/fm.}
\label{fig_tbs-model-aT}
\end{figure}
Fifth, we have checked the sensitivity to the fireball expansion parametrization, Eq.~(\ref{vfb}).
When increasing the transverse acceleration by 20\%, from $a_T$=0.1/fm to 0.12/fm, both the regeneration
contribution and the total $\Upsilon(1S)$ $R_{\rm AA}$ change by
no more than within the typical line thickness of the baseline
curves; cf.~Fig.~\ref{fig_tbs-model-aT}.

\begin{figure}[!t]
\includegraphics[width=0.48\textwidth]{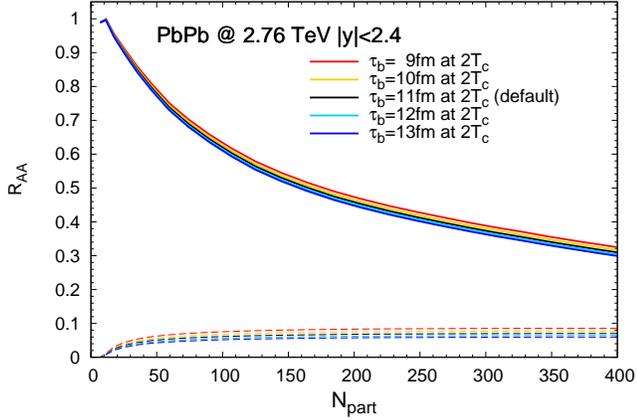}
\caption{Sensitivity of $\Upsilon(1S)$ production to the $b$-quark thermalization time, $\tau_{b}$,
over the range $\tau_{b}$=8-12\,fm at 2$T_{\rm pc}$. Dashed lines are for the regeneration component.}
\label{fig_tbs-model-relax}
\end{figure}
Sixth, we vary the thermal relaxation time of $b$ quarks, $\tau_{b}$, which controls the time scale
for approaching the $Y$ equilibrium limits. Larger relaxation times cause the equilibrium limits
to be recovered slower which reduces the regeneration contributions; recall Eq.~(\ref{eq-rfactor})
for the explicit expression of this implementation. The thermal relaxation time of heavy quarks is
one of the key transport parameters in URHICs, being proportional to the spatial heavy-quark diffusion
coefficient via ${\cal D}_s = \tau_Q (T/m_Q)$. Intense efforts are ongoing to extract this quantity
from open heavy-flavor observables, \ie, from $D$-meson $R_{\rm AA}$'s and $v_2$'s, or, in the future,
and more directly related to the present context, from $B$-meson observables. Our default choice of
$\tau_{b}$=11\,fm at a ``pivot point" of 2\,$T_{\rm c}$, with a mild temperature dependence, approximately
reflects our current knowledge of this quantity (cf.~Ref.~\cite{Prino:2016cni} for a recent review).
Not unexpectedly, the regeneration contribution to the $\Upsilon(1S)$ $R_{\rm AA}$ varies by almost
$\pm$20\% when varying this parameter by $\pm$20\%; see Fig.~\ref{fig_tbs-model-relax}. However, the
relative variation in the total $\Upsilon(1S)$ $R_{\rm AA}$ is much smaller, within $\pm$5\%. Future
analysis of open-bottom observables to extract the temperature-dependent bottom diffusion coefficient
in the QGP will help to reduce this uncertainty.

\subsection{Transverse-momentum dependence for TBS}
\label{ssec:lhc276-pt-tbs}
\begin{figure}[!t]
\includegraphics[width=0.48\textwidth]{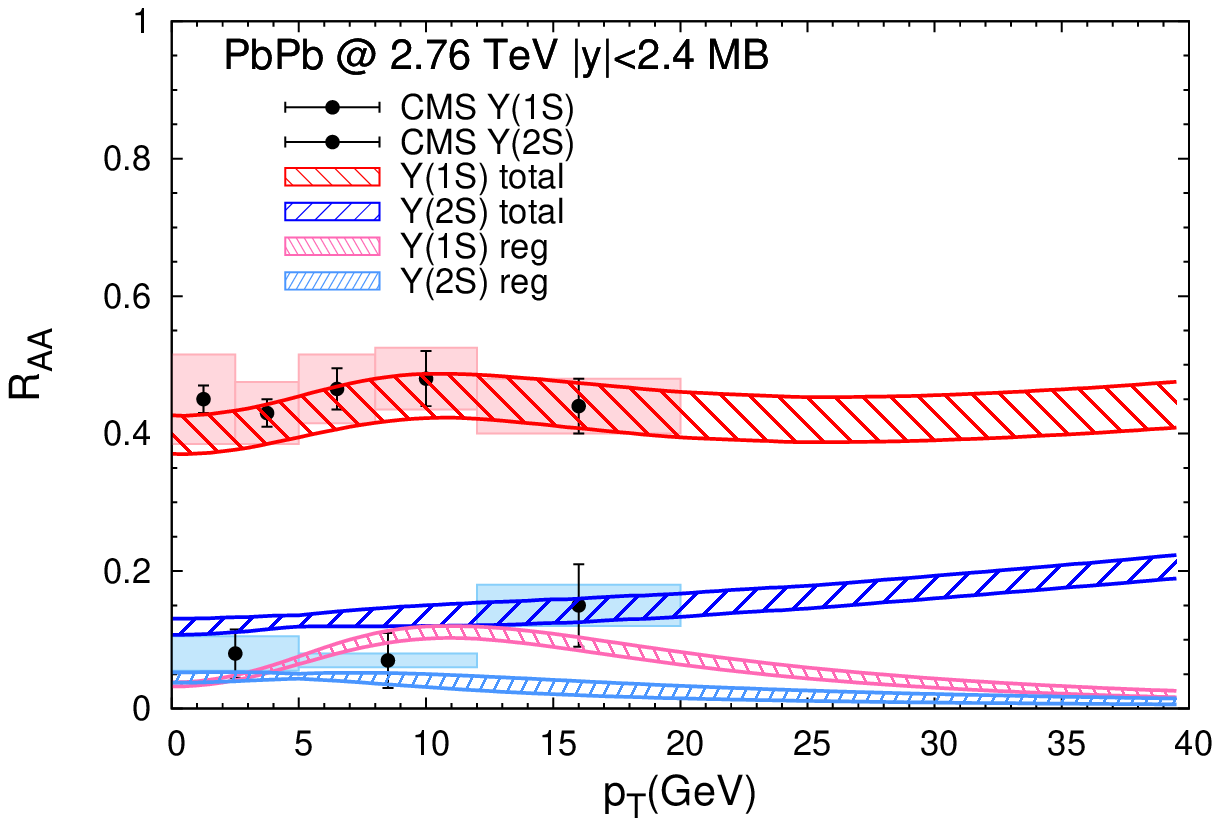}
\includegraphics[width=0.48\textwidth]{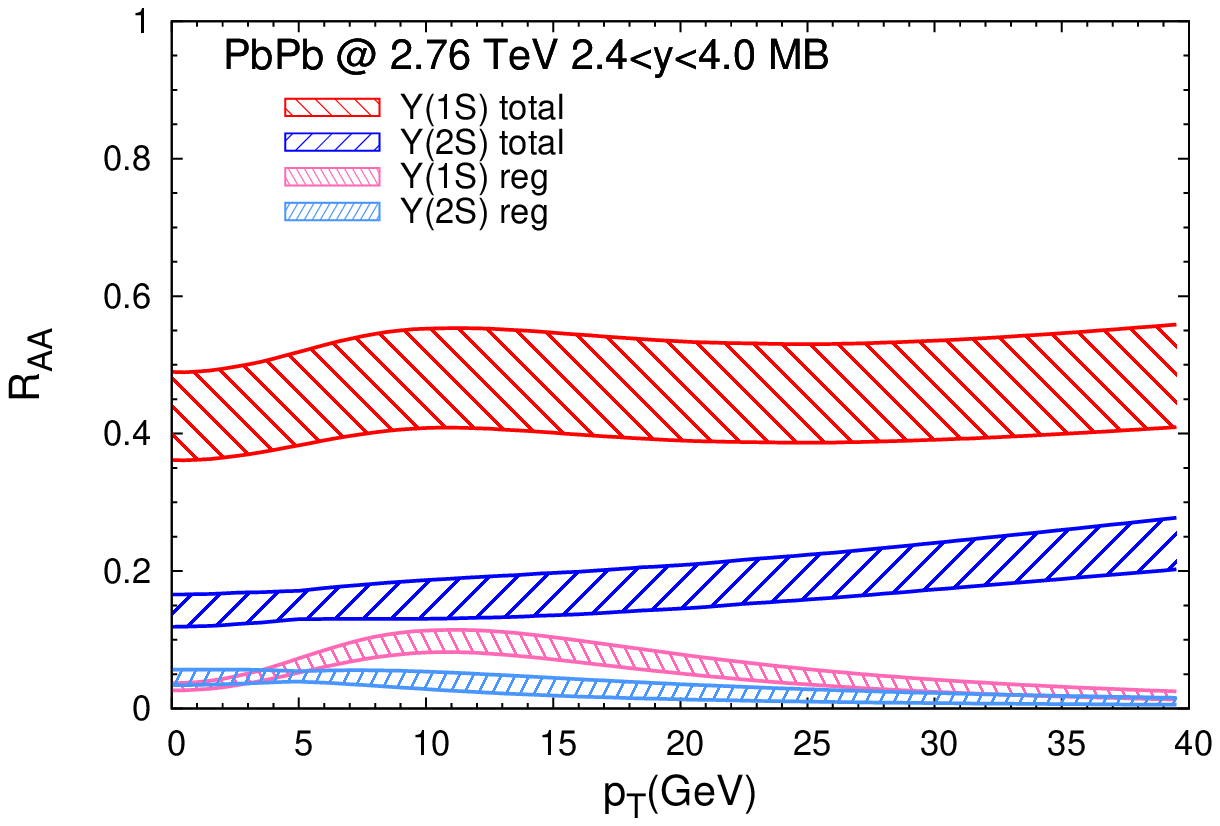}
\caption{Transverse-momentum dependent $R_{\rm AA}$ for inclusive $\Upsilon(1S)$ (red band) and
$\Upsilon(2S)$ (blue band) production and their regeneration component (pink and light blue bands,
respectively) in minimum-bias Pb-Pb(2.76\,TeV) collisions within the TBS for $\eta$=1.0 at mid- and
forward rapidity (upper and lower panel, respectively), compared to CMS data~\cite{Khachatryan:2016xxp}.
The width of the total bands includes a 0-15\% (0-30\%) shadowing suppression for
$|y|<2.4$ ($2.5<y<4.0$), the variation in the temperature window for the regeneration component
(also shown separately) and the uncertainty in the $pp$ baseline spectra.
}
\label{fig_RAAprimCMSw}
\end{figure}
For the $pp$ baseline spectra, which figure into the denominator of the $R_{\rm AA}(p_T)$, we use
the same expression, Eq.~(\ref{pp-pTspec}), as given in Sec.~\ref{ssec:rhic-pt-tbs}, but with
parameters $A$=3.0 (3.0), $D$=5.8 (6.6)\,GeV refitted to  $\Upsilon(1S)$ ($\Upsilon(2S)$) spectra
at 2.76\,TeV, as well as $A$=2.3 (2.3) and $D$=4.9 (5.9)\,GeV at
5.02\,TeV~\cite{Acosta:2001gv,Aad:2012dlq,Khachatryan:2010zg,LHCb:2012aa,Aaij:2014nwa}.

To compute the coalescence component, a temperature range for the hydro hypersurface has to be specified
to evaluate the $b$-quark spectra from the Langevin simulations in the hydrodynamic background. This
range represents the
window over which most of the regeneration of the corresponding bottomonium state occurs. Inspection
of the time (temperature) evolution of the regeneration yields reveals that the relevant temperature
windows are $\bar{T}_{\rm reg}$=220-278\,MeV for the $\Upsilon(1S)$, $\bar{T}_{\rm reg}$=183-201\,MeV for
the $\Upsilon(2S)$ and $\bar{T}_{\rm reg}$=189-212\,MeV for the $\chi_b$ states in minimum-bias (MB) Pb-Pb (2.76\,TeV)
collisions. We use the upper and lower limits of these windows to define the uncertainty band for the
$p_T$ spectra of the regenerated bottomonia.

\begin{figure}[!t]
\includegraphics[width=0.48\textwidth]{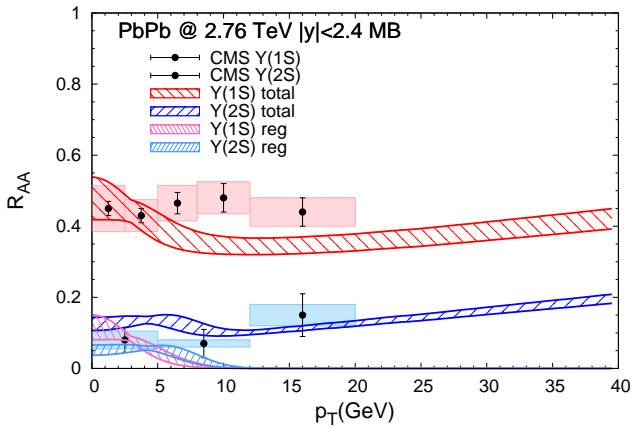}
\caption{Same as upper panel in Fig.~\ref{fig_RAAprimCMSw} but with the regeneration component evaluated
by a thermal blast-wave approximation for the respective $Y$ states.}
\label{fig_RAAprimCMS-thermal}
\end{figure}
The resulting $R_{\rm AA}(p_{T})$'s for $\Upsilon(1S)$ and $\Upsilon(2S)$ are displayed in
Fig.~\ref{fig_RAAprimCMSw} for the TBS with $\eta$=1.0. The interplay of primordial suppression
and coalescence processes results in a total $\Upsilon(1S)$ $R_{\rm AA}$'s with a mild maximum
structure around $p_T$$\simeq$$m_{\Upsilon(1S)}$, caused by the regeneration contribution, in
approximate agreement with CMS data~\cite{Khachatryan:2016xxp}. For the $\Upsilon(2S)$, we find
an over-prediction at low $p_T$, which is not really apparent in the centrality-dependent
$R_{\rm AA}(N_{\rm part})$ in Fig.~\ref{fig_tbs-276}.
However, when instead replacing the regeneration contribution with a thermal blast-wave expression
(corresponding to thermally equilibrated $b$-quark distributions), the low-$p_T$ maximum structure in
the $R_{\rm AA}(p_{T})$ becomes more pronounced and leads to larger deviations from the CMS data,
see Fig.~\ref{fig_RAAprimCMS-thermal}. It thus appears that kinetically not equilibrated $b$-quark
spectra are an important ingredient to properly interpret the bottomonium $p_T$ spectra.

\subsection{Elliptic flow for TBS}
\label{ssec:lhc276-v2-tbs}
\begin{figure}[!t]
\includegraphics[width=0.48\textwidth]{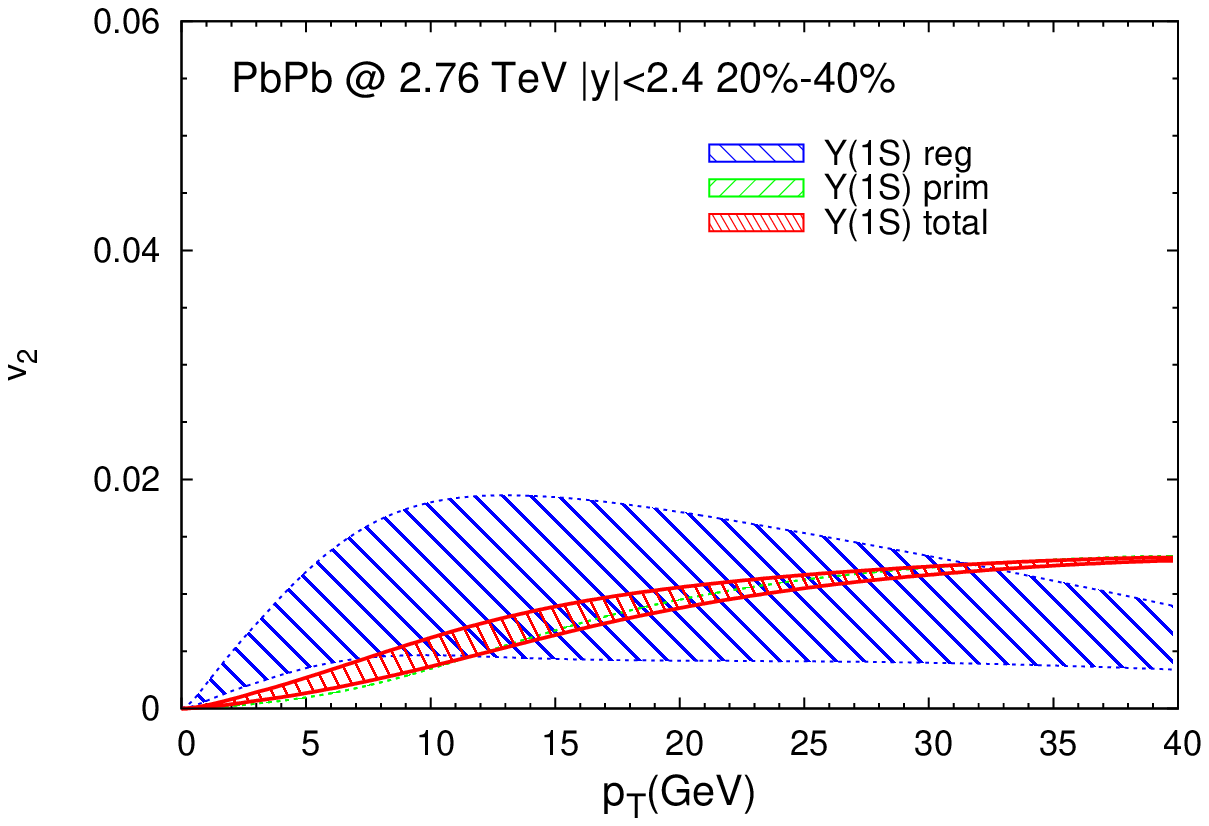}
\includegraphics[width=0.48\textwidth]{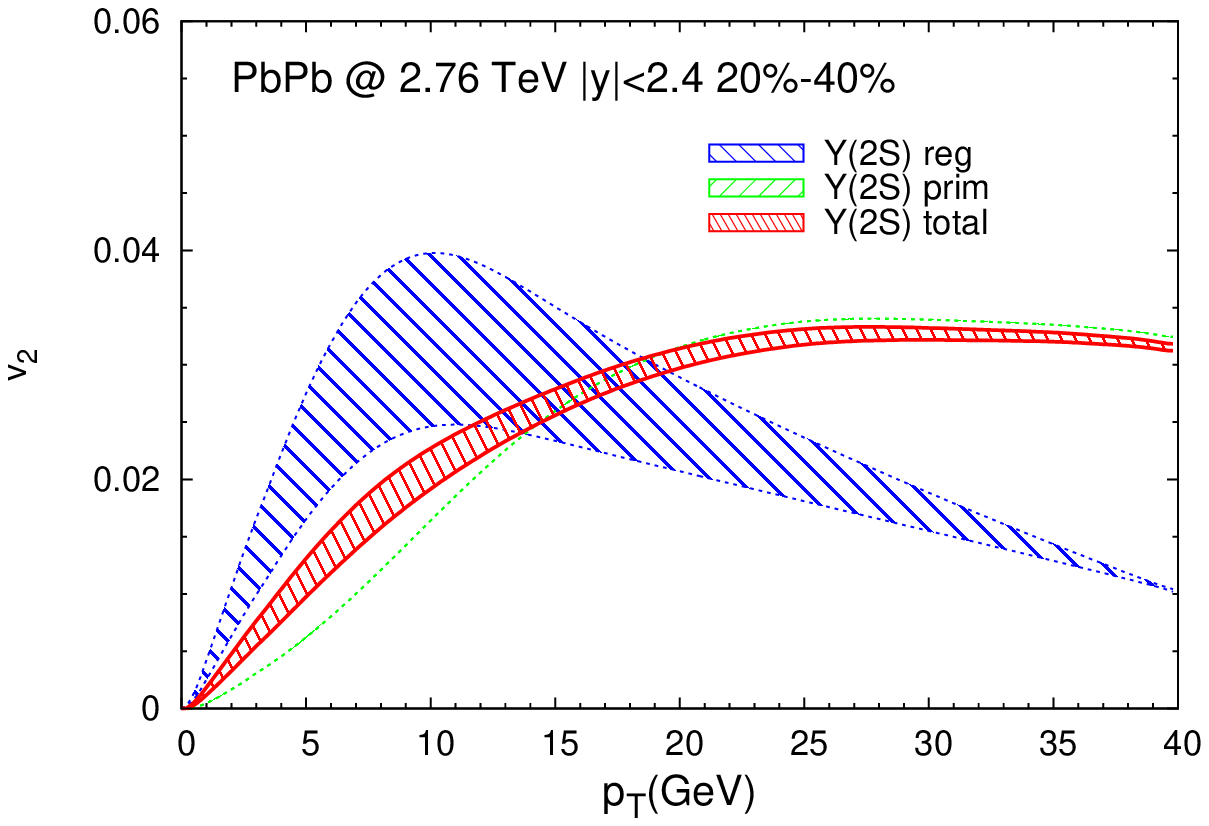}
\caption{The $p_T$ dependence of elliptic flow of $\Upsilon(1S)$ (upper panel) and  $\Upsilon(2S)$
(lower panel) in semi-central Pb-Pb(2.76\,TeV) at mid-rapidity within the TBS ($\eta$=1.0).
In both panels the blue, green, and red curves are for the regeneration component, primordial component
and their weighted sum, respectively, where the band widths reflect uncertainties from varying the
average regeneration temperatures.}
\label{fig_v2-276}
\end{figure}
Based on the bottomonium $p_T$ spectra discussed in the previous section, we provide our
predictions for their elliptic flow within the framework laid out in Sec.~\ref{sssec_v2};
see Fig.~\ref{fig_v2-276}. The same sources of uncertainties apply as encoded in the
bands for the $p_T$ spectra.
The resulting $\Upsilon(1S)$ $v_2$ turns out to be a factor of 2-3 smaller than the one of
the $\Upsilon(2S)$. However, this is not due to the larger relative contribution of the
coalescence yields, since the latter affects the total weighted $v_2$ for both particles
very little: for the $\Upsilon(1S)$ the coalescence contribution has almost no effect on the
total $v_2$, while for the $\Upsilon(2S)$ it increases the total relative to the primordial
by up to a maximum of 1\% at low $p_T$$\simeq$5\,GeV, where, however, the total $v_2$ signal
is not even at 2\%. Thus, at our predicted level of coalescence contributions, and due to their
concentration at low $p_T$ where the absolute signal is small, we conclude that it will be
very challenging at best to discern them from the primordial contributions.
On the other hand, the significantly larger {\em total} $v_2$ of the $\Upsilon(2S)$ compared
to the $\Upsilon(1S)$ is a more robust signal; it is due to the fact that the $\Upsilon(1S)$
suppression occurs earlier in the fireball evolution, where path length differences in the
suppression cannot be sensed as much as they can
for the $\Upsilon(2S)$ where the suppression mechanism is active to lower temperatures, \ie,
later in the fireball evolution. In other words, a temperature-sequential suppression,
which is widely believed to be at the origin of the difference of the $\Upsilon(1S)$ and
$\Upsilon(2S)$ yields, should also manifest itself as a difference in their $v_2$, irrespective
of regeneration.

\section{Bottomonium Production in 5.02\,TeV Pb-Pb collisions}
\label{sec:502}
\begin{figure}[!t]
\includegraphics[width=0.48\textwidth]{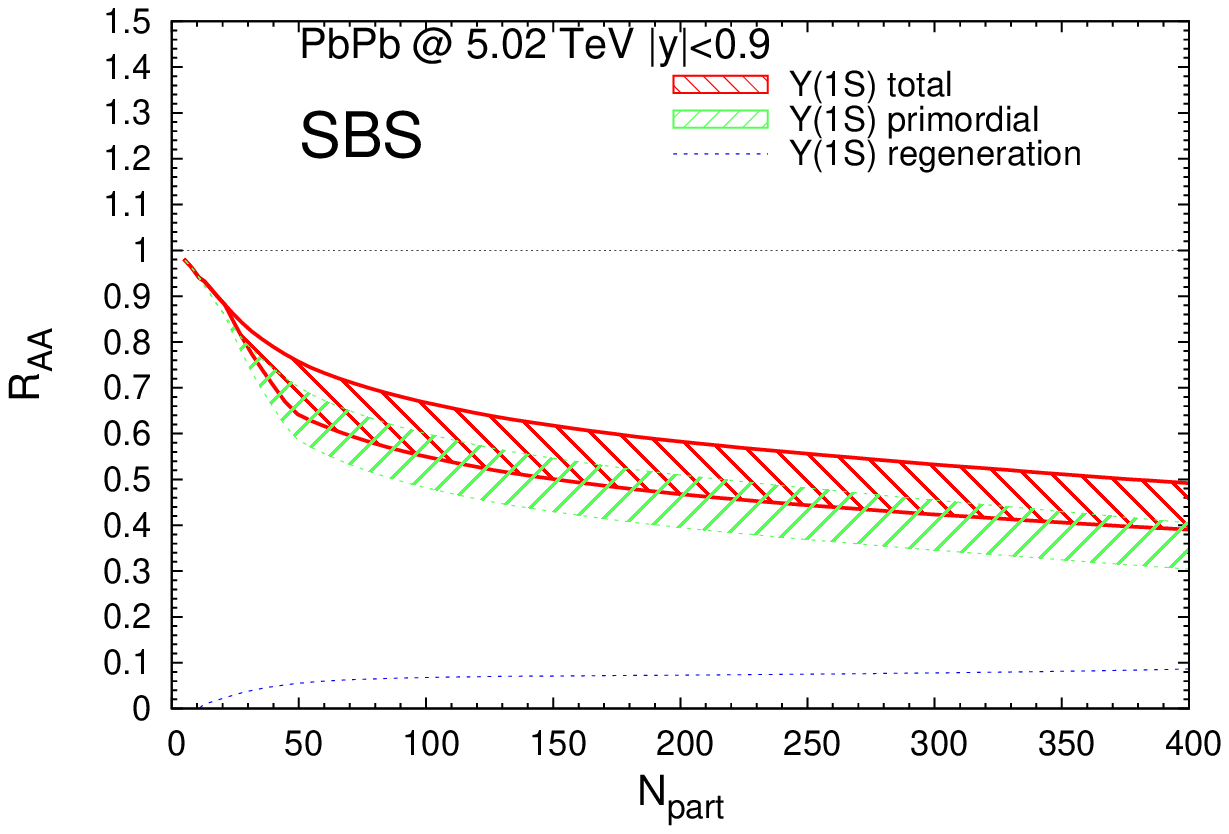}
\vspace{-0.6cm}

\includegraphics[width=0.48\textwidth]{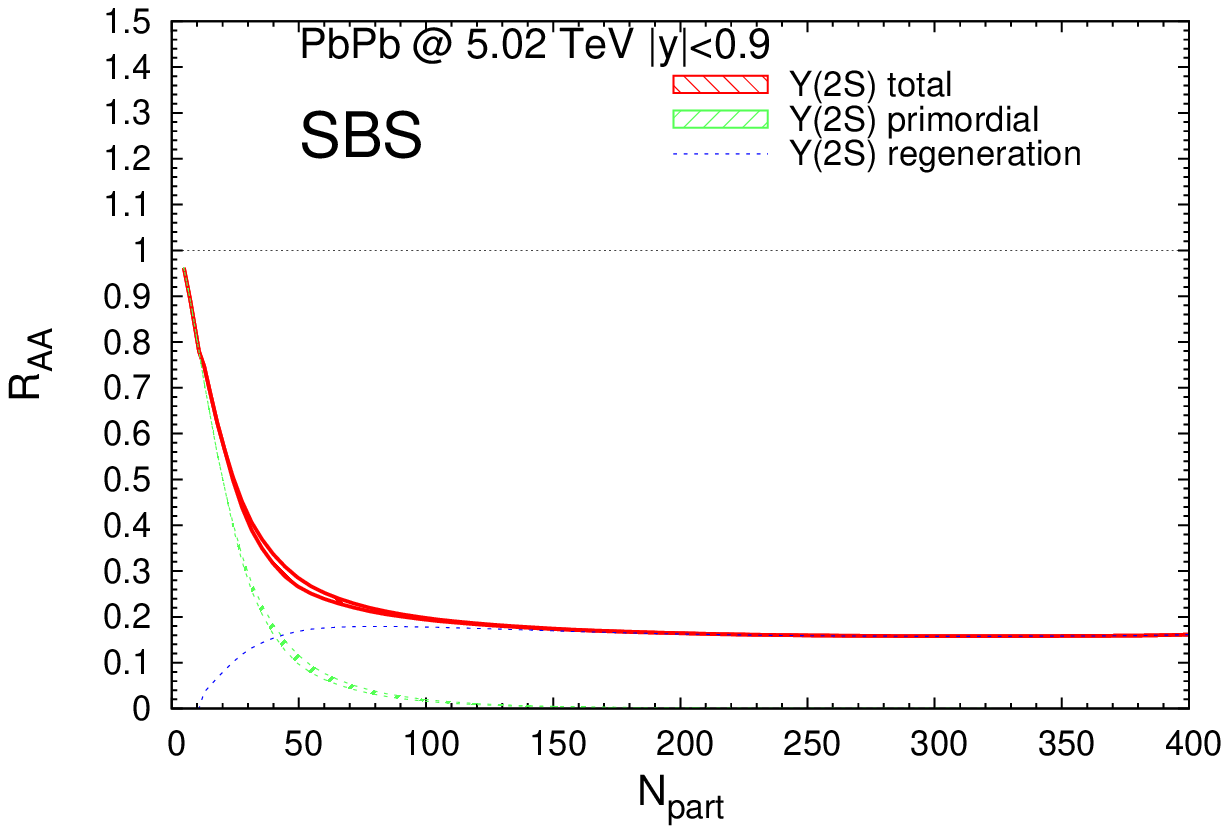}
\vspace{-0.4cm}

\caption{Centrality dependence of $R_{\rm AA}$ for $\Upsilon(1S)$ (upper panel) and $\Upsilon(2S)$ (lower
panel) within the SBS in Pb-Pb(5.02\,TeV) collisions at mid-rapidity. Red, green and blue lines represent the
total, primordial and regeneration contributions, where the bands reflect a 0-25\% shadowing effect.}
\label{fig_sbs-502mid}
\end{figure}
\begin{figure}[!t]
\includegraphics[width=0.48\textwidth]{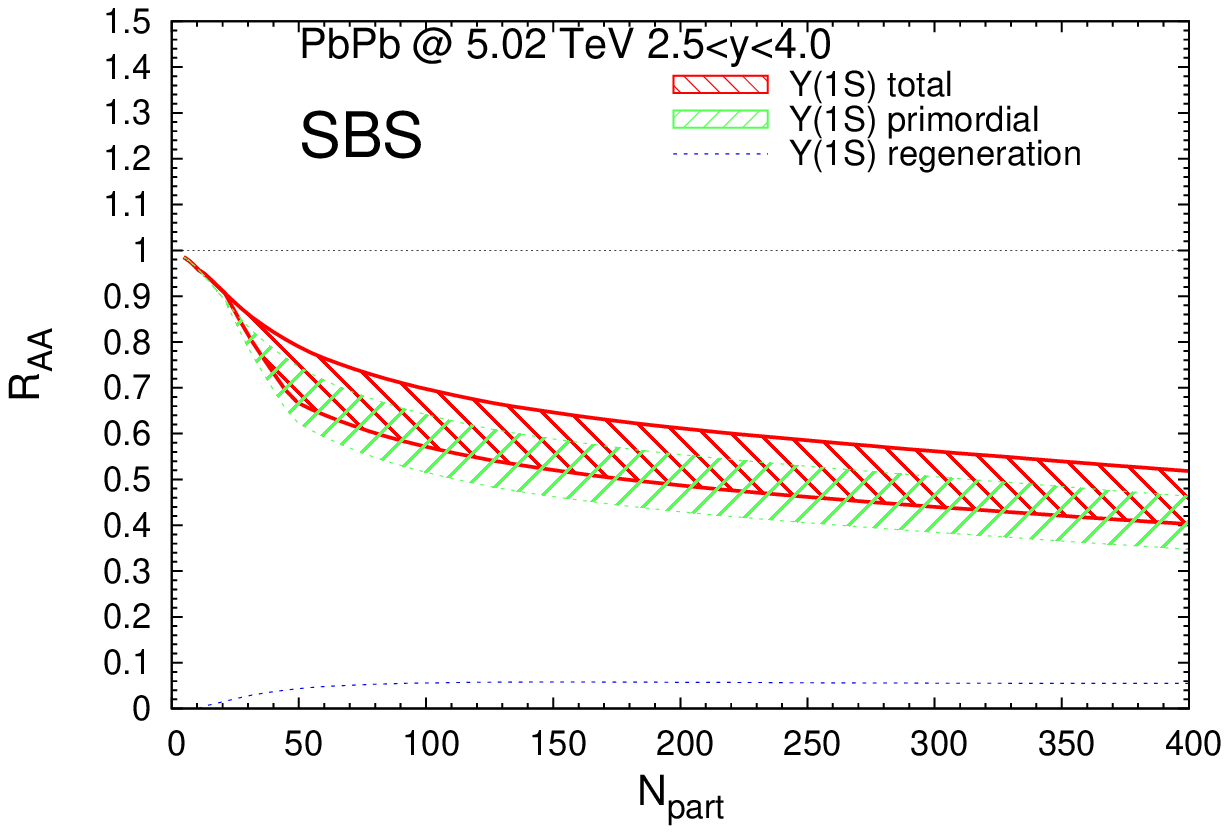}
\vspace{-0.6cm}

\includegraphics[width=0.48\textwidth]{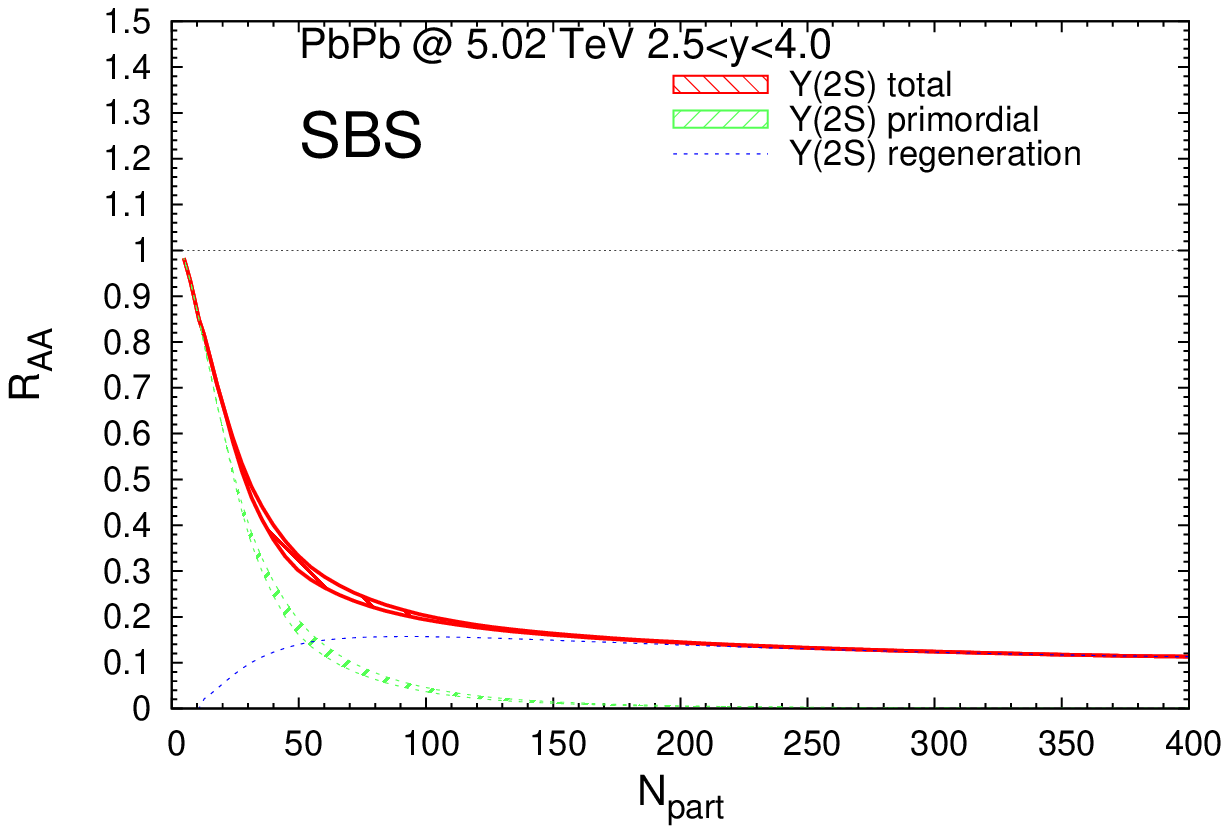}
\vspace{-0.4cm}

\caption{Same as Fig.~\ref{fig_sbs-502mid} but at forward rapidity.}
\label{fig_sbs-502for}
\end{figure}

We now turn to Pb-Pb collisions at 5.02\,TeV which were recently conducted at the LHC.,
Several new bottomonium data from this run have already become available over the course the
present work, and we include those in our discussion.
For the fireball evolution, we have assumed the charged-particle rapidity density,
$\frac{\mathrm{d}N_{ch}}{\mathrm{d}y}$, to
increase by about 22.5\,\%, from 2.76 to 5.02\,TeV, \eg, from 1750 to 2150
in 0-5\% central collisions~\cite{Niemi:2015voa}. This corresponds to an increase of
the total entropy in the fireball from 22000 to almost 27000. With an entropy density of $s\sim T^3$
in the early hot phases, the initial temperature increases by about 7\%. For the charged-particle
rapidity density we implement a reduction of 20\,\% from mid-rapidity, $|y|<2.4$, to forward
rapidity, $2.5<y<4.0$~\cite{Abbas:2013bpa}, as previously done at 2.76\,TeV.
We will start our discussion again by recalling the results from the earlier used SBS in
Sec.~\ref{ssec:lhc502-centrality-sbs}, and then turn to the centrality, $p_T$, and azimuthal-angle
dependencies for the default TBS in Sec.~\ref{ssec:lhc502-tbs}.

\subsection{Centrality Dependence for SBS}
\label{ssec:lhc502-centrality-sbs}
We first display our 5.02\,TeV results for $Y$ production in the previously used
SBS~\cite{Emerick:2011xu} with updated feeddown fractions (but without explicit treatment of
the $3S$ or $2P$ states), at both mid- and forward rapidities, cf.~Figs.~\ref{fig_sbs-502mid}
and \ref{fig_sbs-502for}, respectively.
Compared to the SBS results at 2.76\,TeV, the $\Upsilon(1S)$ suppression slightly increases by up to
$\sim$5\% in central collisions, due to stronger color screening with increased rates at higher
temperature. On the other hand, the $\Upsilon(2S)$ suppression becomes slightly less in central
collisions at 5.02\,TeV due to a small increase in regeneration, while a stronger suppression is
found for peripheral collisions ($N_{\rm part}\lsim50$), where the suppressed primordial contribution
dominates (again for both rapidity regions). This feature is reminiscent of the $J/\psi$ case.

\subsection{Centrality and transverse-momentum dependence for TBS}
\label{ssec:lhc502-tbs}
\begin{figure}[!t]
\includegraphics[width=0.48\textwidth]{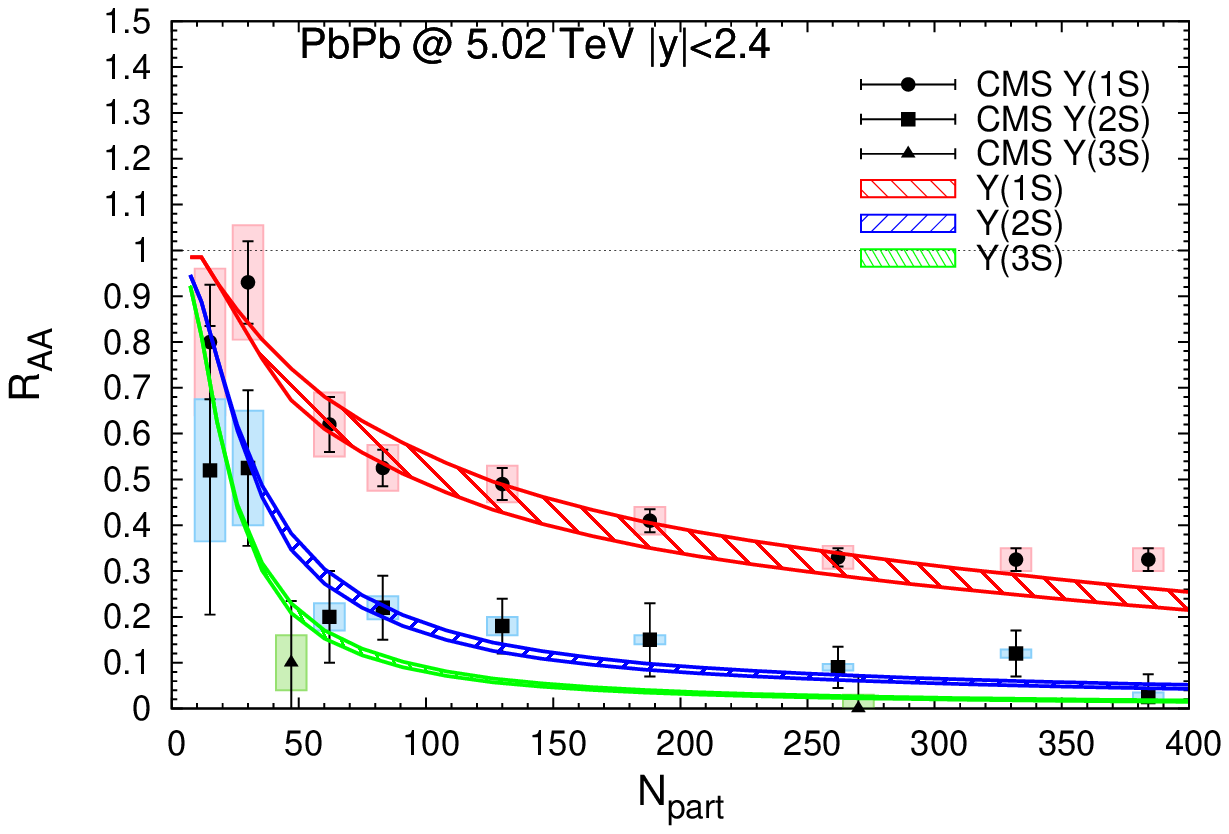}
\vspace{-0.5cm}

\includegraphics[width=0.48\textwidth]{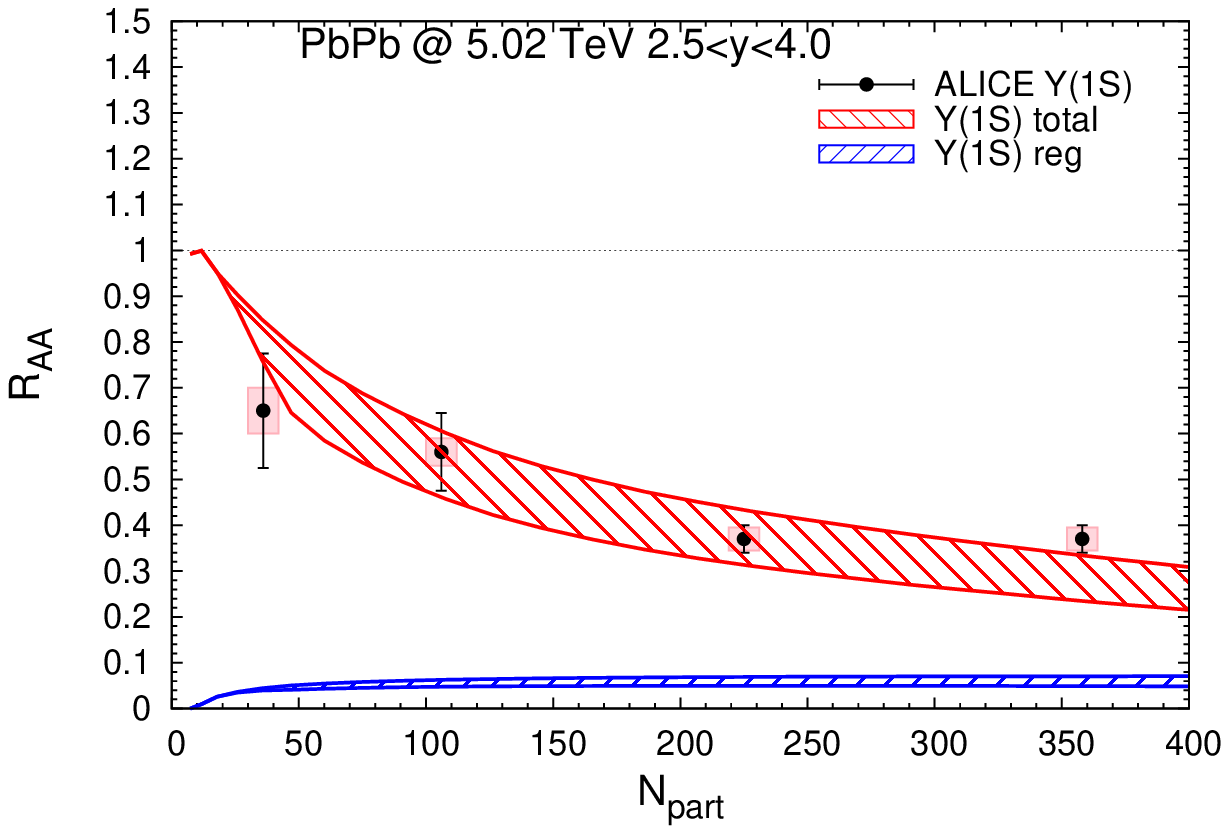}
\vspace{-0.5cm}

\caption{Centrality dependence of bottomonium $R_{\rm AA}$'s in Pb-Pb(5.02\,TeV) collisions within
the TBS ($\eta$=1.0).
Upper panel: mid-rapidity $\Upsilon(1S)$, $\Upsilon(2S)$ and $\Upsilon(3S)$ (red, blue and green bands,
respectively) compared to CMS data~\cite{Flores:2017qmcms,CMS:2017ucd};
the bands are due to a 0-15\% shadowing suppression.
Lower panel: forward rapidity $\Upsilon(1S)$ compared to ALICE data~\cite{Fronze:2016gsr,Das:2017qmalice};
the bands are due to a 0-30\% shadowing suppression.}
\label{fig_tbs-502}
\end{figure}
\begin{figure}[!h]
\includegraphics[width=0.48\textwidth]{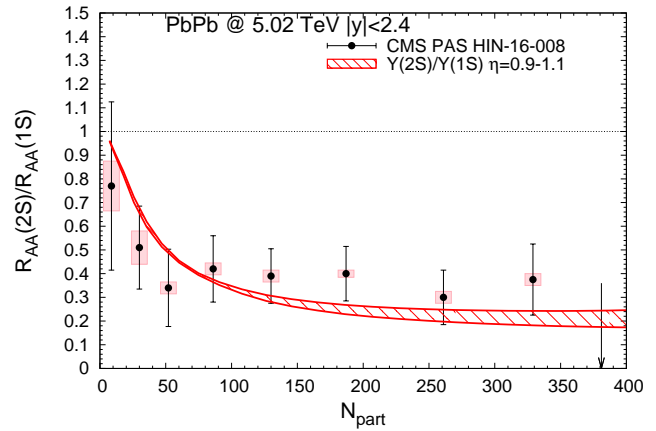}

\vspace{-0.5cm}

\caption{Centrality dependence of the $\Upsilon(2S)$/$\Upsilon(1S)$ $R_{\rm AA}$ double ratio
in 5.02\,TeV Pb-Pb collisions at mid-rapidity within the TBS (with an uncertainty band from
in-medium binding energies for $\eta$=0.9-1.1), compared to CMS data~\cite{Sirunyan:2017lzi}.}
\label{fig_tbs-502drat}
\end{figure}

Next, we turn to the TBS at 5.02\,TeV, encoding our theoretical improvements in the $Y$ transport
approach over the previously used SBS. The centrality dependence of the $R_{\rm AA}$ for
$\Upsilon(1S)$ and $\Upsilon(2S)$ at mid-rapidity is shown in the upper panel of
Fig.~\ref{fig_tbs-502}, and for the $\Upsilon(1S)$ at forward rapidity in the lower panel of
Fig.~\ref{fig_tbs-502}. They are compared to CMS data~\cite{Flores:2017qmcms,CMS:2017ucd} at
mid-rapidity and to ALICE data~\cite{Fronze:2016gsr,Das:2017qmalice} at forward rapidity, respectively.
The in-medium effects lead to a significantly stronger suppression of the $\Upsilon(1S)$ relative
to the SBS discussed in the previous section. At the same time, the $\Upsilon(1S)$ suppression
within the TBS is only slightly increased relative to the 2.76\,TeV results (recall
Fig.~\ref{fig_tbs-276}). The $\Upsilon(2S)$ $R_{\rm AA}$ also shows a small increase in
suppression by about 15\%, amounting, however, to only a $\sim$0.01 change at the absolute
level in the $R_{\rm AA}$ in central collisions. For the latter, the $\Upsilon(3S)$ is
suppressed by another factor of $\sim$2.
At forward rapidity, the comparison to recent ALICE data~\cite{Fronze:2016gsr,Das:2017qmalice},
shown in the lower panel of Fig.~\ref{fig_tbs-502},
is more favorable than it was at 2.76\,TeV.

Next, we compare our calculations for the $\Upsilon(2S)$-over-$\Upsilon(1S)$ double ratio at 5.02\,TeV
to CMS data~\cite{Sirunyan:2017lzi} in Fig.~\ref{fig_tbs-502drat}; as to be expected from the agreement
with the individual $R_{\rm AA}$'s,
the calculated double ratio also agrees fairly well with the observed centrality dependence.

\begin{figure}[!t]
\includegraphics[width=0.48\textwidth]{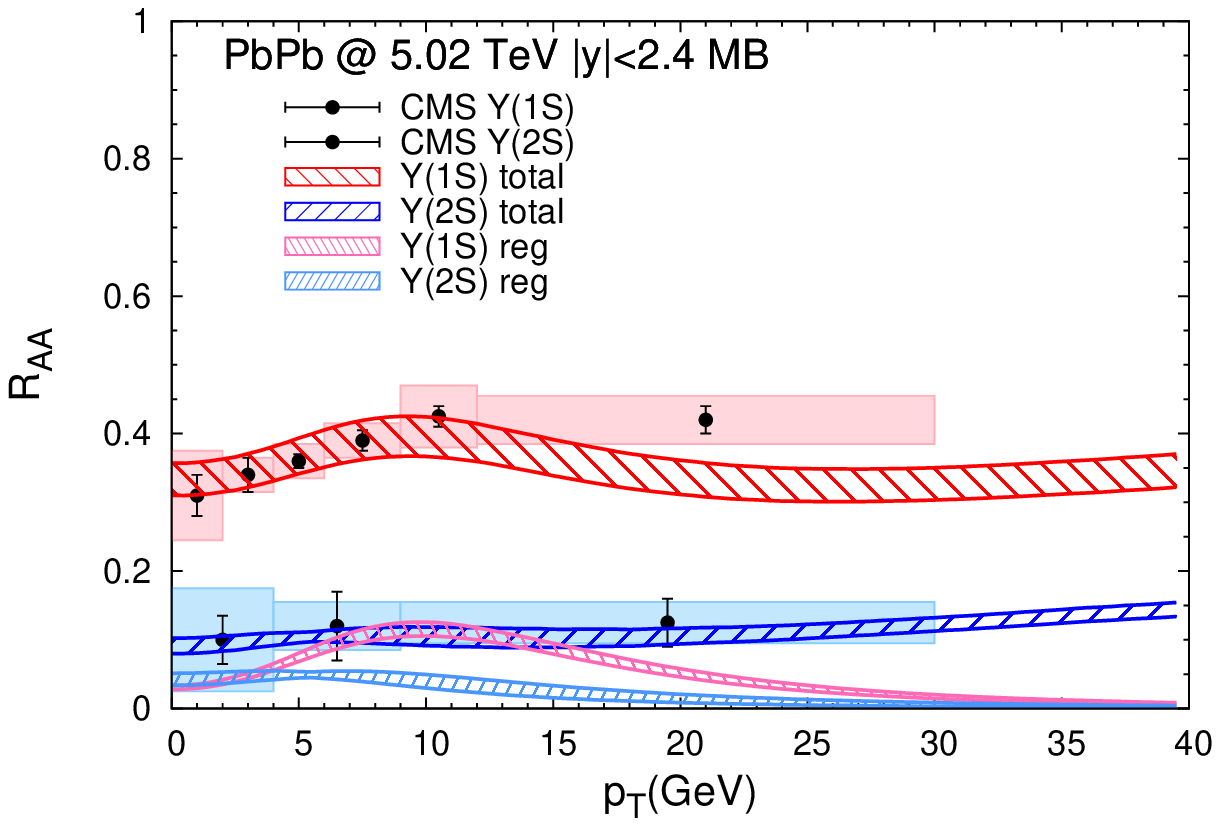}
\vspace{-0.5cm}

\includegraphics[width=0.48\textwidth]{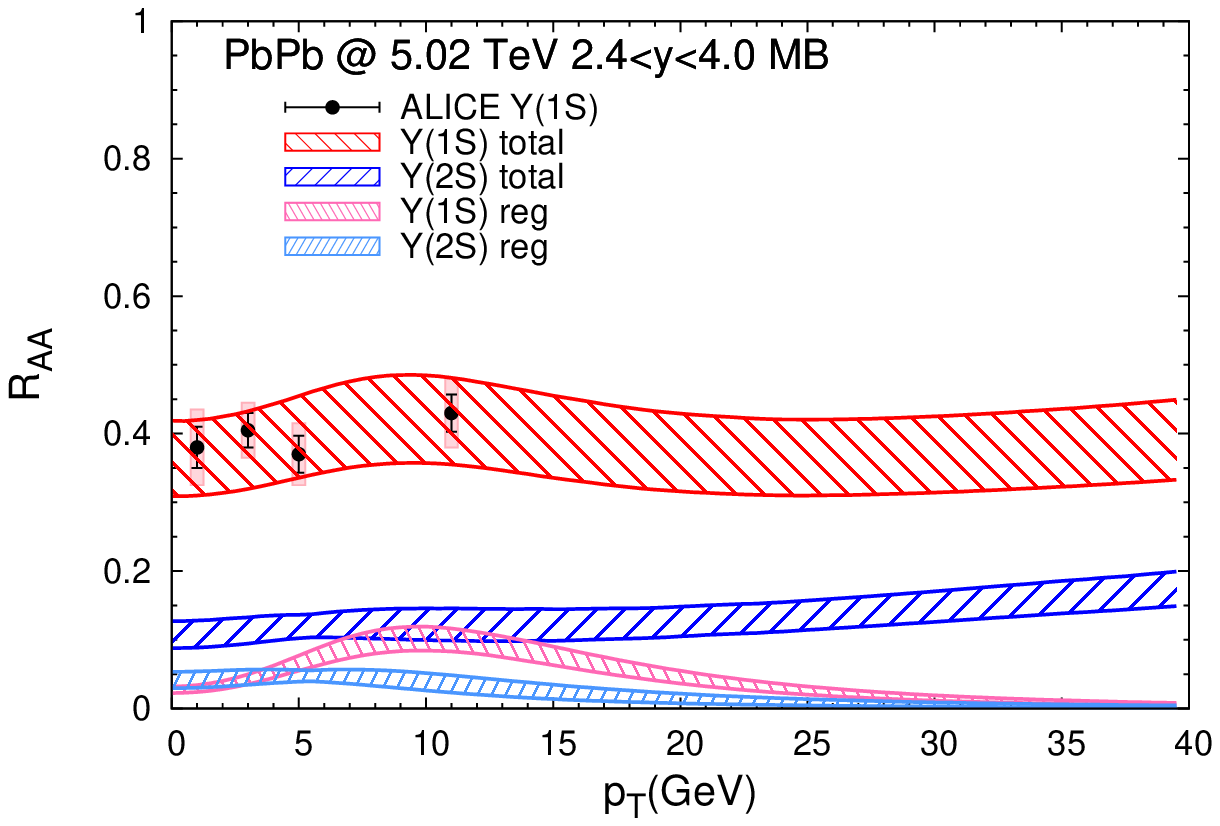}
\vspace{-0.5cm}

\includegraphics[width=0.48\textwidth]{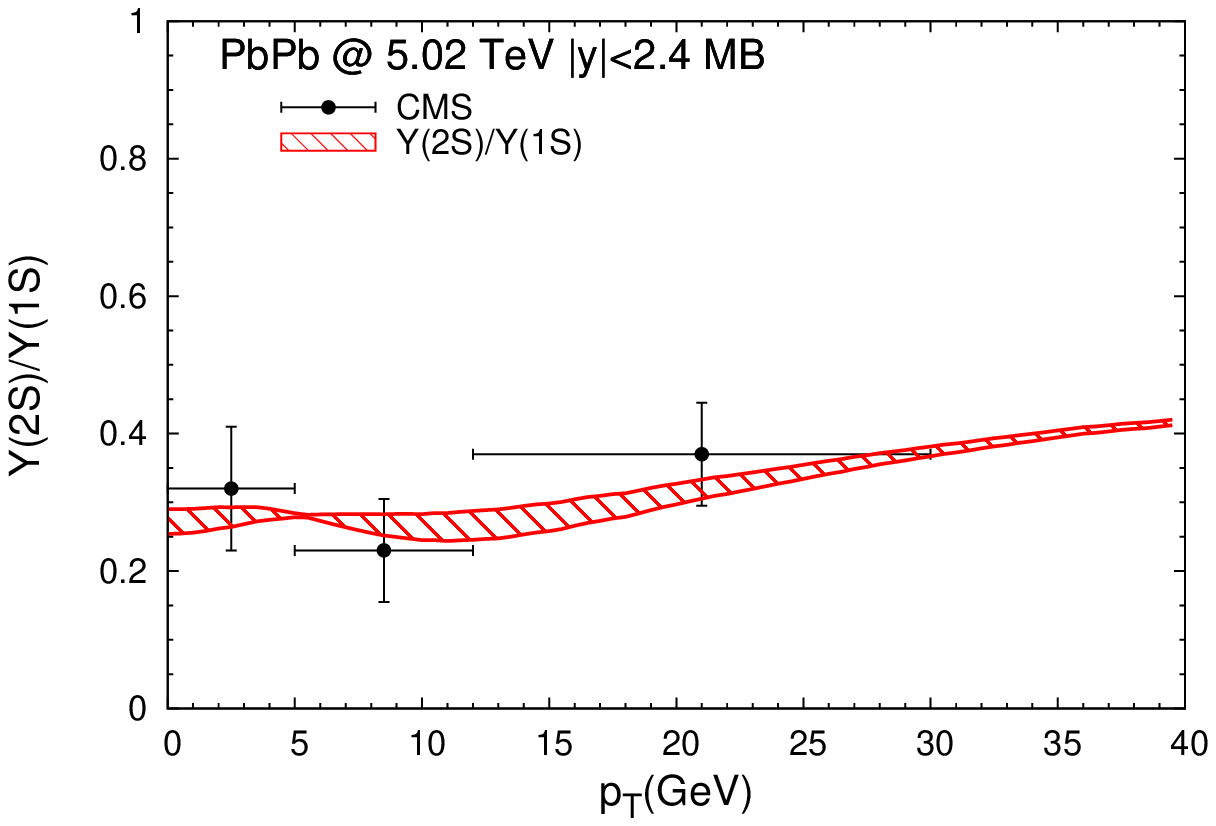}
\caption{The $p_T$ dependence of $\Upsilon(1S)$ and $\Upsilon(2S)$ yields in MB Pb-Pb(5.02\,TeV)
collisions at mid and forward rapidities within the TBS ($\eta$=1.0).
Upper panel: mid-rapidity $\Upsilon(1S)$ and $\Upsilon(2S)$ $R_{\rm AA}$ for total (red and blue curves,
respectively) and regeneration components (pink and light-blue curves, respectively),
compared to CMS data~\cite{Flores:2017qmcms,CMS:2017ucd}; the bands reflect variations
due to a 0-15\% shadowing suppression and the average regeneration temperatures ($\bar{T}_{\rm reg}$)
of the two states.
Middle panel: same as upper panel but at forward rapidity with a 0-30\% shadowing range,
compared to ALICE data~\cite{Das:2017qmalice}.
Lower panel: $\Upsilon(2S)/\Upsilon(1S)$ $R_{\rm AA}$ double ratio compared to CMS data~\cite{Sirunyan:2017lzi};
the band reflects variations in the $\bar{T}_{\rm reg}$'s.
}
\label{fig_tbs-pt502mid}
\end{figure}
Finally, we extract transverse-momentum dependent observables from our calculations, starting
with the $p_T$ dependence of the $R_{\rm AA}$ for $\Upsilon(1S)$ and $\Upsilon(2S)$ at mid- and
forward rapidities; cf.~Fig.~\ref{fig_tbs-pt502mid}. Similar to what we found at 2.76\,TeV, the $\Upsilon(1S)$
$R_{\rm AA}(p_T)$ exhibits a mild maximum structure due to the regeneration contribution computed
with non-thermalized $b$-quark spectra (taken from Langevin transport calculations at 5.02\,TeV),
at both mid- and forward rapidities. The calculations approximately agree with both CMS data at
mid-rapidity (upper panel of Fig.~\ref{fig_tbs-pt502mid}) and ALICE data at forward rapidity
(middle panel of Fig.~\ref{fig_tbs-pt502mid}).
The $\Upsilon(2S)$ $R_{\rm AA}(p_T)$ is also similar to 2.76\,TeV, with a moderate monotonous rise
with $p_T$. The absolute magnitude of the calculated $p_T$ spectra agrees better with the CMS data
than at 2.76\,TeV.
We also plot the $p_T$-dependent double ratio at mid-rapidity in the lower panel of
Fig.~\ref{fig_tbs-pt502mid}; again, based on the agreement with the individual $R_{\rm AA}(p_T)$'s
in the upper panel, no surprises are found.

\begin{figure}[!t]
\includegraphics[width=0.48\textwidth]{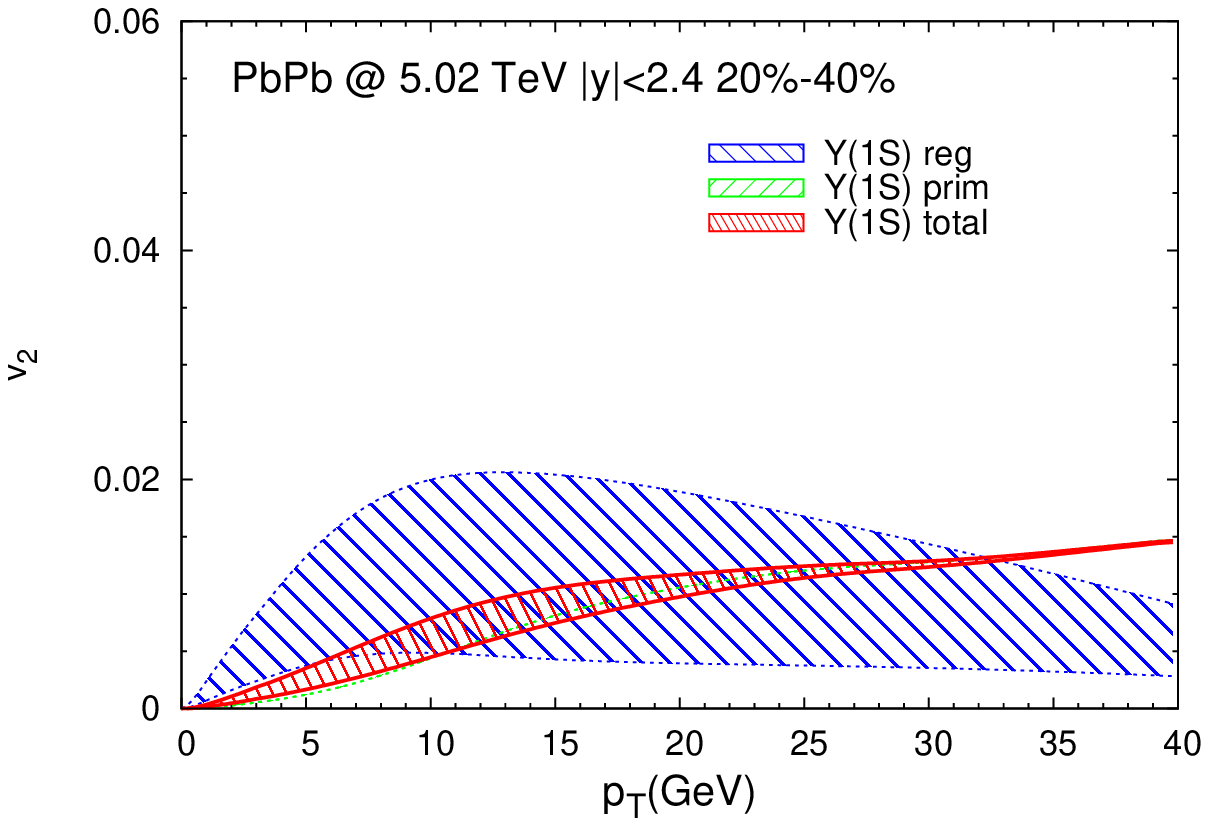}
\includegraphics[width=0.48\textwidth]{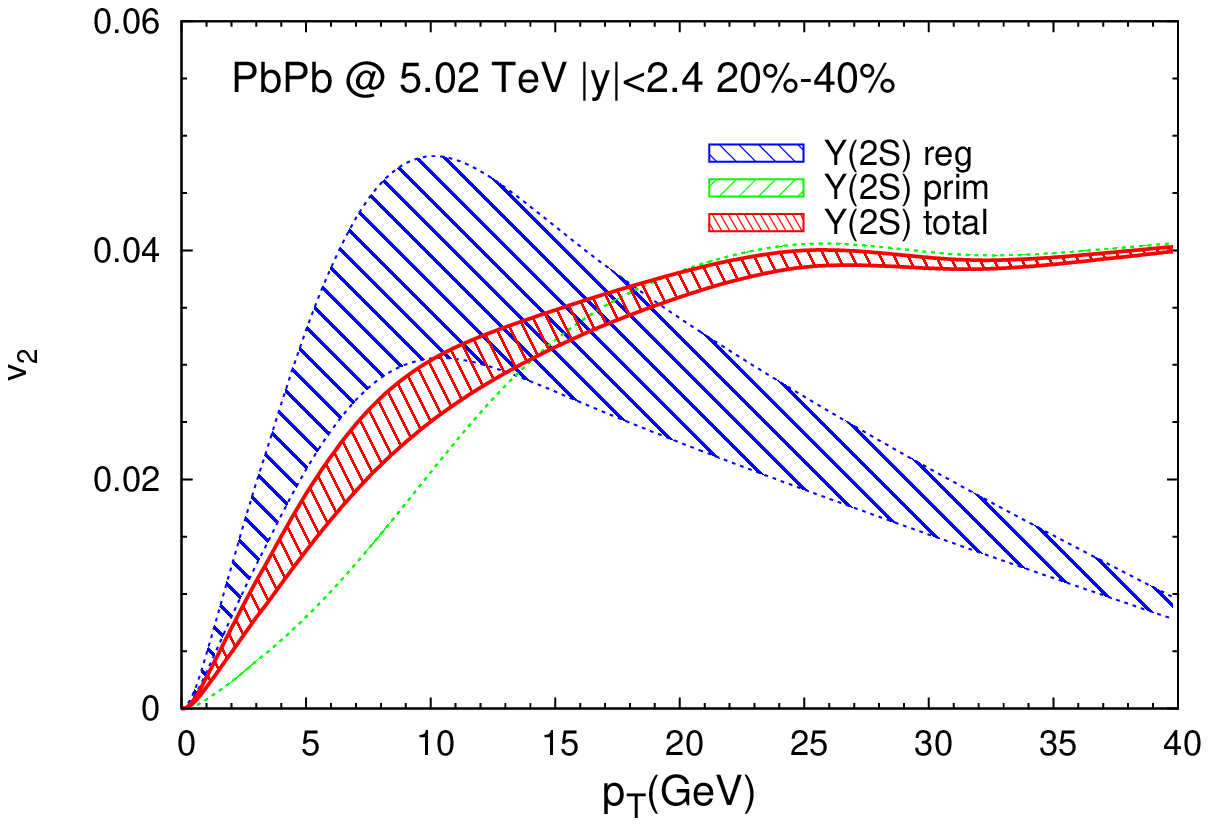}
\caption{The $p_T$ dependence of elliptic flow of $\Upsilon(1S)$ (upper panel) and  $\Upsilon(2S)$
(lower panel) in semi-central Pb-Pb(5.02\,TeV) at mid-rapidity within the TBS ($\eta$=1.0).
In both panels the blue, green, and red curves are for the regeneration component, primordial component,
and their weighted sum, respectively, where the band widths reflect uncertainties from varying
the average regeneration temperatures.}
\label{fig_v2-502}
\end{figure}
The $p_T$ dependence of the elliptic flow for $\Upsilon(1S)$ and $\Upsilon(2S)$ in mid-central Pb-Pb
collisions at mid-rapidity is displayed in Fig.~\ref{fig_v2-502}.
The $v_2$ for both the primordial and regenerated $\Upsilon(1S)$ are small, below 2\%, since both
processes occur early in the fireball evolution -- essentially within the first 2fm/$c$ -- during which
both path length differences and collective-flow anisotropies are limited. The $v_2$ is more than doubled
for the $\Upsilon(2S)$ in both components, which, after an initial rise, levels off at about 4\%.
As was the case at 2.76\,TeV, the $v_2$ does not show a very promising sensitivity to the
regeneration component, since the latter is rather small and concentrated at low $p_T$, where the
mass effect suppresses the signal. However, the difference between $\Upsilon(1S)$ and $\Upsilon(2S)$
$v_2$'s is appreciable; about a factor of $\sim$2.
We note that we did not include initial geometry fluctuations nor elastic rescattering of the $Y$
states in the medium (once they are reasonably tightly bound), which may play a role in generating
a larger $v_2$ of the primordial component at high $p_T$ or the total yields at small and moderate
$p_T$, respectively.

\section{Excitation Function}
\label{sec:excit}
\begin{figure}[!b]
\includegraphics[width=0.48\textwidth]{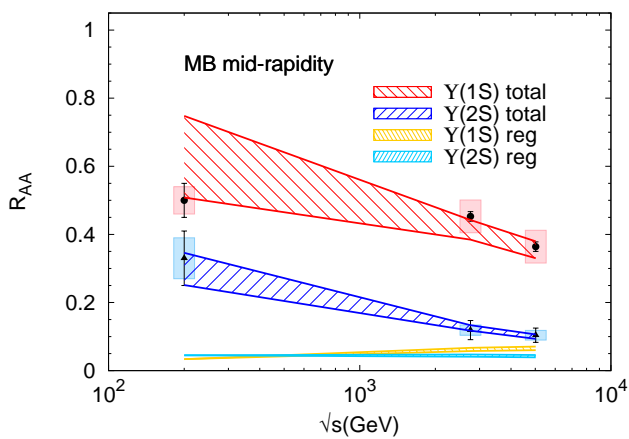}
\caption{Excitation function the MB $R_{\rm AA}$ of $\Upsilon(1S)$ and $\Upsilon(2S)$ with TBS compared to
STAR~\cite{Ye:2017fwv} and CMS~\cite{Khachatryan:2016xxp,Flores:2017qmcms,CMS:2017ucd} data
at mid-rapidity.}
\label{fig_excit}
\end{figure}
In an attempt to combine the information attained within our updated TBS approach to $Y$ production
from RHIC to top LHC energy, we compare our results for the collision energy dependence of the
minimum-bias (MB) $R_{\rm AA}$'s for $\Upsilon(1S)$ and $\Upsilon(2S)$ to STAR and CMS data at mid-rapidity
in Fig.~\ref{fig_excit}. We find a gradual increase in the suppression for both states, with a stronger
absolute suppression of the $\Upsilon(2S)$ than the $\Upsilon(1S)$ especially at the LHC. These features
support a sequential suppression scenario, rather directly reflected in both calculations and data due to
relatively small regeneration contributions. The latter is due to a combination of the small $b\bar b$ cross section
(which at current energies does not produce more than 1 pair per fireball) and the role played by $B$-meson
resonance formation near $T_{\rm pc}$. The possibly most significant indication for the regeneration
contribution is a hint for a flattening  of the $\Upsilon(2S)$ $R_{\rm AA}(\sqrt{s})$ when
going from 2.76 to 5.02\,TeV, in
both data and calculations. The slight increase in the $\Upsilon(1S)$ regeneration, which is subleading
at current energies, is expected to become more significant at collision energies beyond $\sim$10\,TeV.

As pointed out in Ref.~\cite{Rapp:2017chc}, the decreasing $Y$ excitation functions are markedly different
from their $J/\psi$ counter-parts, despite the comparable [or even larger] binding energy of the $\Upsilon(2S)$
[$\Upsilon(1S)$]. This lends considerable support to the overall picture of quarkonium kinetics developed
over the last decade. The relatively large uncertainty in the CNM effects at RHIC clearly calls for an
improved measurement in $p$-A $d$-A collisions at these energies (interesting effects have also been observed
in $p$-Pb at the LHC~\cite{Scomparin:2017pno,Abelev:2014oea,Atlas:2015conf,Chatrchyan:2013nza,Aaij:2014mza}). At face value, the $\Upsilon(1S)$ suppression measured by STAR in
Au-Au ($\sqrt{s}$=0.2\,TeV), which is very similar to the LHC datum at 2.76\,TeV, is not easily
understood from hot-medium
effects alone, while a larger CNM absorption at RHIC could offer a natural explanation for this observation.
An analogous situation is present for the $J/\psi$, where the larger CNM suppression at the lower SPS
energies ($\sqrt{s}$=0.017\,TeV), relative to RHIC, is an important ingredient to interpret
the energy dependence of the hot-medium effects~\cite{Rapp:2017chc}.
In addition, a more precise measurement of the Y excited states at RHIC would go a
long way in improving estimates of the $\Upsilon(2S)$ and $\Upsilon(3S)$ dissociation energies.

%
\section{Conclusions}
\label{sec:concl}
In the present work we have refined our previous Boltzmann/rate-equation approach to bottomonium
transport in heavy-ion collisions. The improvements include the use of in-medium binding energies
and their consequences for dissociation mechanisms and pertinent rates, a lQCD-based equation
of state for the fireball evolution, correlation volume effects for regeneration reactions, and
$B$-meson resonance states appearing close to $T_{pc}$ affecting the regeneration transport parameter.
In this way, the approach has been brought to the same level as employed before for charmonia,
and thus enables interpretations of bottomonium data on an equal footing. In particular, the role
of regeneration contributions, which are essential for charmonia at the LHC, is a priori less obvious
for bottomonia. In an attempt to augment possible signatures of those, we extended our calculations
of the centrality dependence of inclusive bottomonium yields to transverse-momentum spectra and
elliptic flow.

Overall, our improved approach allows for a fair description of existing $\Upsilon(1S)$,
$\Upsilon(2S)$ and $\Upsilon(3S)$ observables at RHIC and the LHC, including new data released
from both facilities very recently. We have found that the suppression level of the
$\Upsilon(1S)$ $R_{\rm AA}$ at the LHC has a significant sensitivity to the in-medium binding
energy used in the calculations and thus can, in principle, serve as a quantitative measure of
the screening of the heavy-quark potential in the QGP.
A similar sensitivity at RHIC energies requires a better control over the CNM effects.
At this point, the LHC data suggest a rather strong heavy-quark potential which supports
$\Upsilon(1S)$ states out to temperatures of $T$$\simeq$500\,MeV.
Inelastic reaction rates can, of course, break up $\Upsilon(1S)$ states at temperatures well
below that. The strong suppression of the $\Upsilon(2S)$, on the other hand, suggests its melting
at much lower temperatures, around $T$$\simeq$$240$\,MeV, implying strong screening effects on
the linear (``confining") part of the heavy-quark potential in this
regime. We also found that the emergence of $B$-mesons near $T_{\rm pc}$, which lowers the
equilibrium limit of the $Y$ states, reduces the regeneration of the $\Upsilon(2S)$, which helps
in quantitatively describing the pertinent CMS data at both 2.76 and 5.02\,TeV.
Some tension persists between our results and the forward-rapidity $\Upsilon(1S)$ ALICE data in
Pb-Pb(2.76\,TeV) collisions, which show a stronger suppression than obtained from our calculations.
The regeneration contributions for both $\Upsilon(1S)$ and $\Upsilon(2S)$ come out at a rather
generic level of around $R_{\rm AA}$$\simeq$\,0.05-0.1 [smaller for the $\Upsilon(3S)$], across
centrality (for $N_{\rm part}\gsim100$),
rapidity and collision energy (even down to RHIC energies). This is mostly a consequence of the
canonical limit, \ie, small open-bottom cross sections which limit the number of $b\bar b$ pairs
to either zero or one in a given fireball.
In the calculations of transverse-momentum spectra, the 3-momentum dependence of the dissociation
rates tends to produce a decrease of the primordial $R_{\rm AA}(p_T)$, which, however, is
counter-balanced by formation time effects at high $p_T$. For the regeneration component, the
inclusion of realistic $b$-quark spectra, taken from Langevin transport simulations which do not
kinetically equilibrate, turns out to be significant. Even though the coalescence contribution is not
large, a thermal blast-wave approximation for regenerated bottomonia produces a low-$p_T$ enhancement
in the $R_{\rm AA}(p_T)$ which is disfavored by the CMS data (in contrast to the $J/\psi$ case, where
a marked low-$p_T$ enhancement is observed). On the other hand, using the transport $b$-quark spectra,
the regeneration component generates a mild maximum structure in the $\Upsilon(1S)$ $R_{\rm AA}(p_T)$
around $p_T\lsim10$\,GeV, which is consistent with experiment. Our predictions for $Y$ elliptic
flow do not exhibit significant discrimination power between primordial and regeneration mechanisms.
However, we predict a factor of $\sim$2 larger total $v_2$ for the $\Upsilon(2S)$ than for the
$\Upsilon(1S)$, since the inelastic reactions for the former remain active to lower temperatures.
This should be helpful in either refuting or corroborating the sequential melting and regeneration,
as opposed to, \eg, statistical production of both particles at the same temperature.

Future work should focus on improving the precision of the approach on several fronts. Our initial
checks of various model components (pertaining to the bulk evolution, correlation volume, $Y$
formation time and $b$-quark thermalization) indicate a promising robustness of our results, in
particular with regards to connecting the observed level of $\Upsilon(1S)$ suppression to the
screening of the fundamental QCD force in the QGP. However, the interplay of
the early bulk medium evolution with quantum effects in the $b\bar b$ wave package deserves
further studies~\cite{Blaizot:2015hya,Katz:2015qja,Brambilla:2016wgg,Kajimoto:2017rel}.
This also applies to nonperturbative interactions in the $Y$ dissociation
mechanisms (\eg, by using explicit $T$-matrix interactions), which, after all, play a central
role in understanding the strong coupling of individual heavy quarks diffusing through the QGP.
These developments will improve our understanding of the systematic errors in the present results
and enable a more controlled assessment of the modifications of the fundamental QCD
force in the QGP.

\acknowledgments
We thank X.~Zhao and A.~Emerick for helpful discussions, and
J.~Fox for his contributions in the early stages of this work.
This work has been supported by the U.S.~National Science Foundation under Grant No.~PHY-1614484.
M. He was supported by NSFC Grant No. 11675079.\\

\vspace{2cm}

\begin{appendix}
\section{Inelastic Bottomonium Cross Sections}
\label{app1}
In this appendix we briefly recollect the expressions used for the cross sections
for the inelastic bottomonium reactions with quarks and gluons.

For gluo-dissociation, $g+Y\to b+{\bar b}$, we employ the cross sections derived from the
operator product expansion for a Coulombic bound state by Bhanot and Peskin~\cite{Peskin:1979},
\begin{equation}
\sigma_{Yg\rightarrow b\bar{b}} = \frac{r_0}{m_b} g_{Y}(x)
\label{sigma_BP}
\end{equation}
where $r_0$ is the ground-state radius and
\begin{eqnarray}
\label{coeff_BP}
g_{Y}(x)=
\begin{cases}
\frac{2}{3}\pi\left(\frac{32}{3}\right)^2 \frac{\left(x-1\right)^{\frac{3}{2}}}
{x^5} \hspace{0.9cm}&{\rm for}~\Upsilon(1S)
\cr
\frac{2}{3}\pi\left(\frac{32}{3}\right)^2 \frac{16\left(x-1\right)^{\frac{3}{2}}\left(x-3\right)^{2}}
{x^7} \hspace{0.3cm}&{\rm for}~\Upsilon(2S)
\cr
\frac{2}{3}\pi\left(\frac{32}{3}\right)^2 \frac{4\left(x-1\right)^{\frac{1}{2}}\left(9x^2-20x+12\right)}
{x^7} \hspace{0.3cm}&{\rm for}~\chi(1P)
\end{cases}
\nonumber\\
\end{eqnarray}
where $x=k_0/E_B$ and $k_0=\frac{s-m_{Y}^2-m_{g}^2}{2m_{Y}}$ is the incident gluon energy in the
quarkonium for a center-of-mass energy squared:
\begin{equation}
s=(p_Y^{(4)}+p_{g}^{(4)})^2=m_{Y}^2+m_{g}^2+2\omega_{Y}\omega_{g}-2\vec{p}_Y\cdot\vec{p}_{g} \ ;
\end{equation}
$p_Y^{(4)}=(\omega_Y,\vec{p}_Y)$ and $p_{g}^{(4)}=(\omega_{g},\vec{p}_{g})$ denote the 4-momenta of
the incoming bottomonium and outgoing gluon, respectively.
The color-Coulomb binding energy and radius follow the hydrogen form,
\begin{eqnarray}
E_0&=&\left(\frac{N^2-1}{2N}\alpha_s\right)^2 \frac{m_b}{4}=\left(\frac{2\alpha_s}{3}\right)^2 m_b
\label{EB_coul}
\\
r_0&=&\frac{2}{m_b\alpha_s}\left(\frac{2N}{N^2-1}\right)=\frac{3}{2m_b\alpha_s}
\label{EBr_coul}
\end{eqnarray}
which slightly differs from the large-$N_c$ limit expressions
\begin{eqnarray}
E_0&=&\left(\frac{N}{2}\alpha_s\right)^2 \frac{m_b}{4}=\left(\frac{3\alpha_s}{4}\right)^2 m_b
\label{EBln_coul}
\\
r_0&=&\frac{2}{m_b\alpha_s}\left(\frac{2}{N}\right)=\frac{4}{3m_b\alpha_s}
\label{EBrln_coul}
\end{eqnarray}
underlying the coefficients in Eq.~(\ref{coeff_BP}).

In previous work~\cite{Emerick:2011xu}, the binding energies in the SBS were taken at the vacuum
values defined by $E_B=2m_{B}-m_{Y}=1.1[0.54]$\,GeV for the $\Upsilon(1S)$ [$\Upsilon(2S)$]
with the coupling constant fixed via the ground-state expression, Eq.~(\ref{EB_coul}).
We here adopt an alternative treatment, which we believe to be more realistic, by eliminating
the $r_0$ dependence and rewriting the cross section as the first power in $\alpha_s$ times a
factor involving the binding energies using the relation of the bottomonium radius to their binding
energies from the hydrogen model expression. With
$\sigma\sim\frac{r_0}{m_b}\sim\alpha_s\cdot r_0^2\sim\frac{\alpha_s}{m_b E_B}$,
the cross sections become
\begin{equation}
\sigma_{Yg\rightarrow b\bar{b}} = \frac{2\alpha_s}{3m_bE_0} g_{Y}(x)
\label{sigma-eq-2}
\end{equation}

Since the in-medium binding energies of the $Y$ states are not necessarily small compared to the
temperature, we also include a phenomenological treatment of interference effects for the
quasifree reaction rates. Starting from the original expression for the quasifree cross section,
schematically given by
$\sigma_{Yp\rightarrow b\bar{b}p}(s)\simeq \int \frac{\mathrm{d}\sigma_{Yp\rightarrow b\bar{b}p}(s,t,u)}{\mathrm{d}t} \mathrm{d}t$,
(which includes the $t$-channel from both quark and gluon as partons, $s$- and $u$-channels from gluon as parton, and their mixed terms)
the interference correction is implemented as
\begin{eqnarray}
\sigma_{Yp\rightarrow b\bar{b}p}(s)\simeq \int \mathrm{d}t \left(\frac{\mathrm{d}\sigma_{Yp\rightarrow b\bar{b}p}(s,t,u)}{\mathrm{d}t}\right)
\nonumber\\
\hspace{3cm}
\times\left(1-\frac{\mathrm{sin}(\sqrt{-t}r)}{\sqrt{-t}r}\right)
\label{sigma_qf}
\end{eqnarray}
(which also influences the $p_T$ dependence).
To relate the dependence on the coordinate $r$ to the in-medium binding energy, $E_B$, we
utilize the Coulomb relations for the two scales $p\sim m_b\alpha_s\sim\frac{1}{r}$ and
$E_B\sim m_b\alpha_s^2$.  One can then either re-express the mass or coupling constant leading to
$E_B\sim \frac{\alpha_s}{r}$ or $E_B\sim \frac{1}{m_br^2}$, respectively. To check which relation is
more realistic, especially in the presence of the nonperturbative string term, we determine the
numerical coefficients for each option from the vacuum binding energy and radius of the
$\Upsilon(1S)$ and then inspect the pertinent prediction for the $\Upsilon(2S)$.
With $\alpha_s$$\simeq$0.3, $m_b$$\simeq$5\,GeV and a vacuum binding and radius of
$E_B(\Upsilon(1S))$$\simeq$\,1\,GeV and $r(\Upsilon(1S))$$\simeq$\,0.2\,fm, we find
$E_B$$\simeq$3.3$\frac{\alpha_s}{r}$ or $E_B$$\simeq$5$\frac{1}{m_br^2}$ for the two options above.
Using $r(\Upsilon(2S))$$\simeq$0.5\,fm then gives $E_B(2S)$$\simeq$3.3$\frac{\alpha_s}{r}$$\simeq$0.4\,GeV
or $E_B(2S)$$\simeq$5$\frac{1}{m_br^2}\simeq$0.16\,GeV. Since the first replacement is closer to the
empirical $\Upsilon(2S)$ binding of $\sim$0.54\,GeV, we will adopt it in our calculations.

\section{Feeddowns and $pp$ baseline cross sections for excited states}
\label{appf}
In this appendix we detail our implementation for updated feeddown fractions. Starting from
Ref.~\cite{Andronic:2015wma}, the direct $\Upsilon(1S)$ cross section is 70\,\%(50\,\%) at low
(high) $p_T$, on average 67\,\%. The feeddowns to the $1S$ state are approximately 17\,\% from $1P$,
9\,\% from $2S$, 1\,\% from $3S$ and 6\,\% from $2P$ and $3S$ together. The main change from the
previous work of Ref.~\cite{Emerick:2011xu} is from the $\chi_b(nP)$ states which now contribute
less at low $p_T$~\cite{Aaij:2014caa}; cf.~Table.~\ref{tab-fd}. Because of the newly included
explicit treatment of the $\Upsilon(3S)$ state, we implement detailed feeddown fractions discussed
below for the TBS calculation. For the SBS calculation, we only include $\Upsilon(1S)$, $\Upsilon(2S)$
and $\chi_b(1P)$ states. A correction from explicitly including $\Upsilon(3S)$ and $\chi_b(2P)$
states in the SBS calculation would result in ca.~$\sim$10\,\% more regeneration for the $\Upsilon(2S)$
and a negligible contribution to the $\Upsilon(1S)$.

From several experimental
data~\cite{Khachatryan:2010zg,Aad:2012dlq,Chatrchyan:2013yna,LHCb:2012aa,Aaij:2013yaa,Adam:2015rta,Aaij:2014nwa},
we conclude $\sigma_{2S}\simeq0.33\sigma_{1S}^{\rm tot}$ and $\sigma_{3S}\simeq0.15\sigma_{1S}^{\rm tot}$.
With the branching ratios $Br(2S\rightarrow1S)=26.7\,\%$ and $Br(3S\rightarrow1S)=6.6\,\%$, we obtain
feeddown fractions of $Fd(2S\to 1S)$=$0.33\cdot26.7\,\%=8.8\,\%$ and $Fd(3S \to 1S)$=$0.15\cdot6.6\,\%=0.99\,\%$,
which are consistent with Ref.~~\cite{Andronic:2015wma}.

The cross section ratio of $\frac{\sigma(\chi_{b2}(1P))}{\sigma(\chi_{b1}(1P))}=0.85$ from
Ref.~\cite{Khachatryan:2014ofa} indicates a smaller production of the heavier $1P$ state.
An assumption for the lighter $\chi_{b0}(1P)$ with a ratio
$\frac{\sigma(\chi_{b0}(1P))}{\sigma(\chi_{b1}(1P))}\simeq1.5$ gives an approximate
branching ratio for the $1P$ state of
\begin{eqnarray}
Br(1P\rightarrow1S)
&\simeq&\frac{Br(\chi_{b0}(1P)\rightarrow1S)\cdot1.5}{3.35}
\nonumber\\
&&+\frac{Br(\chi_{b1}(1P)\rightarrow1S)\cdot1.0}{3.35}
\nonumber\\
&&+\frac{Br(\chi_{b2}(1P)\rightarrow1S)\cdot0.85}{3.35}
\nonumber\\
&=&\frac{1.8\cdot1.5+33.9\cdot1.0+19.1\cdot0.85}{3.35}\,\%
\nonumber\\
&=&15.8\,\% \ .
\end{eqnarray}
We estimate a $1P$ cross section of $\frac{17\,\%}{15.8\,\%}\simeq1.08$ of the inclusive $1S$
cross section, $\sigma_{1P}\simeq1.08\sigma_{1S}^{tot}$.
We estimate the fraction $\frac{\sigma_{2P}}{\sigma_{1P}}\simeq0.8$ from Ref.~\cite{Vogt:2010aa}
so that $\sigma_{2P}\simeq0.864\sigma_{1S}^{tot}$.

Assuming the same ratio between different $\chi_b$ states for the $2P$ multiplet,
we estimate
\begin{eqnarray}
Br(2P\rightarrow1S)
&\simeq&\frac{Br(\chi_{b0}(2P)\rightarrow1S)\cdot1.5}{3.35}
\nonumber\\
&&+\frac{Br(\chi_{b1}(2P)\rightarrow1S)\cdot1.0}{3.35}
\nonumber\\
&&+\frac{Br(\chi_{b2}(2P)\rightarrow1S)\cdot0.85}{3.35}
\nonumber\\
&=&\frac{0.9\cdot1.5+10.8\cdot1.0+8.1\cdot0.85}{3.35}\,\%
\nonumber\\
&=&5.7\,\%.
\end{eqnarray}
and
\begin{eqnarray}
Br(2P\rightarrow2S)
&\simeq&\frac{Br(\chi_{b0}(2P)\rightarrow2S)\cdot1.5}{3.35}
\nonumber\\
&&+\frac{Br(\chi_{b1}(2P)\rightarrow2S)\cdot1.0}{3.35}
\nonumber\\
&&+\frac{Br(\chi_{b2}(2P)\rightarrow2S)\cdot0.85}{3.35}
\nonumber\\
&=&\frac{4.6\cdot1.5+19.9\cdot1.0+10.6\cdot0.85}{3.35}\,\%
\nonumber\\
&=&10.7\,\% \ .
\end{eqnarray}
The latter is almost the same as $Br(3S\rightarrow2S)=10.6\,\%$.
These estimates result in feeddown fractions
\begin{eqnarray}
Fd(2P\rightarrow1S)
&=&\frac{\sigma_{2P}Br(2P\rightarrow1S)}{\sigma_{1S}^{tot}}=0.864\cdot5.7\,\%
\nonumber\\
&=&4.9\,\%
\end{eqnarray}
and
\begin{eqnarray}
Fd(2P\rightarrow2S)
&=&\frac{\sigma_{2P}Br(2P\rightarrow2S)}{\sigma_{2S}}=\frac{0.864\cdot10.7\,\%}{0.33}
\nonumber\\
&=&28\,\% \ ,
\end{eqnarray}
consistent with Ref.~\cite{Andronic:2015wma}.

The above estimate furthermore leads to a total cross section for the higher excited $3S$ and $2P$ states of
about 1.014$\sigma_{1S}^{\rm tot}$. With $Br(3S\rightarrow1S)=6.6\,\%$ and $Br(2P\rightarrow1S)=5.7$, we have
\begin{eqnarray}
&&\hspace{-1.5cm}Fd(3S+2P\rightarrow1S)
\nonumber\\
&=&\frac{\sigma_{3S}Br(3S\rightarrow1S)+\sigma_{2P}Br(2P\rightarrow1S)}{\sigma_{1S}^{tot}}
\nonumber\\
&=&0.15\cdot6.6\,\%+0.864\cdot5.7\,\%=5.9\,\%
\end{eqnarray}
and, with
$Br(3S\rightarrow2S)=10.6\,\%$ and $Br(2P\rightarrow2S)=10.7$,
\begin{eqnarray}
&&\hspace{-1.5cm}Fd(3S+2P\rightarrow2S)
\nonumber\\
&&=\frac{\sigma_{3S}Br(3S\rightarrow2S)+\sigma_{2P}Br(2P\rightarrow2S)}{\sigma_{2S}}
\nonumber\\
&&=\frac{0.15\cdot10.6\,\%+0.864\cdot10.7\,\%}{0.33}=33\,\% \ ,
\end{eqnarray}
consistent with Ref.~\cite{Andronic:2015wma}.

Since the branching ratios from $3S$ or $2S$ to $1P$ are all smaller than 1\,\%, we neglect these
feeddown channels.
\end{appendix}


\end{document}